\documentclass[preprint2]{aastex}
\usepackage{graphicx}
\def\etal{{\it et al}}

\def\P3hat{{\mathaccent 94 P}_3}

\def\eg{{\it e.g.}}
\def\ie{{\it i.e.}}
\def\cf{{\it cf.}}
\def\aap{A\&A}
\def\aaps{A\&A Suppl.}
\def\apj{Ap.J.}
\def\aj{A.J.}
\def\mnras{M.N.R.A.S.}

\def\apjs{{\it Ap.J. Suppl.}}
\def\arep{{\it Astr. Repts.}}
\newcount\notenumber
\def\clearnotenumber{\notenumber=0}
\def\note{\advance\notenumber by1 \footnote{$^{\the\notenumber}$}}
\clearnotenumber
\usepackage{rotating}   
\begin{document}

\title{Toward an Empirical Theory of Pulsar Emission.  IX.  On the Peculiar Properties and Geometric Regularity of Lyne \& Manchester's ``Partial Cone'' Pulsars}

\author{Dipanjan Mitra\altaffilmark{1}}
\affil{National Astronomy and Ionosphere Center, Arecibo Observatory, HC3 Box 53995, PR 00612}
\and
\author{Joanna M. Rankin\altaffilmark{2}}
\affil{Sterrenkundig Instituut ``Anton Pannekoek,'' University of Amsterdam, Science Park 904, 1098 XH Amsterdam, Netherlands}

\altaffiltext{1}{National Centre for Radio Astrophysics, Ganeshkhind, Pune 411 007 India: 
dmitra@ncra.tifr.res.in}
\altaffiltext{2}{Physics Department, 82 University Place, University of Vermont, Burlington, VT 05405 USA: Joanna.Rankin@uvm.edu}
\date{Released 2004 Xxxxx XX}

\label{firstpage}

\begin{abstract}
Lyne \& Manchester (1988) identified a group of some 50 pulsars they called ``partial
cones'' which they found difficult to classify and interpret. They were notable for their
asymmetric average profiles and asymmetric polarization position-angle (PPA)
traverses, wherein the steepest gradient (SG) point fell toward one edge of the total 
intensity profile.  Over the last two decades, this population of pulsars has raised 
cautions regarding the core/cone model of the radio pulsar-emission beam 
which implies a high degree of order, symmetry and geometric regularity.  

In this paper we reinvestigate this population ``partial cone'' pulsars on the basis 
of new single pulse polarimetric observations of 39 of them, observed with the 
Giant Meterwave Radio Telescope in India and the Arecibo Observatory in Puerto 
Rico.  These highly sensitive observations help us to establish that most of these 
``partial cones'' exhibit a core/cone structure just as did the ``normal'' pulsars studied 
in the earlier papers of this series.  In short, we find that many of these ``partial cones'' 
are partial in the sense that the emission above different areas of their polar caps can 
be (highly) asymmetric.  However, when studied closely we find that their emission 
geometries are overall identical to core/double cone structure encountered earlier---that 
is, with specific conal dimensions scaling as the polar cap size.  

Further, the ``partial cone'' population includes a number of stars with conal single 
profiles that are asymmetric at meter wavelengths for unknown reasons (\eg, like those 
of B0809+74 or B0943+10).  We find that aberration-retardation appears to play a 
role in distorting the core/cone emission-beam structure in rapidly 
rotating pulsars.  We also find several additional examples of highly polarized pre- and 
postcursor features that do not appear to be generated at low altitude but rather at high 
altitude, far from the usual polar fluxtube emission sites of the core and conal radiation.
\end{abstract}


\section*{I. Introduction} 
\label{sec:intro}
Early investigators were impressed by the symmetrical emission profiles of 
many pulsars (\eg, Backer 1976) and that these, together with their antisymmetic 
polarization-angle (hereafter PPA) traverses (Radhakrishnan \& Cooke 1969), 
appeared to reflect their polar cap emission geometry directly.  Indeed, the major 
purpose and overall theme of this ``Empirical Theory'' series has been that of 
demonstrating the geometric orderliness of most pulsar emission.  Species of 
profiles were defined in Paper I (\eg, Rankin 1983a; see References for further 
series numbers).  The geometric regularities of core components in relation to 
the polar cap was introduced in Paper IV, and Paper VI then presented a full 
quantitative analysis of pulsar emission geometry using the core-double cone 
model of some 200 stars.\footnote{Commonalities in terms of spectral behavior 
and modulation were studied in Papers II and III.  Three other numbers (Papers 
V, VII and VIII) have respectively discussed circular polarization, radio-to-frequency 
mapping, and edge depolarization.  The results of Paper VI were sufficiently 
surprising that several groups carried out critical studies or independent analyses 
(Bhattacharya \& van den Heuvel 1991, Gil \etal\ 1993; Kramer \etal\ 1994; Mitra 
\& Deshpande 1999), and the core/double cone model of pulsar emission profiles 
was fully vetted.}

Certain ``difficult'' pulsars raised the possibility, even from Paper I, that the polar 
cap emission from some pulsars might be very asymmetric.  Only a few possible examples of one-sided 
``triple'' profiles were mentioned, however, given the difficulty of demonstrating that 
``double'' profiles might sometimes present only a single component.  The term 
``partial cones'' was then introduced by Lyne \& Manchester (hereafter L\&M) in
their 1988 radio-pulsar beamform study to describe a group of profiles that were 
not easily classified as falling into one of their cone- or core-dominated categories.  
They confirmed that the majority of their 200 or so pulsars showed a highly ordered, 
roughly symmetric, quantitatively consistent beam geometry.  By contrast, their largish 
residuum of pulsars with unclassifiable, asymmetric profiles were dubbed ``partial 
cones'', because a number (\eg, B0540+23) had asymmetric profiles reminiscent of 
one side of a classic conal double profile (\eg, B0525+21).  This aberrant group of 
pulsars raised strong cautions---indeed, if some 20-30\% of all profiles cannot be 
classified in terms of cores and cones, is this model not itself suspect?  Given these 
patently inscrutable profiles, often with puzzling asymmetries, they left open the 
possibility that a ``patchy'' pattern of components resulted from ``hot spots'' on the 
polar cap.  

We thus reemphasize that L\&M's work and ours provide highly compatible 
geometrical results for a majority of pulsars in our largely common population, so 
the differing interpretations of the two analyses turn importantly on L\&M's group 
of ``partial cone'' pulsars. 

No further systematic study of L\&M's ``partial cones'' has been carried out over the 
last two decades, so this group of some 60 pulsars remains in many workers minds 
as strong evidence for unsystematic pulsar beaming and perhaps polar-cap ``hot 
spots''.  L\&M's study was based solely on average profiles, most all of them at 
meter wavelengths, and the general weakishness of this population also limited the 
quality of their profiles.  Now, however, not only are much more sensitive observations 
often {possible---and at both higher and lower frequencies---}but pulse-sequence 
(hereafter PS) polarimetry has been carried out for a large fraction of these ``partial 
cones''.  

Surely we concur that many of L\&M's ``partial cone'' pulsars present particular 
difficulties of interpretation.  We now know with certainty that some pulsars do 
illuminate their polar caps very asymmetrically or episodically (Rankin \etal\ 2006a)---producing 
lopsided or distorted profiles---but when investigated in detail these stars also 
exhibit orderly profile dimensions in relation to the polar cap.  

A further set of pulsars with conal single profiles, we now know from detailed studies, 
very often exhibit highly asymmetric profiles (\eg, B0943+10; Deshpande \& Rankin 
2001) despite strong evidence that their emission cones are produced by subbeam 
carousels rotating through our sightline.  Aberration/retardation (hereafter A/R) effects 
have been identified in a number of slower pulsars (\eg, Blaskiewicz \etal\ 1991), 
and clearly may have strong effects in faster pulsars.  Also, recent researches have 
revealed highly polarized profile features---\eg, the ``precursors'' in pulsars B0943+10 
and B1822--09 (see Backus \etal\ 2010)---and even the entire profiles of particular 
stars (\eg, B0656+14, Weltevrede \etal\ 2006a) that exhibit such dissonant properties 
that we are forced to question whether some new non-core/cone emission process 
is entailed! 

Generally the average profile of a radio pulsar has a characteristic steep outer 
edge, which apparently reflects the emitting region along the boundary of the 
``open'' magnetosphere (or polar flux tube) adjacent to the closed field region.   
For a large number of pulsars the polarization-position angle (PPA) across the 
pulse profile is seen to execute a smooth `S-shaped' curve, which according to the 
rotating-vector model (RVM) proposed by Radhakrishnan \& Cooke (1969) is taken 
as evidence for emission arising within the polar flux tube and centered around 
the magnetic axis.  Within the RVM the steepest-gradient (hereafter SG) point (or 
the point of inflection) of the `S-shaped' curve is interpreted as the plane containing 
the magnetic dipole axis, and is often located towards the center of the profile.

Most profiles, however, tend to be asymmetric with the central core component of 
triple or five-component forms seen to lag the centers of their conal-component pairs.
Studies by Malov \& Suleymanova (1998), Gangadhara \& Gupta (2001), Gupta \& 
Gangadhara (2003), Mitra \& Li (1999), and Dyks \etal\ (2004) demonstrate that 
aberration/retardation (hereafter A/R) effects arising due to emission from a finite 
height within the pulsar magnetosphere can give rise to the observed profile 
asymmetries.  Once this A/R effect is properly taken into account, the emission can 
be understood as nested conal emission. 

Partial cones were identified by L\&M as pulsars with profiles having one steeply 
rising edge and another slowly falling edge.  Or, as stars where the steepest 
gradient point of the PPA traverse is located towards one edge of the profile. 
Identification of partial cones thus requires unambiguous determination of the SG 
point of the PPA swing with respect to its total intensity profile.  It is often difficult 
to discern the character of the PPA traverse using only average-profile polarimetry, 
as did L\&M.  This is particularly so due to the presence of the ``orthogonal'' 
polarization modes (hereafter OPMs), which indeed are not always orthogonal 
(\eg, Ramachandran \etal\ 2004).  Departures from modal orthogonality tend to 
produce complex average PPA behaviors, because their relative power often 
varies strongly with pulse longitude, and these can in turn lead to serious 
misinterpretations of a pulsar's PPA traverse.  Hence, polarimetry of individual 
pulses is necessary to distinguish the OPMs and correctly assess the geometrical 
bases of the PPA swings (\eg, Gil \& Lyne 1995).

L\&M suggested that partial cones are perhaps pulsars where only part of the 
polar cap is illuminated.  ``Partial cones'' surely do present difficulties for 
the core-cone beam model.  However, the ubiquity of subpulse modulation 
(\eg, Weltevrede \etal\ 2006b, 2007), implying that cones are generally produced 
by rotating subbeam ``carousels'', also raises strong contradictions in any appeal to 
``hot spots''.\footnote{Even for the recent Karasterigiou \& Johnston (2007) hybrid 
model, where the conal emission ring is illuminated in patches.} 

\begin{figure}
\begin{center}
\includegraphics[width=78mm,angle=0.]{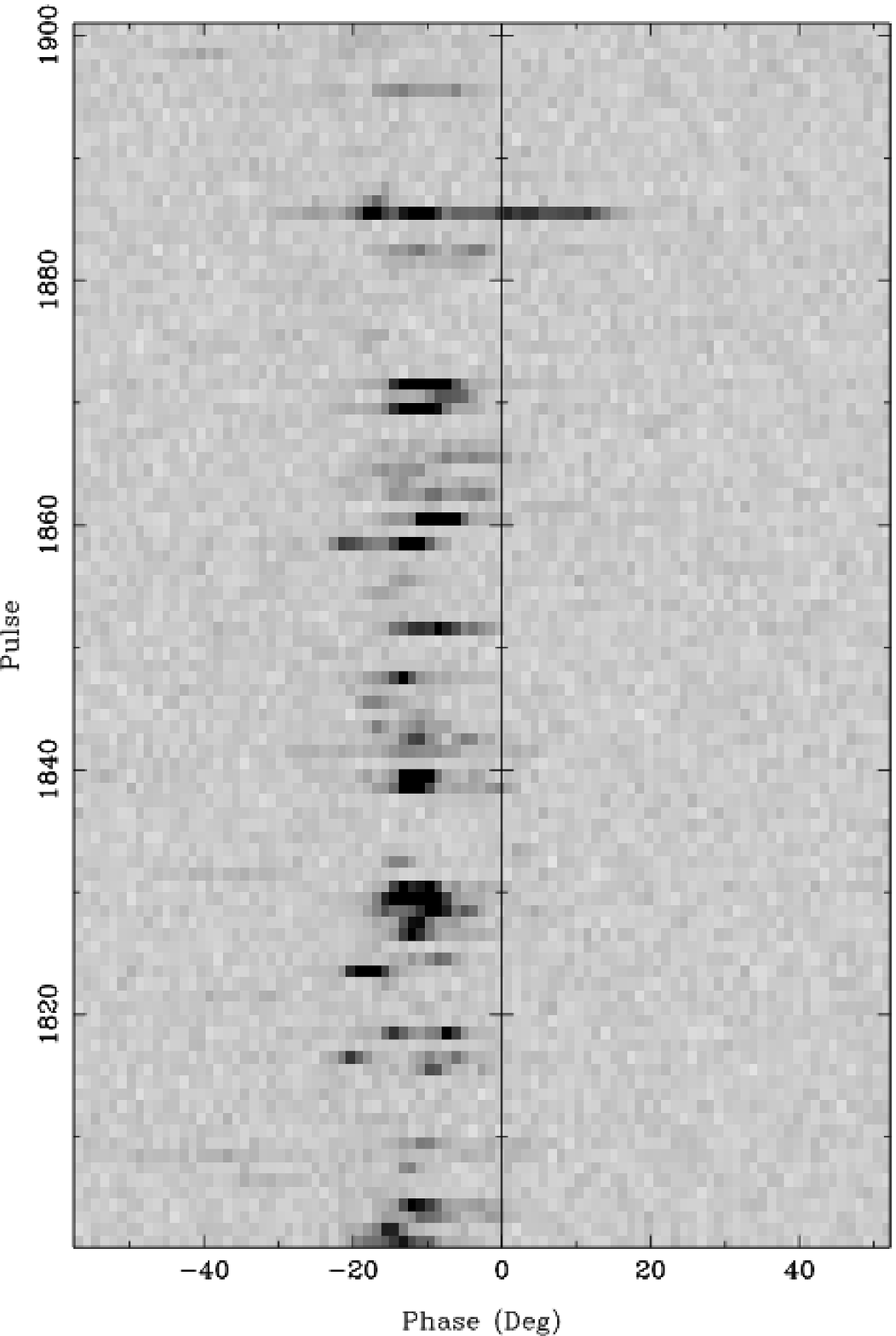}
\end{center}
\end{figure}
\begin{figure}
\begin{center}
\includegraphics[width=78mm,angle=-90.]{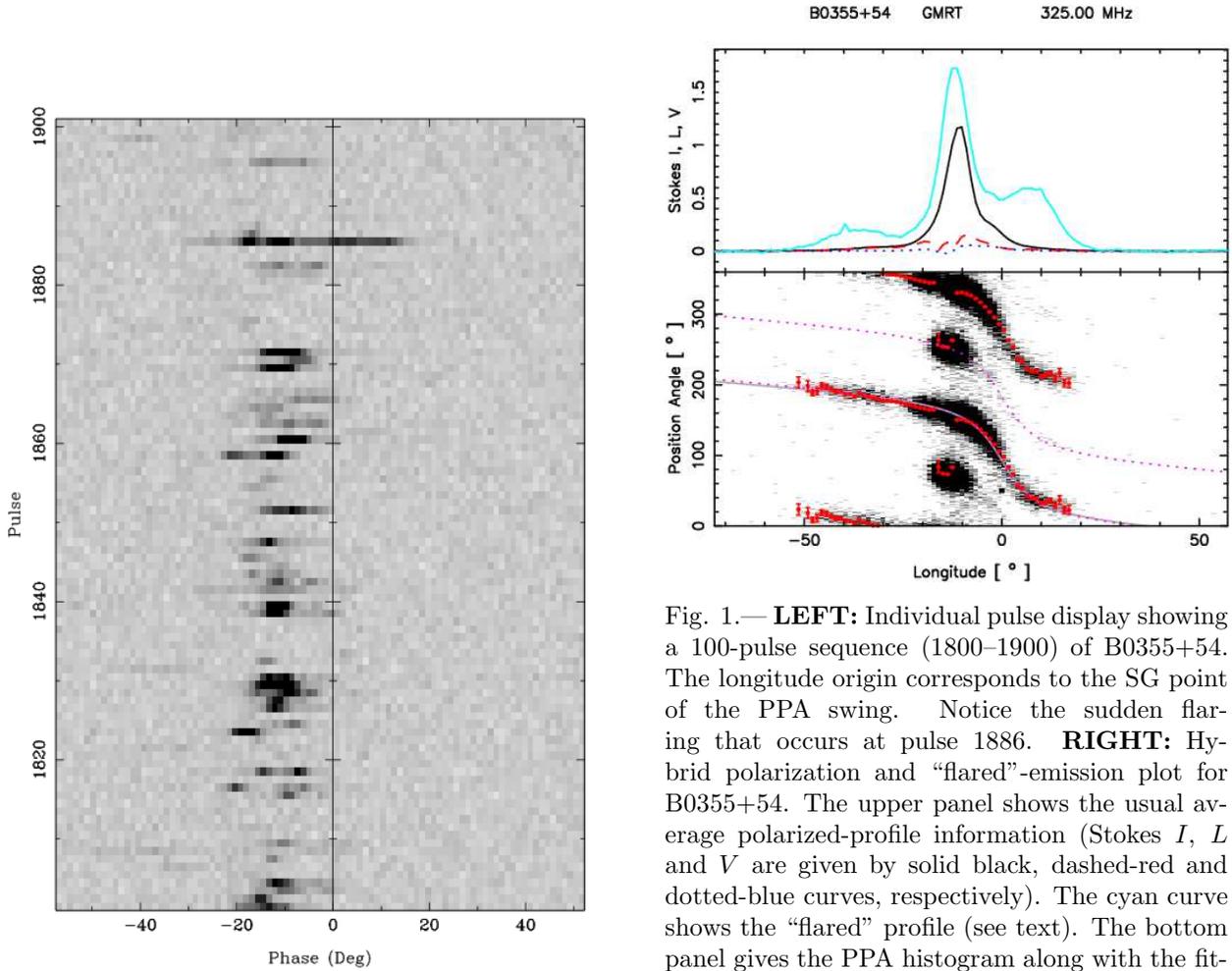}
\caption{{\bf LEFT:} Individual pulse display showing a 100-pulse sequence 
(1800--1900) of B0355+54. The longitude origin corresponds to the SG point of 
the PPA swing.  Notice the sudden flaring that occurs at pulse 1886. {\bf RIGHT:} 
Hybrid polarization and ``flared''-emission plot for B0355+54.  The upper panel 
shows the usual average polarized-profile information (Stokes $I$, $L$ and $V$ 
are given by solid black, dashed-red and dotted-blue curves, respectively).  The 
cyan curve shows the ``flared'' profile (see text).  The bottom panel gives the PPA 
histogram along with the fitted RVM curves; one (dotted magenta, shown for both 
polarization modes) reflects the geometrical models in Table~\ref{tab3}, whereas 
the other (solid grey) corresponds to the fitting results in Table~\ref{tab5}.}
\label{fig1}
\end{center}
\end{figure}

Below we will argue that the emission from a number of ``partial cone'' pulsars 
is indeed partial in the sense that their emission is very asymmetric with respect 
to the longitude of the magnetic axis.  However, we also find that the ``partial cone'' 
pulsars are completely regular in terms of their overall core/double cone emission 
geometry when these asymmetries are accounted for.  Or, said differently, we do 
not yet understand why pulsar radiation is in some cases so beautifully symmetric 
and in other pulsars so utterly asymmetric about the magnetic axis.  However, in 
most cases it is possible to discern {\em some} weak or occasional emission even 
from the dimmer parts of the polar cap---and the core/double cone geometry of this 
emission is identical to that of ``normal'' pulsars.

We then proceed as follows:  \S II describes our GMRT and Arecibo observations, 
and in \S III we discuss our analyses of those pulsars exhibiting ``flared'' or 
episodic emission.  \S IV presents the large subset of ``partial cone'' pulsars with 
narrow conal profiles.  In \S V we introduce new analyses of pulsars with clear 
signatures of A/R in their emission, and in \S VI we discuss the several stars with 
apparently aberrant polarization---components with nearly complete linear and flat 
PPA traverses.  \S VII then presents our overall geometrical analyses, and \S VIII 
gives a summary and discussion of our results.  The Appendix then discusses the 
properties of L\&M's ``partial cone'' population individually.

\section*{II. Observations and Data Analysis} 
\label{sec:obs}
Our observations encompass of 39 of the 50 pulsars identified as ``partial-cone'' 
or likely ``partial-cone'' objects by L\&M (their tables 4 \& 5).  We have observed
these pulsars using the Giant Meterwave Radio Telescope (GMRT) at 325 MHz 
(P band) and the Arecibo (AO) instrument at P and/or L (1100-1700 MHz) band 
in full polarization.

The GMRT (Swarup \etal\ 1991) is an array of 30 45-m antennas, spread over 
a 25-km region 80 km north of Pune, India.  It is primarily an aperture-synthesis 
interferometer but can also be used in a phased-array configuration.  The GMRT 
operates at multiple frequencies (150, 235, 325, 610 and 1000 - 1450 MHz) 
and has a maximum bandwidth of 32 MHz, split into upper and lower sidebands 
of 16 MHz each.  At 325 MHz, which is the frequency of interest here, the feeds
are linearly polarized and converted to circulars using a hybrid.  Our observations
were carried out on 14 February 2006 and 26 October 2007 using the phased-array 
mode (Sirothia 2000; Gupta \etal\ 2000), in which the voltage signals of the upper 
sidebands from each antenna were first added coherently and then fed to the pulsar 
receiver.  The pulsar backends computed the auto- and cross-polarized power 
between the two circularly polarized signals, and these were finally recorded with 
a sampling time of 0.512 msec.  A suitable calibration procedure as described 
in Mitra \etal\ (2005) was applied to the observations to recover the calibrated 
Stokes parameters $I$, $Q$, $U$ and $V$.  The AO observations were carried 
out at both 327 and 1400 MHz in a manner very like that reported in Rankin \etal\ 
(2006a).

The calibrated Stokes parameters were used to compute the total linear
polarization $L (= \sqrt{ U^2 + Q^2}$) and the PPA $\chi (= 0.5 \tan^{-1}(U/Q)$) of 
the several pulse sequences (hereafter PSs).  Table~\ref{tab1} gives the various 
observational parameters for ``partial-cone'' pulsars.  Table~\ref{tab2} then 
reviews some of the properties of these ``partial cone'' pulsars.  

In a number of cases we have fitted the RVM to the PPA $\chi$ traverses using 
the Everett \& Weisberg (2001) convention as follows:
\begin{equation}
\chi= \tan^{-1}\left(\frac{\sin\alpha \sin(\varphi-\varphi_{\circ})}{\sin\xi
\cos\alpha+\cos\xi\sin\alpha\cos(\varphi-\varphi_{\circ})}\right)+\chi_{\circ}
\label{eq0}
\end{equation}
where $\alpha$ is the magnetic latitude, $\beta$ is the sightline impact angle, 
$\xi=\alpha+\beta$ is the sightline-circle radius, and $\chi_{\circ}$ and $\varphi_{\circ}$ 
are the PPA and longitude offsets.  The fits determined four parameters, 
$\alpha$ and $\beta$ as well as the two offsets above in terms of the overall PPA 
as a function of longitude.  As is usual and well known, the $\alpha$ and $\beta$ 
values were usually so poorly determined (large errors) and highly correlated 
(typically 98\%) that they were meaningless. We however use the fitting process to
determine the errors in the fitted parameters by evaluating the amount of change
required for a particular parameter (holding the other parameters fixed)
such that the minimum chi--square value increases by unity (see von Hoensbroech 
\& Xilouris 1997, Everett \& Weisberg 2001, Mitra \& Li 2004).  This way the PPA 
sweep rate $R$ [$=|\Delta \chi/\Delta \phi|_0 = \sin(\alpha)/\sin(\beta)$] and longitude 
offset ``steepest gradient'' (hereafter SG) point $\varphi_{\circ}$ were often well 
determined.  Table~\ref{tab5} gives these latter RVM fitted/computed values and 
their respective errors.  From this fitting exercise, the quantity $R$ is obtained for the 
geometric analyses that are summarized in Table~\ref{tab3} and discussed in \S VII.  
It is to be noted that the errors in $R$ quoted in Table~\ref{tab5} are obtained by 
further fitting linear slopes to the PPAs in restricted regions around the SG points.

Figure~\ref{fig1} gives an example of the polarization displays and fits used throughout 
the paper.   The upper panels show the usual average polarized profile information 
(Stokes $I$, $L$ and $V$ are given by solid black, dashed-red and dotted-blue 
curves, respectively), and the PPA is plotted twice in the lower panel for ease of 
viewing.  As we will see below, our geometric analyses will often provide values 
for $\alpha$ and $\beta$, and the resulting RVM-based PPA traverses are 
indicated in the lower panel by a pair of dotted magneta curves.  The light grey curve 
illustrates the effect of the above RVM-fitted parameters as obtained by fitting 
eq.~\ref{eq0} and given in Table~\ref{tab5}.  The ``flared'' total power profile is shown 
in the upper using a solid cyan curve (see \S III text).

\section*{III. ``Flared'' emission}
Single pulses of pulsars show a great deal of variety.  Generally, subpulses of 
varying intensity are seen to appear and disappear at various pulse longitudes, 
but when averaged together a stable pulse profile is formed.  However, this is 
not always so:  a few pulsars are known for their ``giant'' pulses---most famously 
the Crab pulsar---and in a few others occasional bright pulses can be so very 
strong that the profile form is unstable (\eg, B0656+14; see Weltevrede \etal\ 
2006b).  For a few other pulsars, ``episodic'' illumination has been observed that 
greatly emphasizes parts of a pulsar's profile at the expense of others (Rankin 
\etal\ 2006a).  For these reasons we thought it important to explore whether 
these effects could be active in some of L\&M's ``partial cone'' pulsars.   We 
therefore undertook analyses similar to those of Hankins \& Cordes (1981) and 
Nowakowski (1991).  Almost immediately, we discovered ``flaring'' effects in the 
single pulse emission of some of the ``partial cone'' pulsars.  

In the lefthand panel of Figure~\ref{fig1} we show a GMRT total-power pulse 
sequence (hereafter PS) of the ``partial cone'' pulsar B0355+54 (pulse \#s 
1800-1900).  Notice that most of the bright emission occurs around --10\degr\ 
longitude (where zero longitude corresponds to the SG point of the PPA 
traverse); however, one strong subpulse can be seen extending to +15\degr\ 
(pulse \#1886) and several other fainter subpulses can be discerned around 
--40\degr.  Obviously, this pulsar shows great dynamicity in its pulse-to-pulse 
fluctuations: the core varies dramatically in intensity, often disappearing entirely; 
the leading and trailing conal outriding components are only occasionally 
detectable; and overall the pulsar nulls for some 30\% of the time.  These 
occasional ``flares'' of the conal components are then remarkable---and we 
find that they are very rare in B0355+54---occurring only in 200 pulses within 
a PS of 13000 individual pulses.

We have searched for ``flared'' emission in the entire set of ``partial cone'' pulsars 
available to us.  We used a ``tunable'' window to detect sporadic emission in the 
fainter regions of the average profile where the intensity is close to the noise level.  
Each time the emission exceeded three times the noise level (averaged over the 
window), we marked that pulse and window as having ``flared'' and with adjacent 
windows computed the average ``flared'' profile.  We then repeated this process 
for different window sizes until the ``flared'' profile was stable over a range of 
window widths.  The righthand display of Fig.~\ref{fig1} gives an example of this 
``flared''-profile analysis for B0355+54, and the ``flared'' total power profile is shown 
using a solid cyan curve.  Obviously, this ``flared'' profile shows the contributions 
of the sporadic emission to the far edges of the profile, and it strongly suggests a 
three-component structure.  

We found evidence for ``flaring'' in about half the group of ``partial cone'' pulsars 
under study, and the full results are shown using displays similar to Fig.~\ref{fig1} 
in the Appendix.  Overall, we found little difference between the widths of the 
``flared'' profiles compared to the full discernible widths of the corresponding 
normal average profile; however, the structure was often more scrutable---and 
in some cases we used these ``flared'' widths in the geometric analyses given 
in Table~\ref{tab3}; see the Appendix for discussions of the analyses of the 
individual pulsars.  We also looked for periodicities in the ``flares'' and found 
no evidence for any regular repetitive behavior.  

\begin{figure}
\begin{center}
\includegraphics[height=78mm,angle=-90]{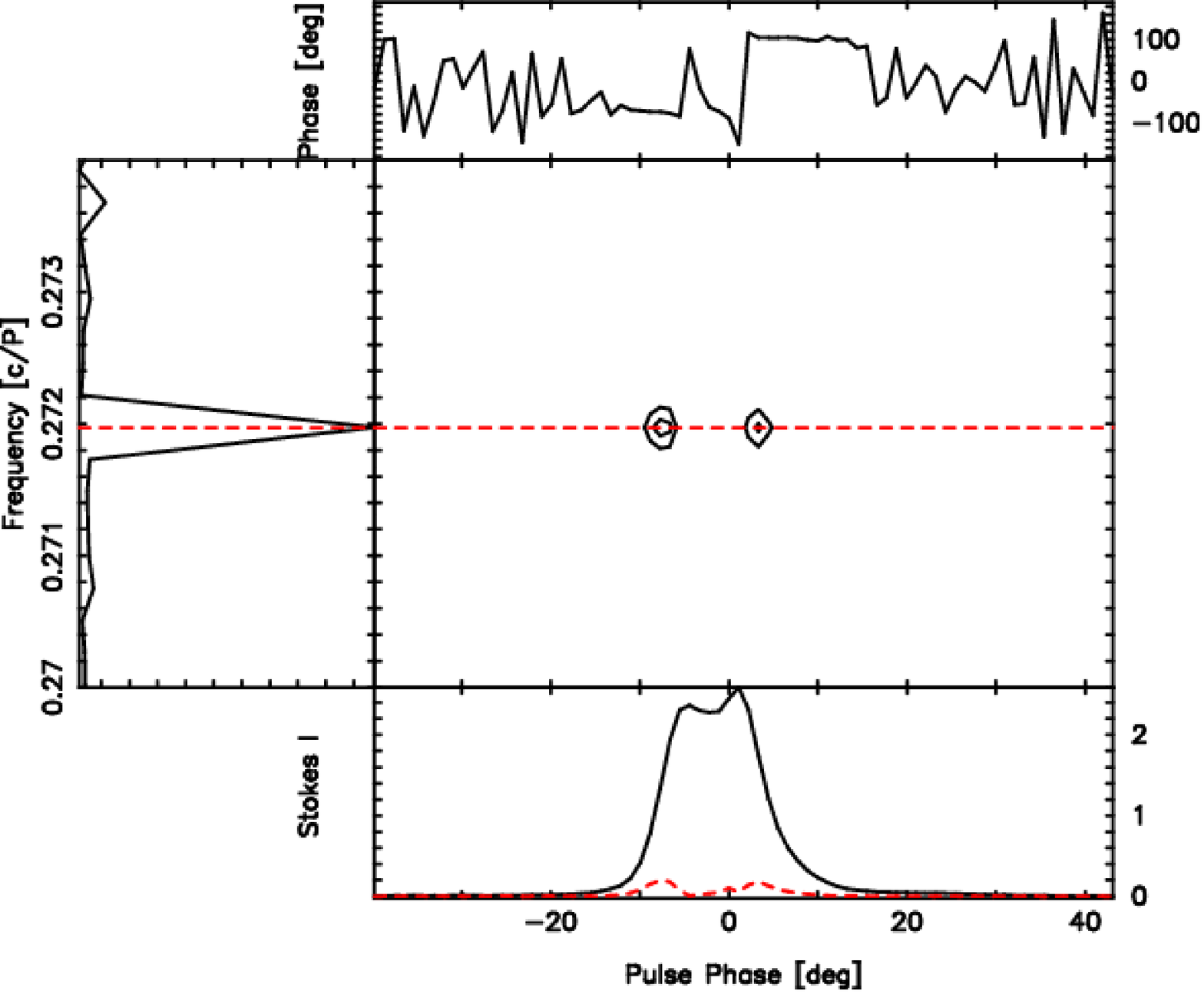}
\includegraphics[height=78mm,angle=-90.]{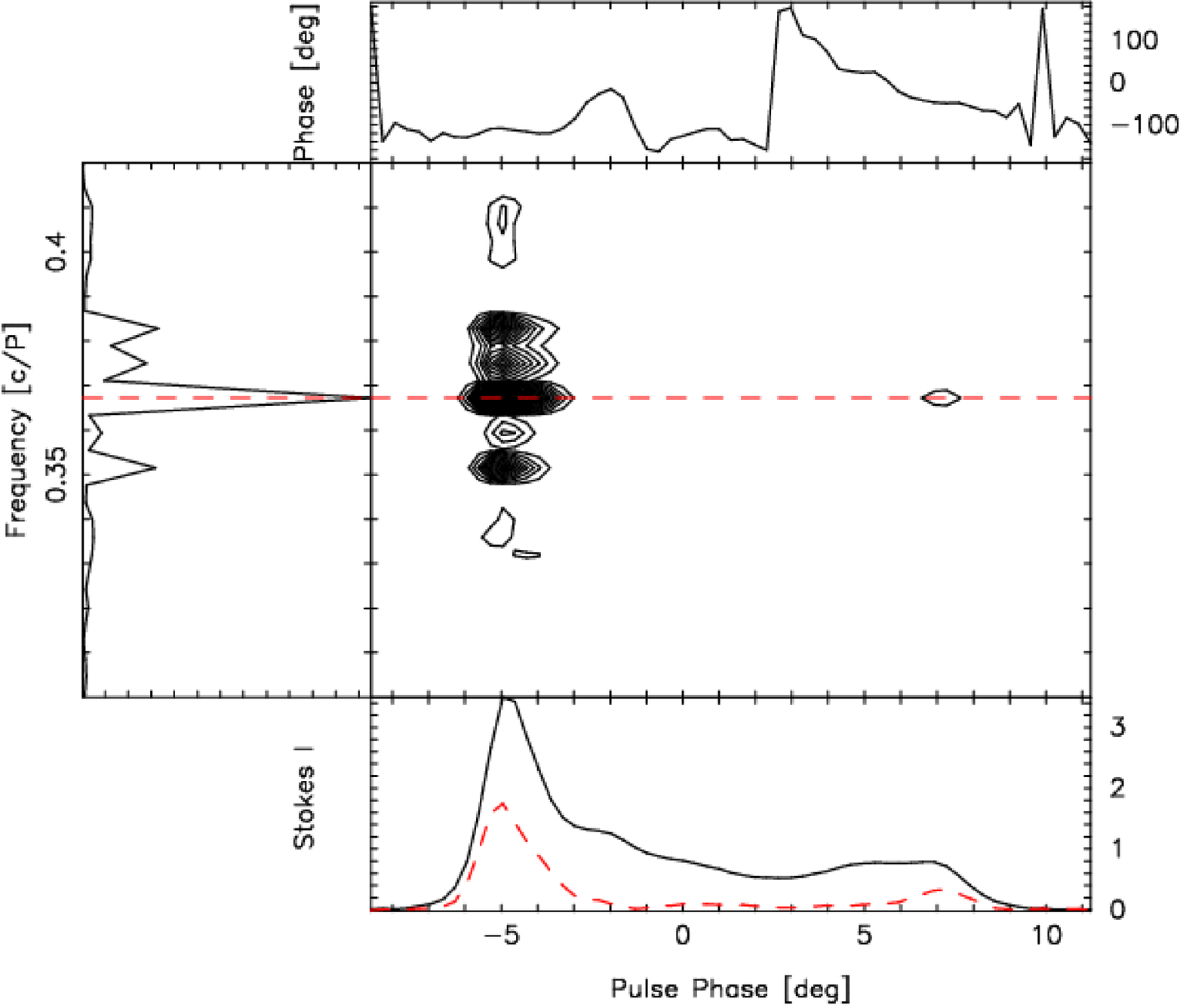}
\caption{Fluctuation spectra for pulsars B0740--28 and B1910+20.  The main 
panels show the fluctuation amplitudes in contours, and the longitude-averaged 
aggregates are given in the lefthand panels.  The bottom panels show the 
total-power average profiles, and the fluctuation phases are given in the top 
panels.}
\label{fig2}
\end{center}
\end{figure}

\section*{IV.  Conal Profile Asymmetry and Symmetry}
Conal single pulsar B0809+74 was listed by L\&M as a partial cone because of 
the strong evidence that its meter-wavelength profiles are asymmetric because 
they are incomplete---or ``absorbed'' (\eg, see Rankin \etal\ 2006b), and we now 
know that a number of other conal single pulsars share this asymmetric property 
(\eg, B0943+10, see Deshpande \& Rankin 2001).  Perhaps the asymmetry is 
due to the circumstance that stars with such profiles entail a highly tangential 
sightline traverse along the outer edges of their conal beams---but although we 
do not understand the cause of these asymmetries adequately, we do now know 
that nearly all conal single ({\bf S}$_d$) profiles---and many narrow inner-cone 
double ({\bf D}) profiles---are asymmetric.  

\begin{figure*}
\begin{center}
\begin{tabular}{@{}lr@{}}
{\mbox{\includegraphics[width=78mm,angle=-90]{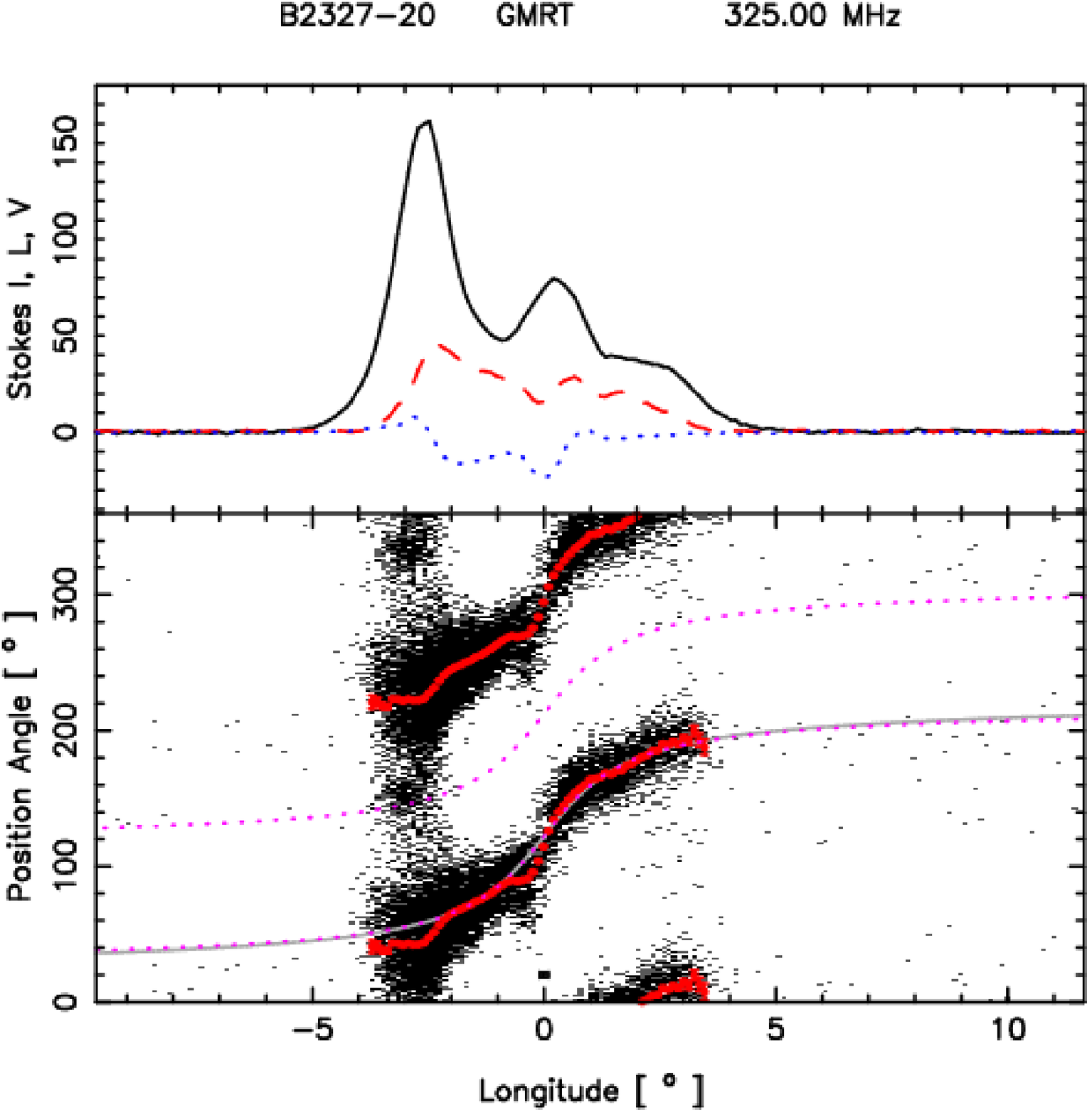}}} & 
{\mbox{\includegraphics[width=78mm,angle=-90.]{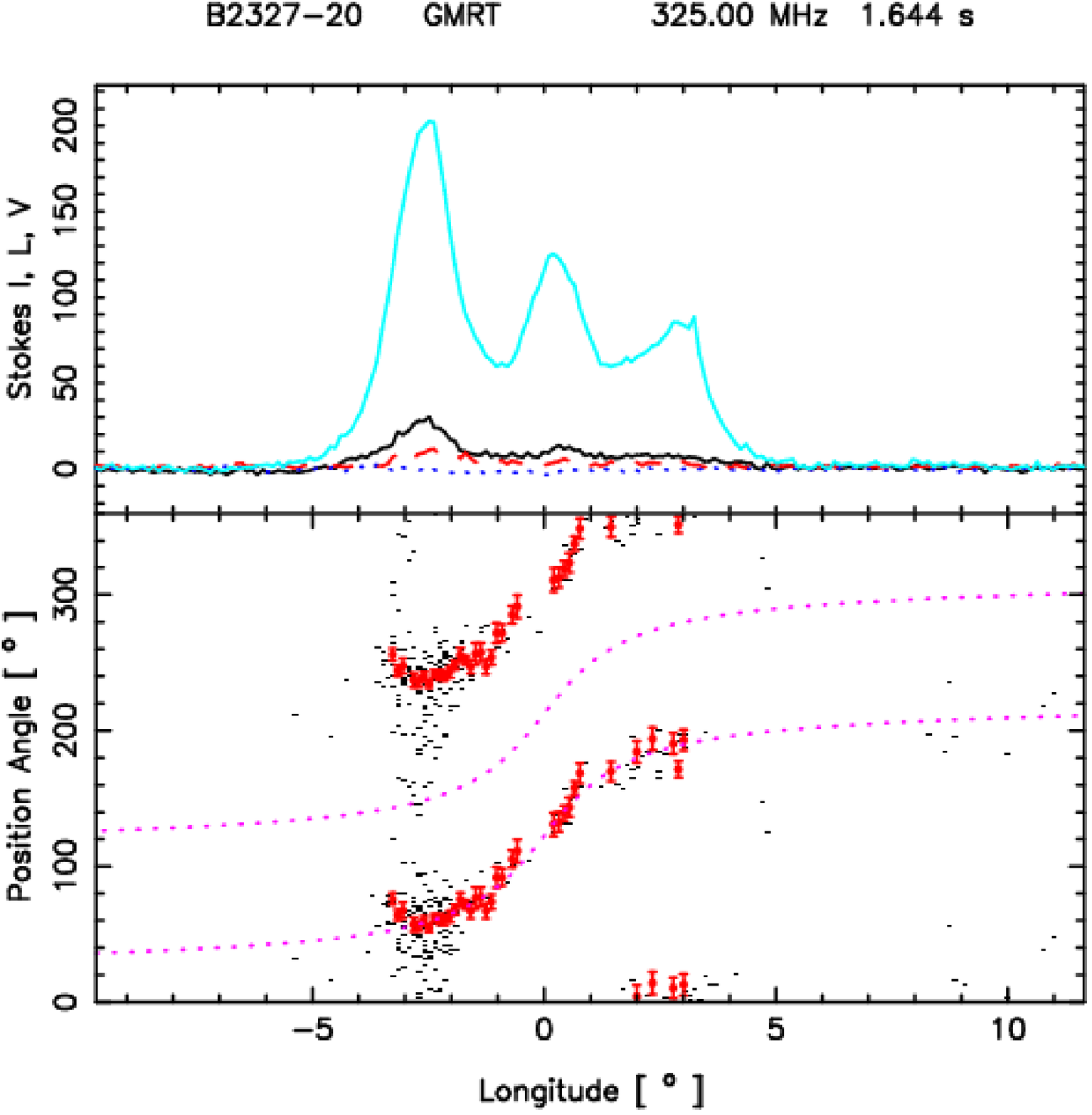}}}\\
\end{tabular}
\caption{PPA histograms for pulsar B2327--20 as in Fig.~\ref{fig1} (right panel).  
The lefthand display shows the full PS, whereas the righthand plot includes only 
pulses having a lower intensity level.  Notice that the latter PPA traverse is much 
smoother and can be used to fit the RVM.  The longitude origin corresponds to 
the SG point of the PPA swing.}
\label{fig3}
\end{center}
\end{figure*}

Perhaps then unsurprisingly, we found that a large proportion of the ``partial cone'' 
pulsars identified by L\&M had conal profiles that were either of the single 
({\bf S}$_d$) or unresolved double ({\bf D}) type.  Many such pulsars exhibit regular 
drifting subpulses and consequently show features indicative of periodic modulation 
in their fluctuation spectra.  Some others, however do not---or do not all of the time 
(as in B0943+10's `Q' mode)---but when such modulation can be detected it argues 
strongly for a conal association.  

We therefore computed fluctuation spectra similar to those in Figure~\ref{fig2} for 
each of the PSs available to us.  Most are not shown, usually because the same 
information was available in the published fluctuation-spectral compendia of 
Weltevrede \etal\ (2006b, 2007; hereafter WES/WSE).  Those dozen or so ``partial 
cone'' pulsars found to have {\bf S}$_d$ or {\bf D} profiles---in a number of cases it was difficult to 
be sure which---are so denoted in Table~\ref{tab3}, and their full analysis is discussed 
in the Appendix.  

For a few other pulsars in our sample we found subpulse modulation features on 
both the leading and trailing edges of their profiles.  Fig.~\ref{fig2} shows fluctuation 
spectra for pulsars B0740--28 and PSR B1910+20, where the same periodicity modulates
both edges of their profiles in a stationary manner.  If their subpulse ``drift'' is produced 
a carousel-beam system rotating about the magnetic axis, then this strongly indicates 
that the emission from these stars does indeed fill most of their polar flux-tube regions.

\section*{IV. The SG point and Profile Symmetry}
We have revisited these symmetry issues for all the ``partial cone'' pulsars for 
which we have high quality single pulse polarimetry.  Our intent has been to determine 
whether the SG point of the PPA traverse, determined using RVM fitting, leads or lags 
the total intensity profile center.  Single pulse polarimetry often helps to identify regions 
of OPM activity which otherwise can complicate the average-PPA traverse, and hence 
cause an inaccurate identification of the SG point.  Thus, we have computed PPA 
histograms for each of the ``partial cone'' pulsars---which are reproduced in the 
Appendix---and wherever possible distinguished their separate PPA traverses before 
fitting the RVM to determine the SG point.  In a few cases PS polarimetry was unnecessary 
to fix the SG point; however for pulsars like B1604--00 or B2043--04 mode 
separation was essential before any sensible RVM fit could be made to their PPA 
traverses.  

For several other pulsars, B1910+20 and B2327--20 among them, the average PPA 
traverse exhibited highly non-RVM behavior.  Mitra \etal\ (2007) noticed for pulsar 
B0329+54 that the PPA traverse can be intensity dependent, and we have used 
their technique of dividing the PS into intensity fractions in order to obtain a smooth 
PPA traverse.  Figure~\ref{fig3} shows for pulsar B2327--20 how an apparently 
orderly RVM behavior can be recovered for fitting even when the total PPA profile 
is distorted by intensity effects.  Here, the SG point obtained by fitting the RVM is 
well constrained.  However, several pulsars in our sample (such as B0906--17, B1742--30 and 
B1112+50) show relatively flat PPA traverses, and hence the SG point is not well 
determined.  Given all these various circumstances, we were able to obtain RVM 
fits for 24 pulsars. The fitted RVM parameters are listed in Table~\ref{tab5} 
and are used to plot the RVM as a grey line in the Figs.~\ref{fig1}, \ref{fig3} and 
many of those in the Appendix. 

Of primary importance is the location of the SG point with respect to the overall extent 
of a pulsar's profile, so as to assess whether A/R is significant. We use the BCW
method of finding the profile center---\ie, measuring the midway point between the 
outer 10\%-intensity points of the profile, and we then compare this with the location 
of the SG point.  For the pulsars with SG points determined via RVM fits, we find that 
for {\em all} of them the SG point either trails or is consistant with fall at the midway point.  Two 
apparent cases of leading SG points were quite interesting, B0138+59 and B2224+65: 
for the former early parts of the profile appear to be missing at all frequencies above 
100 MHz, but SVS's 102.5-MHz Faraday polarimetry shows it well centered; whereas 
for the latter, identification of the trailing component as a postcursor made any such 
argument untenable. Moreover, for several slow stars in the group, like
B2043--04 and B2327--20, the SG point is consistent with being coincident with the 
profile center.  

For several stars the SG point was not well determined and we lacked other 
observations of sufficiently high quality.  The rightmost column of Table~\ref{tab2}
then gives an overview of these SG-point locations with respect the respective 
profile midpoints; here ``T'' refers to the case where the SG point leads the midway 
point, ``L''  (no cases) where the SG point trails, ``U'' refers to unclear cases where the 
midway or the SG point determination fails, and ``---'' to those cases where the 
profile center is consistent with being coincident with the SG point. Justification 
for the SG point determinations for individual stars are found in the Appendix.
  
\section*{V. Abberation/Retardation Effects}
We saw just above that the PPA-traverse SG points falling on or after the 
profile centers all the cases.  A natural explanation for this circumstance is the 
abberation/retardation (A/R) effects first studied by Blaskeiwicz \etal\ (1991; 
BCW).  This BCW model provided a substantial improvement over the RVM 
model, by incorporating these (A/R) relativistic effects on pulsar-emission 
properties.  For emission arising from a finite height $r_{em}$ above the center 
of a rotating neutron star with period $P_1$, they derived an expression for the 
PPA $\chi$ as a function of pulse longitude $\varphi$ as,
\begin{equation}
\chi= \tan^{-1}\left(\frac{\sin\alpha \sin(\varphi-\varphi_{\circ})-3\hat{r}\sin\xi}{\sin\xi
\cos\alpha+\cos\xi\sin\alpha\cos(\varphi-\varphi_{\circ})}\right)+\chi_{\circ}
\label{eq1}
\end{equation}
where $\hat{r}$=$r_{em}/r_c$ is the emission height $r_{em}$ in terms of the 
light-cylinder radius $r_c$=$cP_1/2\pi$, and $c$ is the speed of light.  Note that 
eq.~\ref{eq1} reduces to the RVM (eq.~\ref{eq0}) for $r_{em}$ tending to zero.

In short, the BCW model predicts that for radio emission arising from a constant 
finite height, the overall PPA traverse will lag the total intensity profile. 
To first order, particularly for slowly rotating pulsars, this shift is a simple translation of
the PPA traverse towards the trailing parts of the profile---hence no change is 
required to fit the RVM to the PPA traverse.  However, the SG point will now be 
found to lag the profile center by an amount $\Delta\varphi$= $4\pi r_{em}/P_1c$, a 
shift which has been observed in several pulsars (\eg, BCW, von Hoensbroech 
\& Xilouris 1997, Mitra \& Li 2004) and then used to estimate the relevant radio 
emission heights---giving typical values of a few hundred kilometers. In some studies,
(see e.g. Malov \& Suleimanova 1998; Gangadhara \& Gupta 2001, Krzeszowski \etal\ 2009),
Srostlik \& Rankin(2005) and Force \& Rankin (2010) this shift is also seen with respect to 
the central core component.

Here we want to justify our conclusion that for the majority of ``partial cone'' pulsars,
the lagging of the SG point with respect to the profile center is primarily due to A/R 
effects.  On the one hand, the ``flared'' profile analysis provides a valuable method of 
assessing the full emission width in longitude---that is, the total extent of emission 
activity within the polar flux-tube region---and this in turn permits us to be more 
certain about the position of a profile's center and thus the relative placement of 
the SG point.  Then, on the other hand, the RVM fitting often permits us to be sure 
about the symmetry properties of the PPA traverse and thus its placement relative 
to the profile center.  The shift $\Delta\varphi$ is hence found as the difference 
between the (conal component-pair) profile center and the SG point, and it is from 
this quantity that an emission height can be computed.  

Among our ``partial cone'' population, we found 13 cases for which an A/R emission 
height could be computed as above, and these are tabulated in Table~\ref{tab4}.
For two stars B1732--07 and B1742--30 we have used the central core--component 
peak with respect to which the shifts have been computed for finding the emission heights.  
Note that the values obtained are roughly 200-400 km---therefore, not very different 
from those height estimates computed for normal (non-``partial cone'') pulsars.  This 
result strongly supports the conclusion that the majority of ``partial cone'' pulsars are 
very similar to other ``normal'' pulsars---that is, their emission arises from similar 
heights and (at least sometimes) involves most of the polar cap region.  For the few 
slower pulsars in the ``partial cone'' population, PSR B1910+20, B2043--04 and 
B2327--20, the measured A/R shifts are small, such that the profile centers and SG 
points are almost coincident. Although one expects that A/R shifts should be 
inversely related to pulsar period, such that faster pulsars should show larger 
shifts, none of the A/R studies in the literature has cleanly demonstrated this 
effect.  Failure to see this effect systematically could be due to a number of factors 
[see the detailed discussion by Mitra\& Li (2004) on factors affecting A/R effects].
%

We note that similar effects have been found by Karastergiou \& Johnston (2006) 
in B1054--62 and B1356--60, the latter of which is discussed below in the Appendix 
with the other ``partial cones''.  Several other cases where A/R appears to affect the 
profile structure are denoted by ``ar'' in their Table~\ref{tab3} classifications.

\section*{VI. Aberrant Linear Polarization Signatures?}
Our recent analyses (Backus \etal\ 2010) on the precursor components of pulsars 
B0943+10 and B1822--09, strongly suggest that these features are ``other''---that 
is, they are not emitted at low altitude in the polar flux tube as is the conal and 
core emission with which we are familiar.  We argued that the precursors were 
aberrant largely on the basis of their nearly complete linear polarization and flat 
PPA traverses.  Among L\&M's ``partial cone'' grouping, we encounter B1822--09 
again, and the geometric analysis in Table~\ref{tab3} (see also Fig.~\ref{figA6}) 
reflects the conclusions of the above study in that we do not regard the star's 
precursor component as a part of its main pulse.  

Three other such objects were found among L\&M's ``partial cones'', B1322+83, 
B1530+27 and B2224+65.  In the first case seen in Fig.~\ref{figA3}, the highly 
polarized feature is a precursor to what otherwise is probably a conal single main 
pulse; whereas, for the latter two in Fig.~\ref{figA3} and \ref{figA10} the aberrant 
features fall as a postcursors to what seem to be a conal single and core-single 
main pulses, respectively.  A number of other such features can now been found 
in the published polarimetry, but at the time of L\&M's study, very few were known, 
so it is not surprising that they regarded them as outstanding in core/core terms.  
Indeed, they yet remain so, but we now know of enough that they represent something 
of a distinct phenomenon.  

For B1322+83, we note also that were this star an asymmetric conal double (which 
is {\em not} what we conclude), the putative profile midpoint at about --4\degr\ falls 
far ahead of the SG point under the trailing feature.  Following this interpretation we 
can compute an A/R emission height of some 3700 km, which is very large for any 
pulsar.  Therefore this interpretation is almost certainly incorrect.

\section*{VII. Analyses of the Emission Geometry}
Paper VI of this series gave an extensive analysis of the emission geometry of 
some 200 pulsars.  The core-component width $W_{core}$ was often used to 
determine the magnetic latitude $\alpha$ using the relationship 
$W_{core}=P_1^{-1/2}/\sin{\alpha}$ (Paper IV).  The sightline impact angle $\beta$ 
could then be fixed (within a sign\footnote{All have been taken positive, as the 
poleward or equatorward sense of the sightline traverse cannot easily be known.  
In fact, the sense of $\beta$ usually makes little difference in these modeling 
computations.}) using $\alpha$ and the PPA sweep rate $R$, which 
can be determined empirically at the SG point as $|\Delta\chi/\Delta\varphi|_o$ (and 
within the RVM is also $\sin(\alpha)/\sin(\beta)$; see eq.~\ref{eq0}).  Finally, the 
conal radii were computed using the dimensions of the conal components or 
pairs as in Paper VI; see eqs. (2-6).  Several of the ``partial cone'' pulsars under 
study here were also included in this Paper VI analysis, but the results---based 
entirely on average-profile dimensions---were disappointing---just as they were 
for L\&M and for virtually the same reasons.  Detailed geometric models for a 
few others have been developed elsewhere; see Table~\ref{tab3}, footnotes b-f.    

\begin{figure}
\begin{center}
\includegraphics[height=78mm,angle=-90]{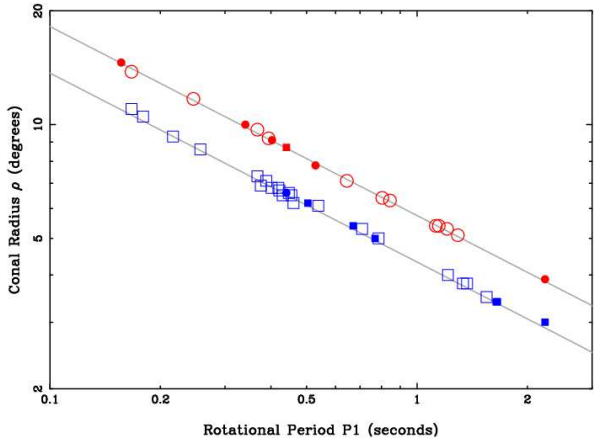}
\caption{Plot showing (outside half-power) conal emission radius $\rho$ vs. 
rotation period $P_1$ for the ``partial cone'' population of pulsars in Table~\ref{tab3}.  
The red symbols represent the outer cones and the blue the inner cones, 
respectively.  The filled symbols reflect a full analysis where the magnetic 
latitude $\alpha$ can be determined from the core width and $\beta$ then 
determined from the fitted PPA sweep rate $R$; whereas, the open symbols 
indicate pulsars for which compatible geometrical configurations could be 
identified despite having no means of estimating $\alpha$ independently---all 
as in Paper VI (see text for details).  The lower and the upper grey lines correspond 
to the characteristic inner and outer conal radii of 4.33 $P{_1}^{-1/2}$ and 5.75 
$P{_1}^{-1/2}$, respectively}.
\label{fig6}
\end{center}
\end{figure}

Here we are now in a position to reinvestigate the emission geometry of L\&M's 
``partial cone'' population with much more information and thus a greatly enhanced 
expectation of success.  The pulsar-by-pulsar discussions in the Appendix 
together with Table~\ref{tab5} summarize 
the RVM-fitting results for all those stars for which it was possible, and Table~\ref{tab3} 
gives the PPA sweep rate $R$ in boldface when determined by this fitting.  Similarly, 
this table shows $\alpha$ in boldface when it was possible to estimate it from a core 
width or by other means.  Then, the conal dimensions are computed from the profile 
width information just as was done previously in Paper VI.  The outside half-power 
widths of the respective inner and outer conal beams are given along with the 
estimated emission heights---and the parameter $\beta/\rho$ is also tabulated for 
many pulsars as an indication of the expected profile form.  

The overwhelming results of Paper VI, confirmed by later studies, is that pulsar 
emission cones come in two types, inner and outer, with outside half-power 1-GHz 
dimensions of 4.33 and 5.75\degr, respectively.  Somewhat surprisingly, this implies 
nominal 1-GHz emission heights of some 130 and 220 km, respectively.  Therefore, 
just as in Paper VI, we here endeavor to demonstrate two distinct propositions:  a) 
when all the above information is available, we show that a specific ``partial cone'' 
pulsar will have compatible conal radii and thus nominal emission heights; however, 
b) when this ``full solution'' is impossible, we use the available information to model 
the emission geometry to achieve appropriate radii and heights, thus resulting in a 
useful estimate of the magnetic latitude $\alpha$.  When multifrequency profiles are 
available, it is usually clear whether a specific star has an inner or outer cone (or in 
a few cases both) because outer cones spread strongly with wavelength and inner 
ones do not.  Finally, we have taken $\beta$ positive in all cases, given that its sign 
cannot usually be determined.

One might worry that the conal dimension and emission heights for pulsars 
in category b) above are meaningless because the former have been 
constrained to values near the characteristic inner or outer conal radius.  
However, this exercise is hardly arbitrary:  Paper VI showed that there 
were two distinct types of cones, inner and outer, with specific angular dimensions 
and therefore nominal emission heights, and other work (\eg, Paper VII) demonstrated 
that the two conal species could usually be distinguished by whether or not their 
dimensions increased significantly at low frequency.  For each such pulsar we have 
used all available evidence to make this determination (as discussed for each star in 
the Appendix), and when successful---as it was in most cases---appropriate angular 
radii could be determined.  Then, we explored whether a value of $\alpha$ existed 
that was compatible with each star's conal radius together with its profile's conal width 
and PPA sweep rate $R$ (so as to determine $\beta$) using Paper VI: eq.(4).  Usually 
such an $\alpha$ could be found providing a plausible geometric model, and these 
values are given in Table~\ref{tab3} and the conal radii plotted as the open symbols 
in Figure~\ref{fig6}.\footnote{Note that the geometrical assumptions behind eq.(4) 
above are only satisfied within certain parameter ranges, and in a few cases (\ie, 
B1930+22) no model could be constructed despite some suggestion of a core/cone 
structure.}  Such simple models have proven to be quite reliable---\eg, as many of the 
values in Bhattacharya \& van den Heuvel's (1991) comparison were estimated by this 
means.

For several reasons the geometrical error indications in Table~\ref{tab3} are 
only approximate.  Our purpose here is to demonstrate the overall geometric 
``normality'' of L\&M's ``partial cone'' population.  A definitive geometric analysis 
of the individual stars is far beyond the possible scope of this effort.  Often, for 
instance, no pair of quality high and low frequency observations was available, so we could 
not extrapolate to 1 GHz as in Paper VI.  More importantly, the difficult character of 
this population has entailed using a variety of methods with different assumptions 
and uncertainties.  Therefore, where an $R$ value could be determined by fitting, 
we show it in bold in Table~\ref{tab3} with its errors given in Table~\ref{tab5}, 
and where an $\alpha$ could be estimated by a core width, it is also so indicated 
by bold type.  The conal dimensions in Table~\ref{tab3} are typically accurate to 
about a degree or so of longitude; when they could be measured more precisely 
(and extrapolated to 1 GHz), 
we give them with a decimal, or when very approximate, we show them with a 
question mark.  Finally, when the conal radii and emission heights could be fully 
determined, the latter are shown in bold.  These values near 130 and 220 km are 
only nominal, rather than physical, quantities, so the accuracy of their determination 
is of secondary importance.

These geometric results are then plotted in Fig.~\ref{fig6}, and the results 
are quite dramatic.  The values fall on two parallel tracks representing the 
outer and inner conal radii, respectively.  The solid symbols indicate the full 
solutions above, and the open ones compatibility where $\alpha$ could not 
be determined independently.  Overall we see that there is no geometric 
distinction at all between L\&M's ``partial cone'' pulsars and those with more 
ordinary and symmetrical profiles.  Of course, all of the ``partial cone'' pulsars 
for which we have observations are not represented in Fig.~\ref{fig6}---some 
of them are very difficult to understand as we have seen in the previous 
section---but here we see clearly that the great majority exhibit the same 
orderly conal and core dimensions as was found earlier in Paper VI.

\section*{VIII. Summary and Conclusions} 
In the foregoing sections we restudied the population of ``partial cone'' pulsars 
so identified by Lyne \& Manchester in 1988.  Using a combination of recent 
GMRT and Arecibo polarimetry, we have based our analyses predominantly 
on sensitive PS observations.  These analyses have attempted to identify 
regions of ``flared'' and A/Red emission as well as searching for the more 
usual periodic subpulse modulation.  

Through this analysis we have been able to show that---
\begin{itemize}
\item In general, L\&M's ``partial cone'' pulsars exhibit no particular property 
or difference as compared to the rest of the slow pulsar population.  Rather, 
they exhibit a range of characteristics, many of which are well understood and 
some of which are not.
\item Overall the ``partial cones'' exhibit cone and double cone profile structures 
just as the ``normal'' pulsars do.  To a significant extent PS analyses are needed 
to establish this regularity, because many of the ``partial cones'' do preferentially 
illuminate only a part of their polar flux-tube emission regions.  However, when 
these small difficulties are accounted for, the emission geometry of most ``partial 
cones'' is remarkably regular in the terms established in Paper VI---that is, both 
the cores and cones have particular angular dimensions that scale with the size 
of a pulsar's polar cap.  
\item We find several further examples among the ``partial cones'' of highly 
polarized pre- or postcursors with flat PPA traverses.  Following our analysis of 
such features in B0943+10 and B1822--09 (Backus \etal\ 2010), we argued that 
these features cannot be emitted at low altitude within the usual polar fluxtube 
region.  Such features are important, because they provide clues to the 
electrodynamics of the larger magnetosphere.  Clearly this emission is coherent 
(highly polarized), beamed and likely emitted at very high altitude.  
\item Among the ``partial cone'' pulsars we find {\em no} good examples of profiles 
where the SG  point leads the profile center.  Surely this can be taken as strong 
evidence that A/R operates to some degree in all pulsar profiles.
\item A number of examples of A/R shifts, both in the PPA traverses and component 
positions, were encountered among the ``partial cones''.  It seems likely that A/R is 
an important factor in distorting the core/cone structure of pulsars that rotate quite 
rapidly.  
\end{itemize}

It is hardly surprising that a study as ambitous at that of Lyne \& Manchester's 
would encounter a residuum of objects that were difficult to categorize and study.  
Indeed, the ``partial cone'' population were overall among the faster, weaker and 
more highly dispersed stars available to them in 1988.  A variety of both technical 
and scientific factors now permit us to understand that most of these ``partial cone'' 
pulsars are as ``normal'' in their beaming geometry as the many studied in earlier 
papers in this series.  Technically, quality PS polarimetry of highly dispersed stars 
with periods down to 100 msec are no longer challenging.  Scientifically, a great deal 
has been learned about cones and their constituent subbeam carousels---and that 
the emission from such systems rarely produces symmetrical profiles.  Single pulse 
observations permit us to identify emission centers that are too weak or irregular to 
show in up in average profiles.  Similarly, BCW introduced the A/R discussion only 
three years after L\&M's study, but it has taken fully these two decades for workers 
to begin to identify A/R regularly and confidently in pulsar emission, given the many 
other factors that tend to obscure its full effect.  Finally, we now see that the ``normal'' 
core/cone emission from the slow pulsar population is regular enough in its properties 
that some aberrant features can be identified.  We cannot yet be sure where and how 
the highly polarized pre- and postcursor features are emitted, but it seems ever less 
likely that they come from the usual low altitude polar fluxtube region.

The other effect that A/R predicts---that the intensity of the 
leading conal regions of the profile will be brighter than the trailing parts---this we 
do not see in our analyses.  Rather it appears that the probability of radio emission 
across the pulse profile (or within the polar flux tube) varies strongly.  For example, 
the ``flared'' emission we see in several stars (\eg, PSR B0355+54 in Fig.~\ref{fig1}) 
is overall rare, occurring within only 1-5\% of all active pulse longitudes, whereas in 
many other such regions the emission is virtually continuous, occurring essentially 
100\% of the time.  Of course, this implies that the shape of a pulsar's total-intensity 
profile varies strongly across the ``active'' window because the several different processes 
entailed in this emission also vary strong with longitude.  The PPA traverse, on the
other hand, closely follows the RVM (particularly when complications due to OPM 
and A/R effects can be accounted for).

Hence, the phenomenological model that emerges from our overall analyses is that 
pulsar coherent radio emission almost always arises from open dipolar field lines, at 
several hundred km above the pulsar polar cap.  Within this polar flux-tube region, 
there is an underlying double cone/core structure of the pulsar radio-emission beams, 
although the pulse shape itself depends on the probability of coherent radio emission 
($P_{cre}$), which varies strongly with magnetic colatitude and azimuth and thus with 
pulse longitude along the sightline trajectory of a given star and viewing geometry.  
For example, under the vacuum-gap model and assuming curvature radiation as the 
radio emission mechanism, $P_{cre}$ should be viewed as a combined probability of 
$P_{cre} = P_{ppc} \times P_{spc} \times P_{ccr}$.  Here $P_{ppc}$ is the probability 
of primary pair creation in the vacuum gap, $P_{spc}$ is the probability of secondary 
pair creation and $P_{ccr}$ would be the criteria for exciting coherent curvature radiation 
(RS75, Sturrock 1971, Gil, Lyubarsky \& Melikidze 2004).  Further since average pulse 
profiles are stable, $P_{cre}$ fluctuates around a mean value, indicating the presence 
of a stable physical quantity at every pulse longitude.  We conjecture that $P_{cre}$ is 
primarily guided by the underlying multipolar magnetic field across the polar cap.  Such 
a structure causes the field to vary in magnitude and curvature radius across the polar 
cap (see Gil, Melikidze \& Mitra 2002).  The field at the region where radio emission 
arises is however significantly dipolar.

\acknowledgments
We thank our anonymous referee and the editor for critical 
comments which helped in improving the manuscript significantly.
We are pleased to acknowledge S. Sarala for important assistance with 
aspects of the observations and analysis of GMRT data. 
We thank our first referee Aris Karastergiou for
encouraging us to enlarge our study to include as many of L\&M's
``partial-cone'' pulsars as possible. 
We thank the staff of the GMRT that made these observations possible. 
We also wish to thank Joel Weisberg
for his assistance with the ionospheric Faraday rotation corrections.  One 
of us (JMR) thanks the Anton Pannekoek Astronomical Institute of the 
University of Amsterdam for their generous hospitality and both Netherlands 
National Science Foundation and ASTRON for their Visitor Grants.  
DM thanks National Astronomy and Ionosphere Center, Arecibo Observatory, Puerto Rico,
for supporting a visiting scientist position during which this work was completed.
Portions 
of this work were carried out with support from US National Science Foundation 
Grants AST 99-87654 and 08-07691. Arecibo Observatory is operated by 
Cornell University under contract to the US NSF.  
GMRT is run by the National Centre for Radio Astrophysics of the Tata 
Institute of Fundamental Research.
This work used the NASA 
ADS system.

The following is a list of the pulsars mentioned to comply with AAS\TeX \S2.15.3. It doesn't print in the right place.
\objectname{B0138+59}
\objectname{B0254$-$53}
\objectname{B0355+54}
\objectname{B0450+55}
\objectname{B0540+23}
\objectname{B0643+80}
\objectname{B0740$-$28}
\objectname{B0809+74}
\objectname{B0906$-$17}
\objectname{B0919+06}
\objectname{B1055$-$52}
\objectname{B1112+50}
\objectname{B1221$-$63}
\objectname{B1240$-$64}
\objectname{B1322+83}
\objectname{B1356$-$60}
\objectname{B1426$-$66}
\objectname{B1449$-$64}
\objectname{B1530+27}
\objectname{B1530$-$53}
\objectname{B1540$-$06}
\objectname{B1556$-$44}
\objectname{B1604$-$00}
\objectname{B1612+07}
\objectname{B1641$-$45}
\objectname{B1648$-$42}
\objectname{B1700$-$18}
\objectname{B1732$-$07}
\objectname{B1742$-$30}
\objectname{B1745$-$12}
\objectname{B1756$-$22}
\objectname{B1822$-$09}
\objectname{B1842+14}
\objectname{B1851$-$14}
\objectname{B1859+07}
\objectname{B1900+05}
\objectname{B1907$-$03}
\objectname{B1910+20}
\objectname{B1913+10}
\objectname{B1915+13}
\objectname{B1924+16}
\objectname{B1930+22}
\objectname{B1937$-$26}
\objectname{B1944+17}
\objectname{B1944+22}
\objectname{B2021+51}
\objectname{B2043$-$04}
\objectname{B2053+36}
\objectname{B2217+47}
\objectname{B2224+65}
\objectname{B2327$-$20}

\label{lastpage}

\appendix
\setcounter{figure}{0} 
\renewcommand{\thefigure}{A\arabic{figure}}
\renewcommand{\thetable}{A\arabic{table}}
\setcounter{table}{0} 
\renewcommand{\thefootnote}{A\arabic{footnote}}
\setcounter{footnote}{0} 
\twocolumn
\subsection*{APPENDIX: Results for Individual Pulsars}
\label{sec:results}



\begin{table*}
\caption{Observational Parameters}
\begin{center}
\begin{tabular} {lllll}
\hline \hline
{\bf PSR } & \bf{Obs'y} & \bf{MJD} & \bf{length}/ & \bf{Fig.}\\
  Bname    &  \bf{Band} &                 &  \bf{res (\degr)}& \\
\hline
B0138+59 & GM:P &54399& 1961/0.15 &\ref{figA1}\\
B0355+54 & GM:P &53780& 13144/1.18 &\ref{figA1}\\
B0450+55 & GM:P &53245& 2671/0.54 &\ref{figA1}\\
B0540+23 & AO:P &54015& 2440/0.66 &\ref{figA1}\\
B0740--28 & GM:P &53781& 3649/1.10 &\ref{figA2}\\
 \vspace{3pt}
B0809+74 & GM:P &54399& 940/0.14 &\ref{figA2}\\
B0906--17 & GM:P &54399& 2256/0.46 &\ref{figA2}\\
B0919+06 & AO:L & 52854& 1115/0.43 &\ref{figA2}\\
B1055--52I & GM:P &54537& 16571/0.93 &\ref{figA3}\\
{\it B1112+50} & GM:P &54399& 2002/0.11 &\ref{figA3}\\
 \vspace{3pt}
{\it B1322+83m} & GM:P &54399& 2700/0.28 &\ref{figA3}\\
B1530+27 & AO:P  &53994& 1032/0.33 &\ref{figA3}\\
B1540--06 & GM:P &54399& 2129/0.26 &\ref{figA4}\\
B1556--44 & GM:P &53781& 3629/0.72 &\ref{figA4}\\
B1604--00 & AO:L  &53372& 1605/0.22 &\ref{figA4}\\
\vspace{3pt}
B1612+07 & AO:P &53378& 1094/0.31 &\ref{figA4}\\
{\it B1700--18} & GM:P & 54399 & 1917/0.23 &\ref{figA5}\\
{\it B1732--07} & GM:P &54399& 2015/0.42 &\ref{figA5}\\
{\it B1742--30} & GM:P &54399& 1975/0.50 &\ref{figA5}\\
{\it B1745--12} & GM:P &53781& 2740/0.47 &\ref{figA5}\\
 \vspace{3pt}
B1822--09 & GM:P &54399& 1962/0.24 &\ref{figA6}\\
{\it B1842+14} & AO:P  &53378& 1600/0.46 &\ref{figA6}\\
{\it B1851--14} & GM:P &54399& 1079/0.16 &\ref{figA6}\\
{\it B1900+05} & AO:L & 54842 & 1045/0.39 &\ref{figA6}\\
{\it B1907--03} & GM:P &54399& 2035/0.37 &\ref{figA7}\\
 \vspace{3pt}
B1910+20 & AO:L  &53372& 906/0.33 &\ref{figA7}\\
{\it B1913+10} & AO:L & 54538 & 2077/0.26 &\ref{figA7}\\
B1915+13 & AO:L  &48918& 4000/0.33 &\ref{figA7}\\
B1924+16 & AO:L  &54538& 2522/0.26 &\ref{figA8}\\
{\it B1930+22} & AO:L  &54540& 4151/0.64 &\ref{figA8}\\
 \vspace{3pt}
B1937--26 & GM:P &54399& 1965/0.46 &\ref{figA8}\\
{\it B1944+17} & AO:P & 53966 & 7038/0.31 &\ref{figA8}\\
{\it B1944+22} & AO:P & 55276 & 932/0.35 &\ref{figA9}\\
B2021+51 & GM:P &54399& 2282/0.35 &\ref{figA9}\\
{\it B2043--04} & GM:P &54399& 1993/0.12 &\ref{figA9}\\
 \vspace{3pt}
B2053+36 & AO:L &52837& 52837/0.42 &\ref{figA9}\\
B2217+47 & GM:P &54399& 2243/0.34 &\ref{figA10}\\
B2224+65m & GM:P &54399& 2101/0.27 &\ref{figA10}\\
B2327--20 & GM:P &54399& 1865/0.11 &\ref{figA10}\\
\hline
\end{tabular}
\end{center}
\footnotesize 
{Notes: Pulsars with Bnames in normal type appear in L\&M's Table 4; those
in italics are denoted as ``Partial cones?'' in their Table 5}
\label{tab1}
\end{table*}

\begin{table*}
\caption{Parameters for Lyne \& Manchester's ``Partial Cone'' Pulsars.}
\label{tab2}
\begin{center}
\begin{tabular} {llllllll}\hline \hline
PSR & PSR& $P_1$ & {$\log$($\tau$)} & {$\log$($B$)} & {$\log$(\.E)}& Fig. &Remarks \\
  Bname    &  Jname   &     (s)         &   (yrs)                 &   (G)              &    (erg s$^{-1}$) &&         \\
\hline
B0138+59 & J0141+6009 & 1.223 & 7.69 & 11.85 & 30.93 &\ref{figA1}&---\\
B0254--53 & J0255--5304 & 0.448 & 8.37 & 11.07 & 31.11 &---&U\\
B0355+54 & J0358+5413 & 0.156 & 5.75 & 11.92 & 34.66 &\ref{figA1}&T\\
B0450+55 & J0454+5543 & 0.341 & 6.36 & 11.96 & 33.38 &\ref{figA1}&T\\
B0540+23 & J0543+2329 & 0.246 & 5.40 & 12.29 & 34.61 &\ref{figA1}&T\\
{\it B0643+80} & J0653+8051 & 1.214 & 6.71 & 12.34 & 31.92  &---&U\\
\vspace{2pt}
B0740--28 & J0742--2822 & 0.167 & 5.20 & 12.23 & 35.16 &\ref{figA2}&T\\
B0809+74 & J0814+7429 & 1.292 & 8.09 & 11.67 & 30.49 &\ref{figA2}&T\\
B0906--17 & J0908--1739 & 0.402 & 6.98 & 11.72 & 32.61 &\ref{figA2}&U\\
B0919+06 & J0922+0638 & 0.431 & 5.70 & 12.39 & 33.83 &\ref{figA2}&T\\
B1055--52I & J1057--5226 & 0.198 & 5.73 & 12.04 & 34.48 & \ref{figA3}&T\\
\vspace{2pt}
{\it B1112+50} & J1115+5030 & 1.656 & 7.02 & 12.31 & 31.34 &\ref{figA3}&U\\
B1221--63 & J1224--6407 & 0.216 & 5.84 & 12.02 & 34.28 &---& U \\
B1240--64 & J1243--6423 & 0.388 & 6.14 & 12.13 & 33.48 &---& U \\
{\it B1322+83m} & J1321+8323 & 0.670 & 7.27 & 11.80 & 31.87 &\ref{figA3}&---\\
B1356--60 & J1359--6038 & 0.128 & 5.50 & 11.96 & 35.08 &---& U \\
B1426--66 & J1430--6623 & 0.785 & 6.65 & 12.17 & 32.36 &---& U \\
B1449--64 & J1453--6413 & 0.179 & 6.02 & 11.85 & 34.28 &---& U \\
\vspace{2pt}
B1530+27 & J1532+2745 & 1.125 & 7.36 & 11.98 & 31.33 &\ref{figA3}&U\\
B1530--53 & J1534--5334 & 1.369 & 7.18 & 12.15 & 31.34 &---& --- \\
{\it B1540--06} & J1543--0620 & 0.709 & 7.11 & 11.90 & 31.99 &\ref{figA4}&U\\
B1556--44 & J1559--4438 & 0.257 & 6.60 & 11.71 & 33.38 &\ref{figA4}&T\\
B1604--00 & J1607--0032 & 0.422 & 7.34 & 11.56 & 32.21 &\ref{figA4}&T\\
B1612+07 & J1614+0737 & 1.207 & 6.91 & 12.23 & 31.72 &\ref{figA4}&U\\
\vspace{2pt}
B1641--45 & J1644--4559 & 0.455 & 5.56 & 12.49 & 33.92 &---& U \\
B1648--42 & J1651--4246 & 0.844 & 6.44 & 12.31 & 32.51 &---& U \\
{\it B1700--18} & J1703--1846 & 0.804 & 6.87 & 12.08 & 32.12 &\ref{figA5}&T\\
{\it B1732--07} & J1735--0724 & 0.419 & 6.74 & 11.86 & 32.81 &\ref{figA5}&U\\
{\it B1742--30} & J1745--3040 & 0.367 & 5.74 & 12.30 & 33.93 &\ref{figA5}&U\\
{\it B1745--12} & J1748--1300 & 0.394 & 6.71 & 11.85 & 32.89 &\ref{figA5}&T\\
{\it B1756--22} & J1759--2205 & 0.461 & 5.83 & 12.36 & 33.64 &---&U\\
B1822--09 & J1825--0935 & 0.769 & 5.37 & 12.81 & 33.66 &\ref{figA6}&---\\
{\it B1842+14} & J1844+1454 & 0.375 & 6.50 & 11.93 & 33.15 &\ref{figA6}&U\\
{\it B1851--14} & J1854--1421 & 1.147 & 6.64 & 12.34 & 32.04 &\ref{figA6}&U\\
(B1859+07) & J1901+0716 & 0.644 & 6.65 & 12.09 & 32.53 &---&\\
{\it B1900+05} & J1902+0556 & 0.747 & 5.96 & 12.50 & 33.08 &\ref{figA6}& U\\
{\it B1907--03} & J1910--0309 & 0.505 & 6.56 & 12.03 & 32.83 &\ref{figA7}&U\\
B1910+20 & J1912+2104 & 2.233 & 6.54 & 12.68 & 31.56 &\ref{figA7}&T\\
{\it B1913+10} & J1915+1009 & 0.405 & 5.62 & 12.40 & 33.96 &\ref{figA7}&U\\
B1915+13 & J1917+1353 & 0.195 & 5.63 & 12.08 & 34.59 &\ref{figA7}&T\\
\vspace{2pt}
B1924+16 & J1926+1648 & 0.580 & 5.71 & 12.52 & 33.56 &\ref{figA8}&T\\
{\it B1930+22} & J1932+2220 & 0.144 & 4.60 & 12.47 & 35.88 &\ref{figA8}&T\\
B1937--26 & J1941--2602 & 0.403 & 6.82 & 11.80 & 32.76 &\ref{figA8}&T\\
{\it B1944+17} & J1946+1805 & 0.441 & 8.46 & 11.02 & 31.041 &\ref{figA8}&U\\
{\it B1944+22} & J1946+2244 & 1.334 & 7.38 & 12.04 & 31.18 &\ref{figA9}&U\\
B2021+51 & J2022+5154 & 0.529 & 6.44 & 12.11 & 32.91 &\ref{figA9}&T\\
{\it B2043--04} & J2046--0421 & 1.547 & 7.22 & 12.18 & 31.20 &\ref{figA9}&---\\
B2053+36 & J2055+3630 & 0.222 & 6.98 & 11.46 & 33.13 &\ref{figA9}&U\\
B2217+47 & J2219+4754 & 0.538 & 6.49 & 12.09 & 32.84 &\ref{figA10}&---\\
\vspace{2pt}
B2224+65m & J2225+6535 & 0.683 & 6.05 & 12.41 & 33.08 &\ref{figA10}&U\\
B2327--20 & J2330--2005 & 1.644 & 6.75 & 12.45 & 31.61 &\ref{figA10}&---\\
\hline
\end{tabular}
\end{center}
\end{table*}

{\footnotesize\noindent Table 2 Notes:  Pulsars with Bnames in normal type appear in L\&M's Table 4; those 
in italics are denoted as ``Partial cones?'' in their Table 5; and one other star, 
B1859+07 (parentheses) is included from our own work.  The periods ($P_1$), 
age ($\tau$=$P_1/2\dot{P}_1$), magnetic field ($B$) and energy are taken from 
the ATNF pulsar catalogue.  The referenced figures appear in the Appendix, 
and the last column specifies if the SG point is either trailing (T) or leading (L) 
or unclear (U) {\it wrt} the pulsar profile, and the (---) refers to cases where
the SG point is consistent with being coincident with the profile center.}

\begin{table*}
\begin{center}
\caption{Emission-Beam Geometry of ``Partial Cone'' Pulsars}
\label{tab3}
\begin{tabular} {lcccccccccccc}\hline \hline
&&&$|\Delta\chi/\Delta\varphi|_o$&&\_\_\_\_\_\_\_&Inner&\_\_\_\_\_\_\_&\_\_\_\_\_\_\_&Outer&\_\_\_\_\_\_\_&\_\_\_\_\_r&(km)\_\_\_\_\_ \\
PSR&Class&$\alpha$&(\degr/\degr)&$\beta$&$\Delta\Psi$&$\rho$&$\beta/\rho$&$\Delta\Psi$&$\rho$&$\beta/\rho$&Inner&Outer \\
\hline
\small
B0138+59&M/cQ?&\bf{20}&\bf{--11.2}&1.7&$\sim$ ---& ---& ---&27&5.1&0.34&  ---&211 \\
B0254--53&D?&55&--8?&5.9&7&6.6&0.89& ---& ---& ---&129?& --- \\
B0355+54&arT/M&\bf{42}&\bf{--9.2}&4.2& ---& ---& ---&40&14.6&0.29& ---&221 \\
B0450+55&arT&\bf{32}&\bf{--8.5}&3.5& ---& ---& ---&34&10.0&0.35& ---&226 \\
B0540+23&D/T?&30?&\bf{--3.4}&8.5& ---& ---& ---& 29?& 11.7& 0.72& ---& 224\\
 \vspace{2pt}
 {\it B0643+80}&Sd/D?&22&+6&3.6&9&4.0&0.89& ---& ---& ---&130& --- \\
B0740--28&arM?&90?&\bf{--5.5}&10.5&7?&11.0&0.95&18?&13.8&0.76&135&211 \\
B0809+74$^b$&Sd&8.8&\bf{--1.8}&4.9& ---& ---& ---&17.0&5.1&0.95& ---&227 \\
B0906--17&arT&31&--6?&4.9&17?&6.8&0.73& ---& ---& ---&124& --- \\
B0919+06$^c$&T&\bf{53}&+9&5.1&10&6.5&0.78& ---& ---& ---&122& --- \\
B1055--52i$^d$&M?&\bf{22}&\bf{+9.1}&2.4& ---& ---& ---&65?&13.1&0.18&--&224 \\
 \vspace{2pt}
{\it B1112+50}&St?&\bf{30}&\bf{+10.1}&2.8&7&3.4&0.84& ---& ---& ---&126& --- \\
B1221--63&T?&61&+7&7.2&13&9.3&0.77& ---& ---& ---&125& --- \\
B1240--64&St&69&+15&3.6&13?&7.1&0.50& ---& ---& ---&131& --- \\
{\it B1322+83m}&Sd?&\bf{14}&\bf{+2.8}&5.1&12?&5.4&0.95& ---& ---& ---&130& --- \\
B1356--60&St&\bf{79}&+3$^a$&19.1& ---& ---& ---& ---& ---& ---& ---& --- \\
B1426--66&T&\bf{54}&--50$^a$&0.9&12&5.0&0.19& ---& ---& ---&131& --- \\
B1449--64&St&\bf{43}&+7$^a$&5.6&25?&10.5&0.53& ---& ---& ---&132& --- \\
 \vspace{2pt}
B1530+27m&Sd/D?&30&\bf{+5.8}&4.9& ---& ---& ---&9&5.5&0.90& ---&225 \\
B1530--53&D?&22&--18$^a$&1.2&19?&3.8&0.31& ---& ---& ---&128& --- \\
{\it B1540--06}&Sd&59&--14?&3.5&9?&5.3&0.67& ---& ---& ---&131& --- \\
B1556--44&St/T&\bf{32}&--9$^a$&3.4&28&8.6&0.40& ---& ---& ---&125& --- \\
B1604--00&cT&50&--8?&5.5&9.8&6.7&0.82& ---& ---& ---&128& --- \\
B1612+07&Sd&25&--4.6&5.2& ---& ---& ---&4.5&5.3&0.98& ---&224 \\
 \vspace{2pt}
B1641--45&St&\bf{33}&$\infty$&0.0&24&6.5&0.0& ---& ---& ---&128& --- \\
B1648--42&D/cT&6.5&--4$^a$&1.6& ---& ---& ---&100&6.3&0.26& ---&226 \\
{\it B1700--18}&Sd&44&\bf{--8.2}&4.7& ---& ---& ---&12&6.4&0.74& ---&221 \\
{\it B1732--07}&T?&\bf{54}&$\infty$&0.0&17&6.8&0.0& ---& ---& ---&131& --- \\
{\it B1742--30}&M&\bf{24}&--3.6&6.4&15?&7.3&0.89&32&9.7&0.67&129&228 \\
{\it B1745--12}&T/cQ?&75&\bf{--11.7}&4.9& ---& ---& ---&16?&9.2&0.53& ---&222 \\
{\it B1756--22}&St/T?&$\sim$90?&$\infty$&0.0&12?&6.2&0.0& ---& ---& ---&118& --- \\
B1822--09m$^e$&T&\bf{86}&$\infty$&0.0&10.0&5.0&0.0& ---& ---& ---&128& --- \\
{\it B1842+14}&St?&63&+12&4.2&12&6.9&0.62& ---& ---& ---&119& --- \\
{\it B1851--14}&Sd?&34?&\bf{--7.8}&4.1& ---& ---& ---&12?&5.4&0.76& ---&224 \\
(B1859+07)$^c$&T/M?&30&+6&4.8& ---& ---& ---&20&7.1&0.67& ---&219 \\
{\it B1900+05}&St?&59?&$\infty$&0.0&12?&5.1&0.0& ---& ---& ---&132& --- \\
{\it B1907--03}&St/T&\bf{44}&$\infty$&0.0&18&6.2&0.0& ---& ---& ---&129& --- \\
B1910+20&cQ/M&\bf{32}&\bf{+30}&1.0&$\sim$10.5&3.0&0.34&14&3.9&0.26&133&225 \\
{\it B1913+10}&St?&\bf{64}&?& ---& ---& ---& ---& ---& ---& ---& ---& --- \\
B1915+13&arSt&68&\bf{--9.8}&5.4& ---& ---& ---& ---& ---& ---& ---& --- \\
 \vspace{2pt}
B1924+16&arSt&\bf{34}&\bf{+5.2}&6.2& ---& ---& ---& ---& ---& ---& ---& --- \\
{\it B1930+22}&arSt?&---&\bf{+8.6}&---&---&---&---& ---& ---& ---&---& --- \\
B1937--26&T?&\bf{42}&\bf{--4.5}&8.5& ---& ---& ---&9&9.1&0.94& ---&222 \\
{\it B1944+17}$^f$&cT/cQ&\bf{5}&\bf{+0.8}&6.3&$\sim$95?&8.7&0.72&30&6.6&0.95& 222&126 \\
{\it B1944+22}&Sd/D?&40&--12?&3.1&7?&3.8&0.80& ---& ---& ---&132& --- \\
B2021+51&Sd?&\bf{30}&\bf{+3.9}&7.3& ---& ---& ---&10.3&7.8&0.93& ---&216 \\
B2043--04&Sd/D&73&\bf{+27.1}&2.0&6?&3.5&0.57& ---& ---& ---&128& --- \\
B2053+36&St&\bf{34}&$\infty$& ---& ---& ---& ---& ---& ---& ---& ---& --- \\
B2217+47&St&\bf{42}&+8.5&4.5&12.0&6.1&.73& ---& ---& ---&135& --- \\
 \vspace{2pt}
B2224+65m&St?&\bf{27}&\bf{--3.6}& 4.9& ---& ---& ---& ---& ---& ---& ---& --- \\
B2327--20&T?&\bf{66}&\bf{+43}&1.2&7.0&3.4&0.35&---&--&--&128& --- \\
\hline
\end{tabular}
\end{center}
\end{table*}

{\footnotesize\noindent Table 3 Notes:  Pulsars with Bnames in normal type appear in 
L\&M's table 4; those 
in italics are denoted as ``Partial cones?'' in their Table 5; and one other star, 
B1859+07 (parentheses) is included from our own work.  The $\alpha$ values 
in boldface were determined using the core-width method; while 
the others were estimated from profile dimensions.  The $|\Delta\chi/\Delta\varphi|_o$ 
values in boldface were determined by PPA fitting; the others were taken from 
Paper VI or the $^a$ values from L\&M.  Other geometric solutions as follows: 
$^b$RRS/RRvLS; $^c$RRW; $^d$WW09; $^e$BMR; $^f$KL10.}

\begin{table*}
\caption{Table of A/R height estimates}
\label{tab4}
\begin{center}
\begin{tabular} {lllllll}\hline \hline
{\bf PSR } &  {\bf $P_1$}  & \bf{Left width} & \bf{Right Width} & $\sigma_{\phi_\circ}$ & \bf{Shift}/ & \bf{Height}\\
  Bname    &   {\bf (sec)}  & \bf{(\degr)}     &  \bf{(\degr)}&  \bf{(\degr)}& \bf{(\degr)}& \bf (km) \\
\hline
B0355+54  & 0.156 & --39.7 $\pm$0.5  & 9.3 $\pm$0.4  & 0.5  & 15.2 $\pm$0.6  & 494 $\pm$ 19 \\
B0450+55  & 0.340 & --18.5 $\pm$0.2  & 10.8$\pm$0.2  & 0.1  & 3.9  $\pm$0.2  & 275 $\pm$ 12 \\
B1700--18$^*$ & 0.804 & --6.6  $\pm$0.1  & 3.3 $\pm$0.1  & 1.2  & 1.6  $\pm$1.2  & 279 $\pm$ 201 \\
B1732--07$^*$ & 0.419 & --10.0 $\pm$0.2  & 5.3 $\pm$0.2  & 0.1  & 2.3  $\pm$0.2  & 205 $\pm$ 15 \\
B1742--30 & 0.367 & --16.2 $\pm$0.1  & 13.9$\pm$0.5  & 0.1  & 1.1  $\pm$0.3  & 87 $\pm$  20 \\
B1745--12 & 0.394 & --12.0 $\pm$0.5  & 6.8 $\pm$0.5  & 0.8  & 2.6  $\pm$0.8  & 215 $\pm$ 71 \\
B1910+20  & 2.232 & --6.4  $\pm$0.1  & 5.4 $\pm$0.1  & 0.1  & 0.5  $\pm$0.1  & 228 $\pm$ 56 \\
B1924+16  & 0.579 & --17.0 $\pm$0.3  & 8.6 $\pm$0.3  & 0.7  & 4.2  $\pm$0.7  & 506 $\pm$ 88 \\
B1930+22  & 0.144 & --14.5 $\pm$0.3  & --2.2 $\pm$0.3  & 1.5  & 8.3  $\pm$1.5  & 250  $\pm$ 45\\
B1937--26 & 0.402 & --3.9  $\pm$0.2  & 2.2 $\pm$0.2  & 1.5  & 0.8  $\pm$1.5  & 71  $\pm$ 126\\
B2021+51  & 0.529 & --18.3 $\pm$0.2  & 16.2$\pm$0.2  & 0.6  & 1.0  $\pm$0.6  & 113 $\pm$ 67 \\
B2043--04 & 1.546 & --3.5  $\pm$0.1  & 4.1 $\pm$0.1  & 1.6  & --0.2 $\pm$1.6  & --96 $\pm$ 515 \\
B2327--20 & 1.643 & --2.6  $\pm$0.1  & 3.0 $\pm$0.1  & 0.4  & --0.1 $\pm$0.4  & --68 $\pm$ 139 \\
\hline
\end{tabular}
\end{center}
\footnotesize 
{Notes: The table gives the pulsar name, period, measured outer conal left and right widths
and the shift w.r.t the SG point ($^*$or core component) which is taken as the longitude 
origin. The estimated A/R heights are given in the last column.}
\end{table*}

\begin{figure*}
\begin{center}
\begin{tabular}{@{}lr@{}}
{\mbox{\includegraphics[width=78mm,angle=-90.]{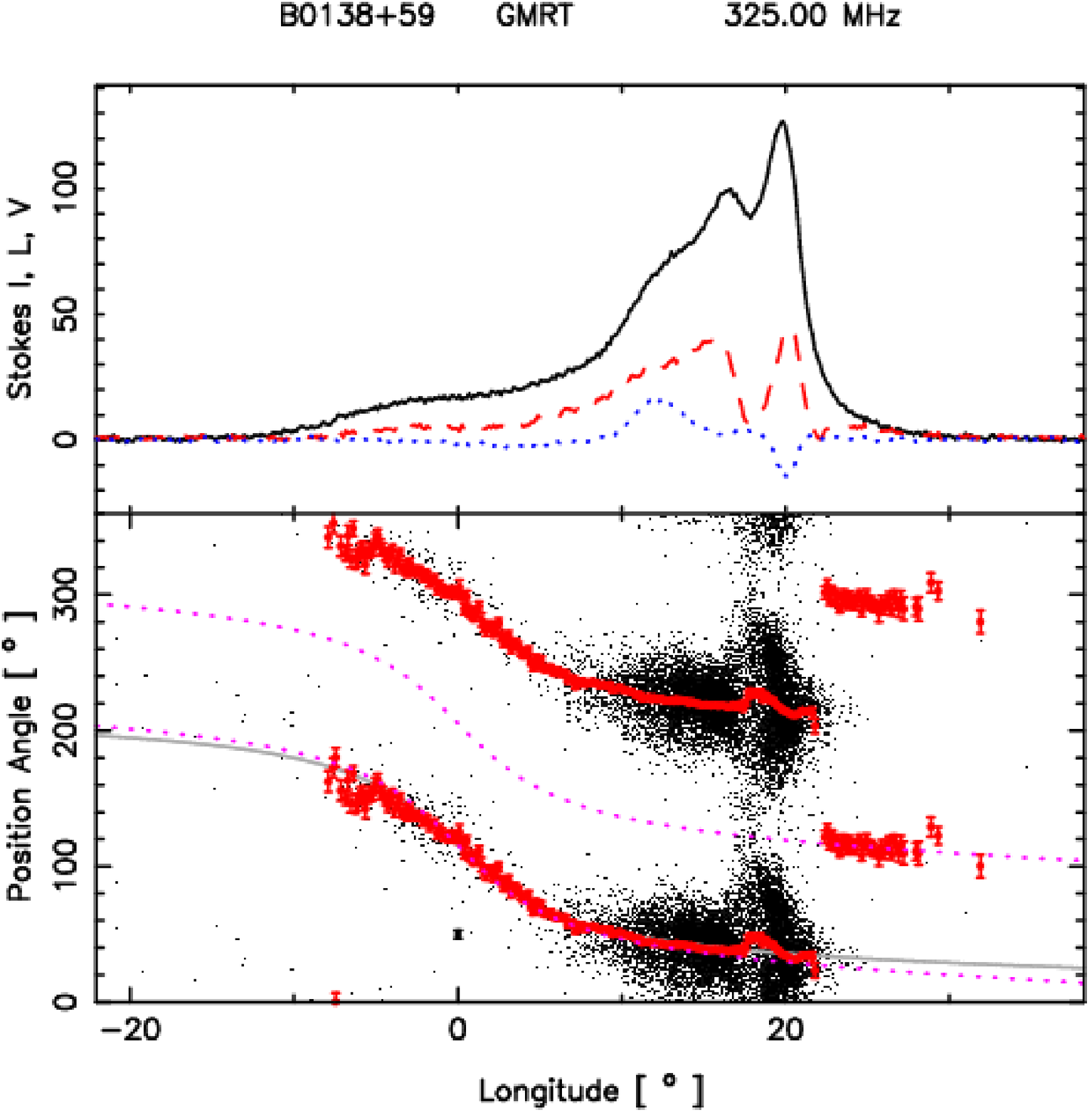}}}&
{\mbox{\includegraphics[width=78mm,angle=-90.]{B0355+54_PAHIST.ps}}}\\
{\mbox{\includegraphics[width=78mm,angle=-90.]{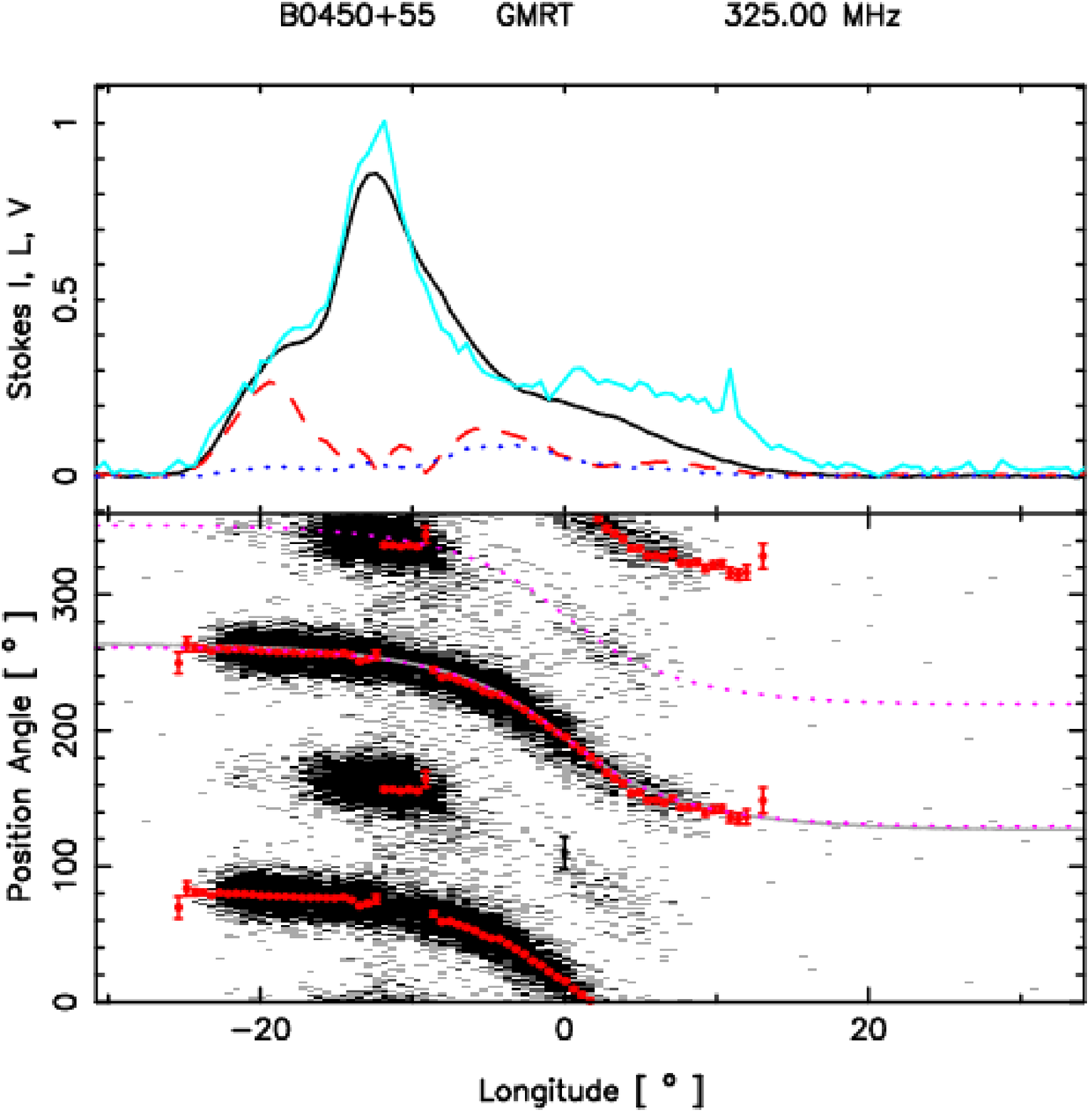}}}&
{\mbox{\includegraphics[width=78mm,angle=-90.]{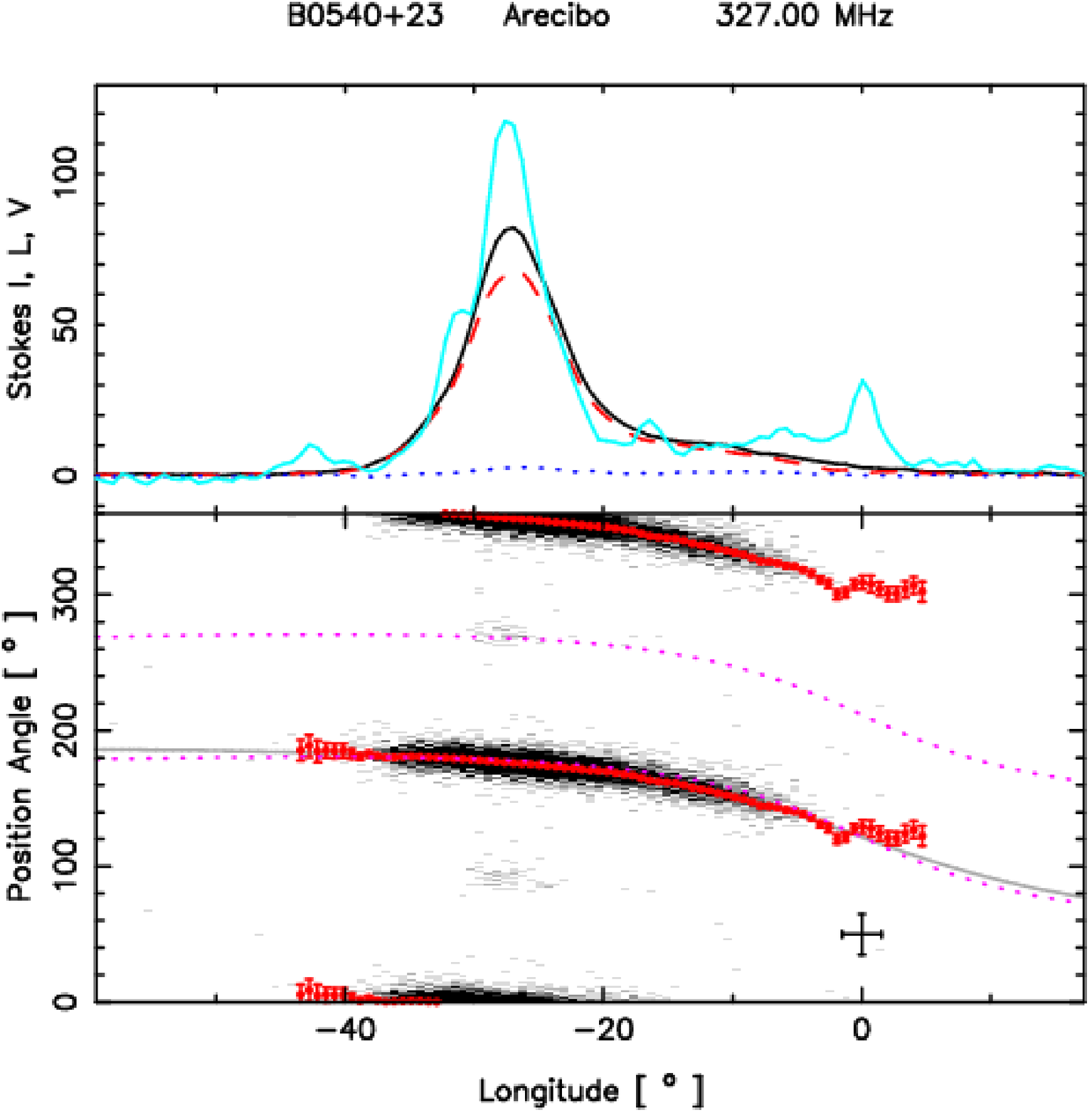}}}\\
\end{tabular}
\caption{PPA histograms and ``flared''-emission profiles for ``partial cone'' 
pulsars B0138+59, B0355+54, B0450+55 and B0540+23, where the instrument 
and band is indicated above each plot.  The respective upper panels give the
total power (black), total linear (red) and circular polarization LH-RH (blue).  
The cyan curve (if plotted) correspond to the computed flared profile (see 
text for details).  The lower panels give the polarization-angle (hereafter PPA) 
density, plotted twice for clarity.  In cases where PPA fits were possible, a 
solid grey curve depicts the results of Table~\ref{tab5}, the longitude origin is 
taken at the corresponding steepest gradient point (hereafter SG), and error 
bars show its uncertainty; whereas the two dotted (magenta) curves indicate 
the primary- and secondary-mode (hereafter PPM and SPM) PPA traverses 
corresponding to the geometric models in Table~\ref{tab3}.  Otherwise, when 
no RVM fitting information was available, the zero longitude was usually chosen 
as the peak of the profile (unless mentioned otherwise in notes on each pulsar 
in this Appendix).}
\label{figA1}
\end{center}
\end{figure*}

\begin{figure*}
\begin{center}
\begin{tabular}{@{}lr@{}}
{\mbox{\includegraphics[width=78mm,angle=-90.]{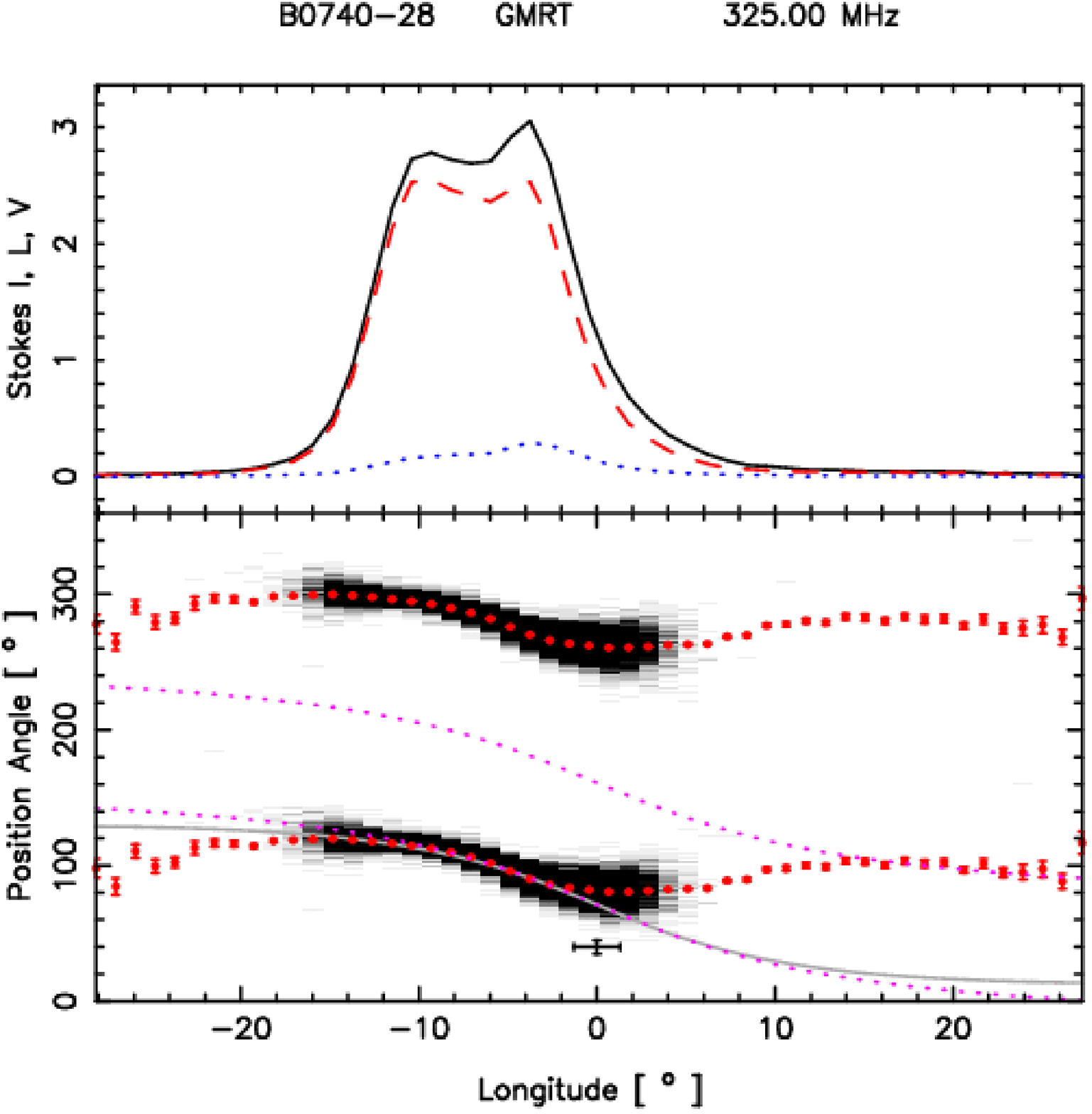}}}& \ \ \ \ \ \ 
{\mbox{\includegraphics[width=78mm,angle=-90.]{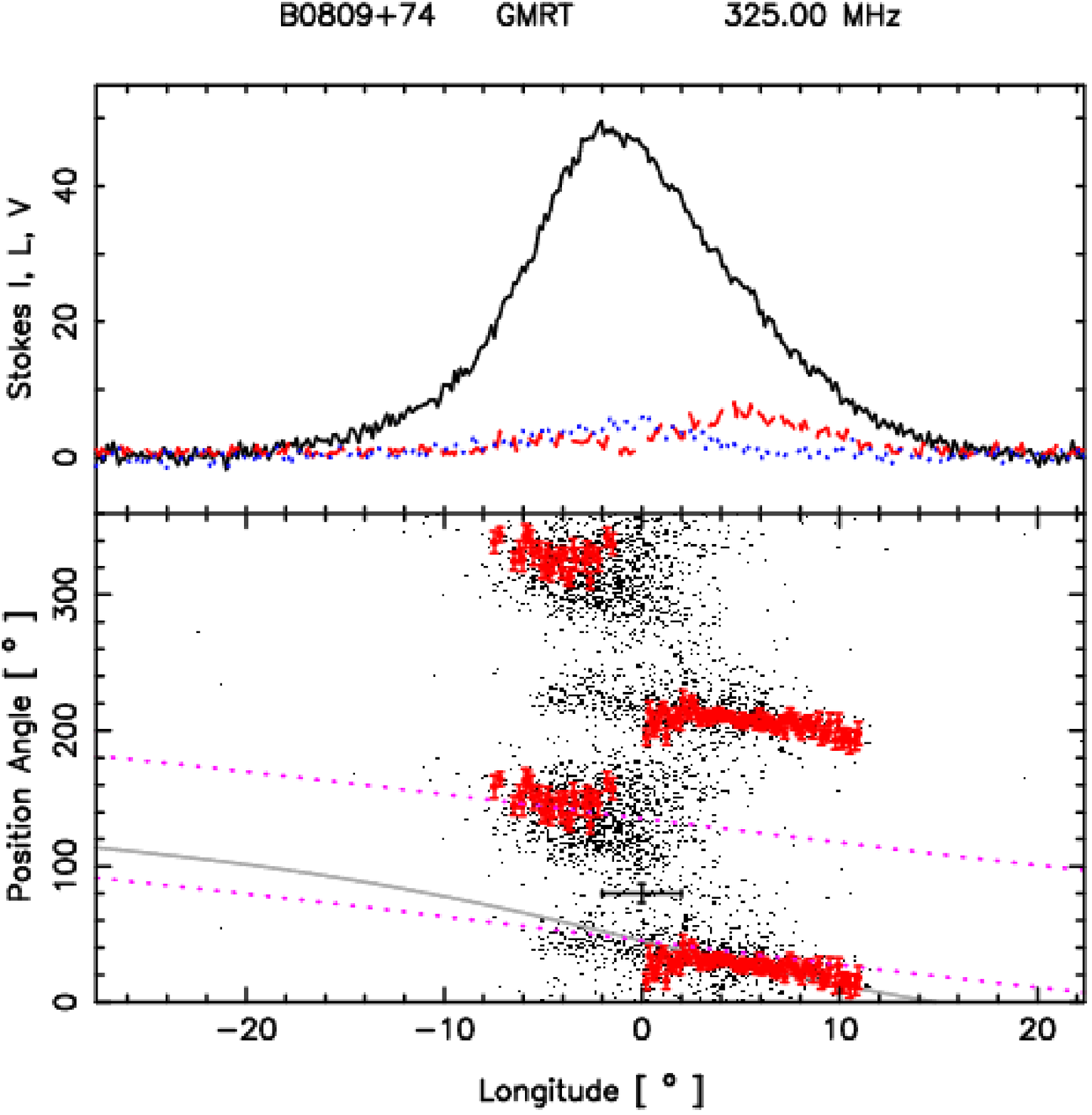}}}\\
{\mbox{\includegraphics[width=78mm,angle=-90.]{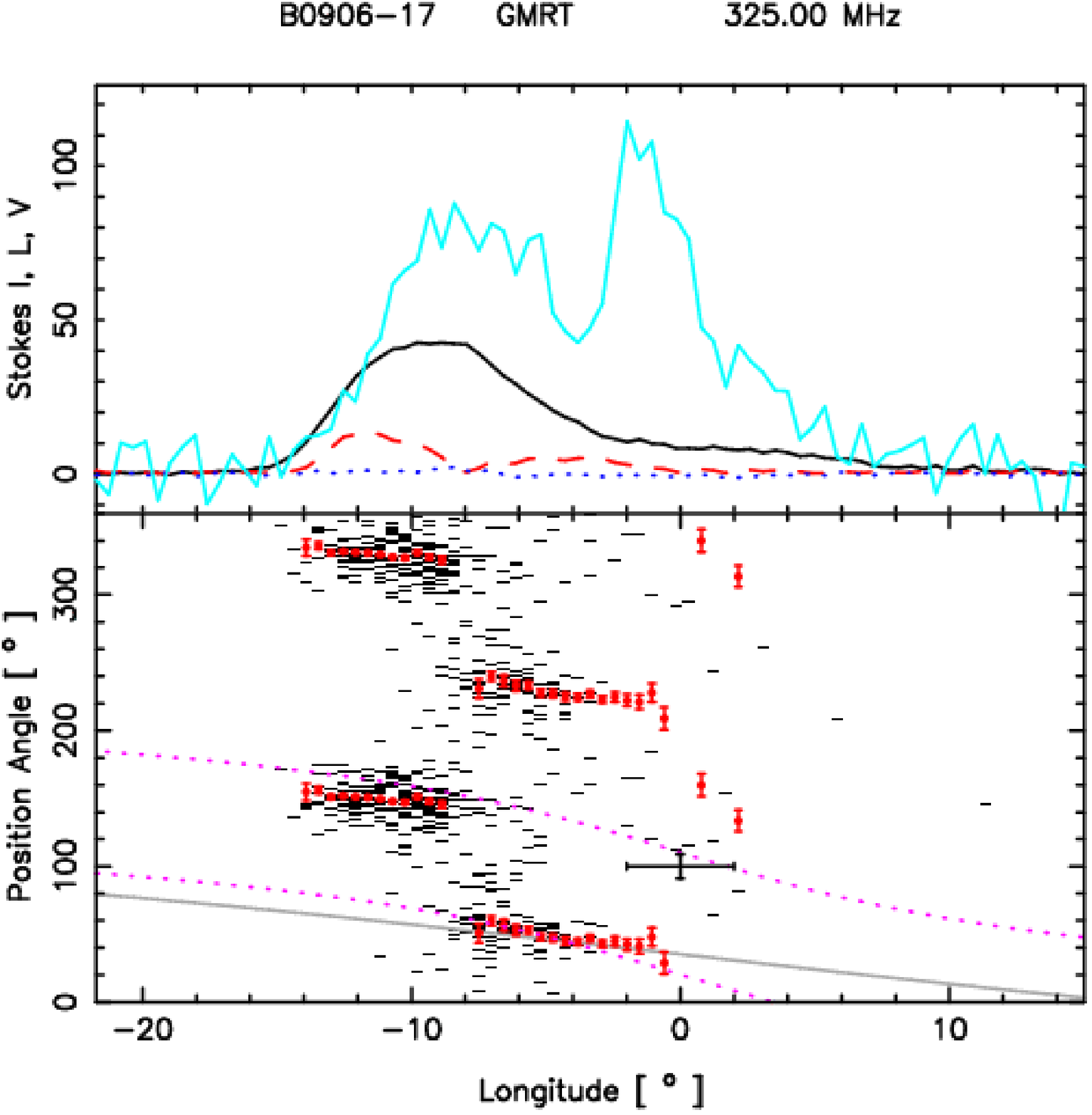}}}& \ \ \ \ \ \ 
{\mbox{\includegraphics[width=78mm,angle=-90.]{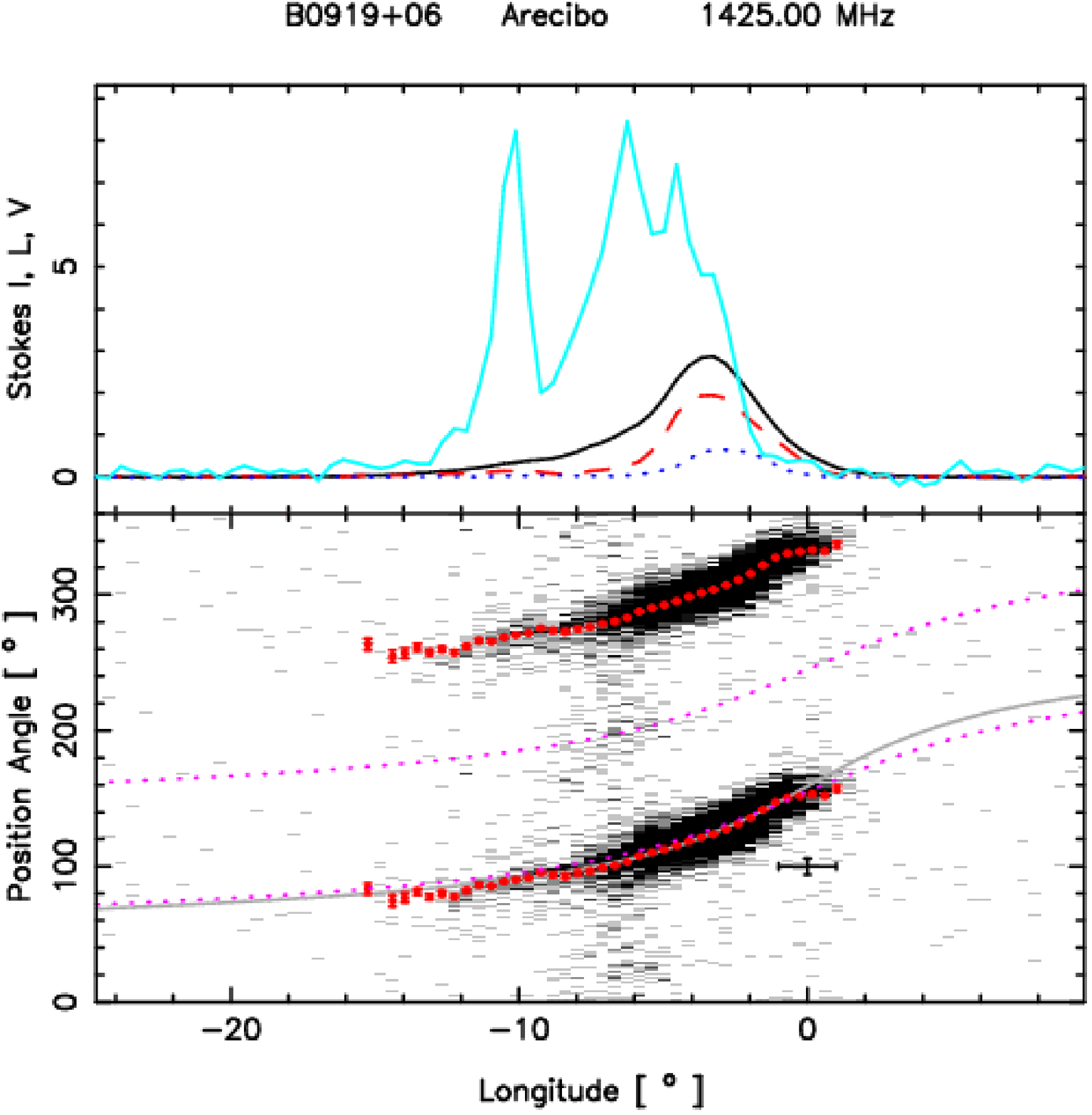}}}\\
\end{tabular}
\caption{PPA histograms and ``flared''-emission profiles as in Fig.~\ref{figA1} 
for pulsars B0740--28, B0809+74, B0906--17 and B0919+06.}
\label{figA2}
\end{center}
\end{figure*}

\begin{figure*}
\begin{center}
\begin{tabular}{@{}lr@{}}
{\mbox{\includegraphics[width=78mm,angle=-90.]{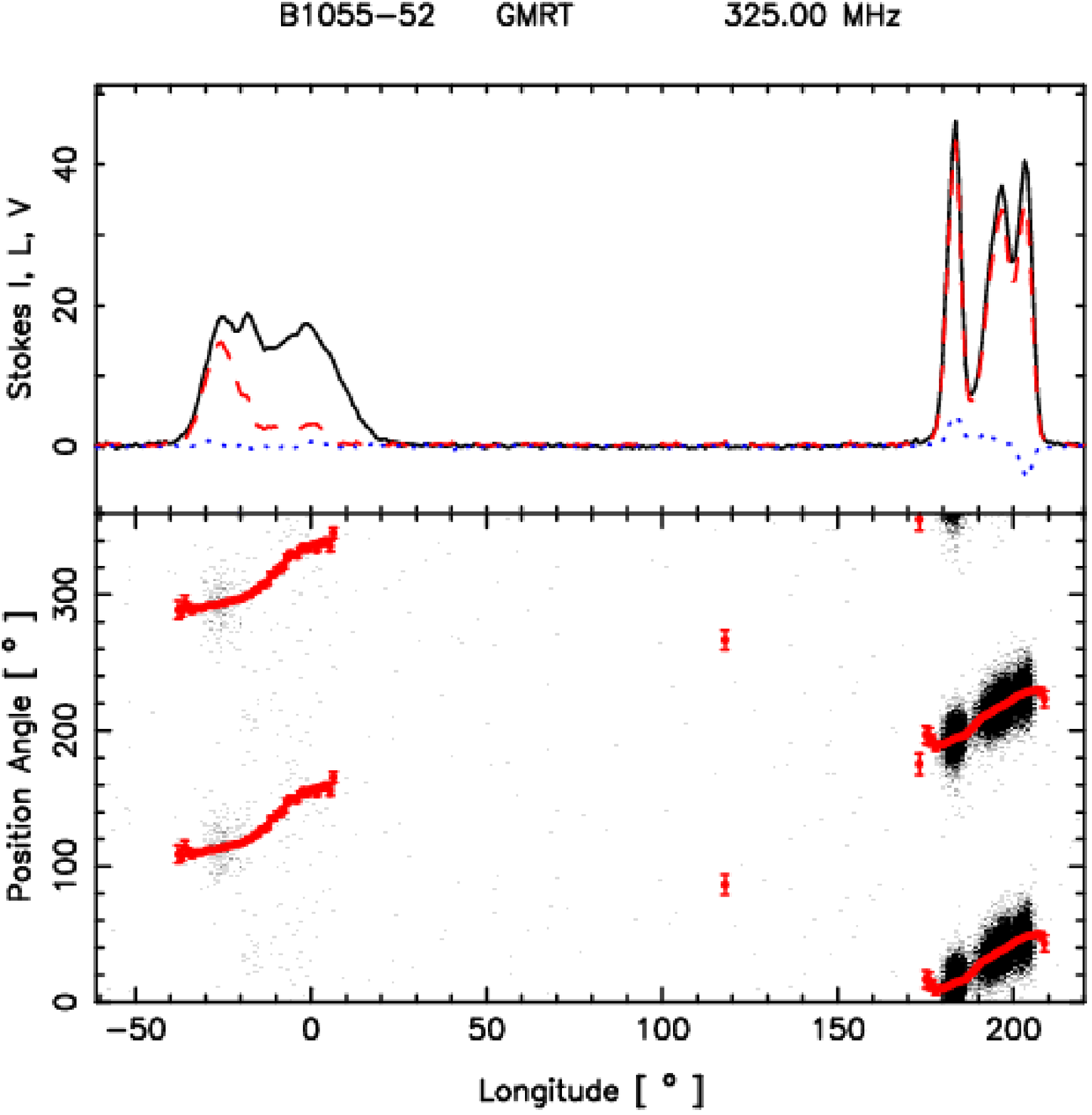}}}& \ \ \ \ \ \ 
{\mbox{\includegraphics[width=78mm,angle=-90.]{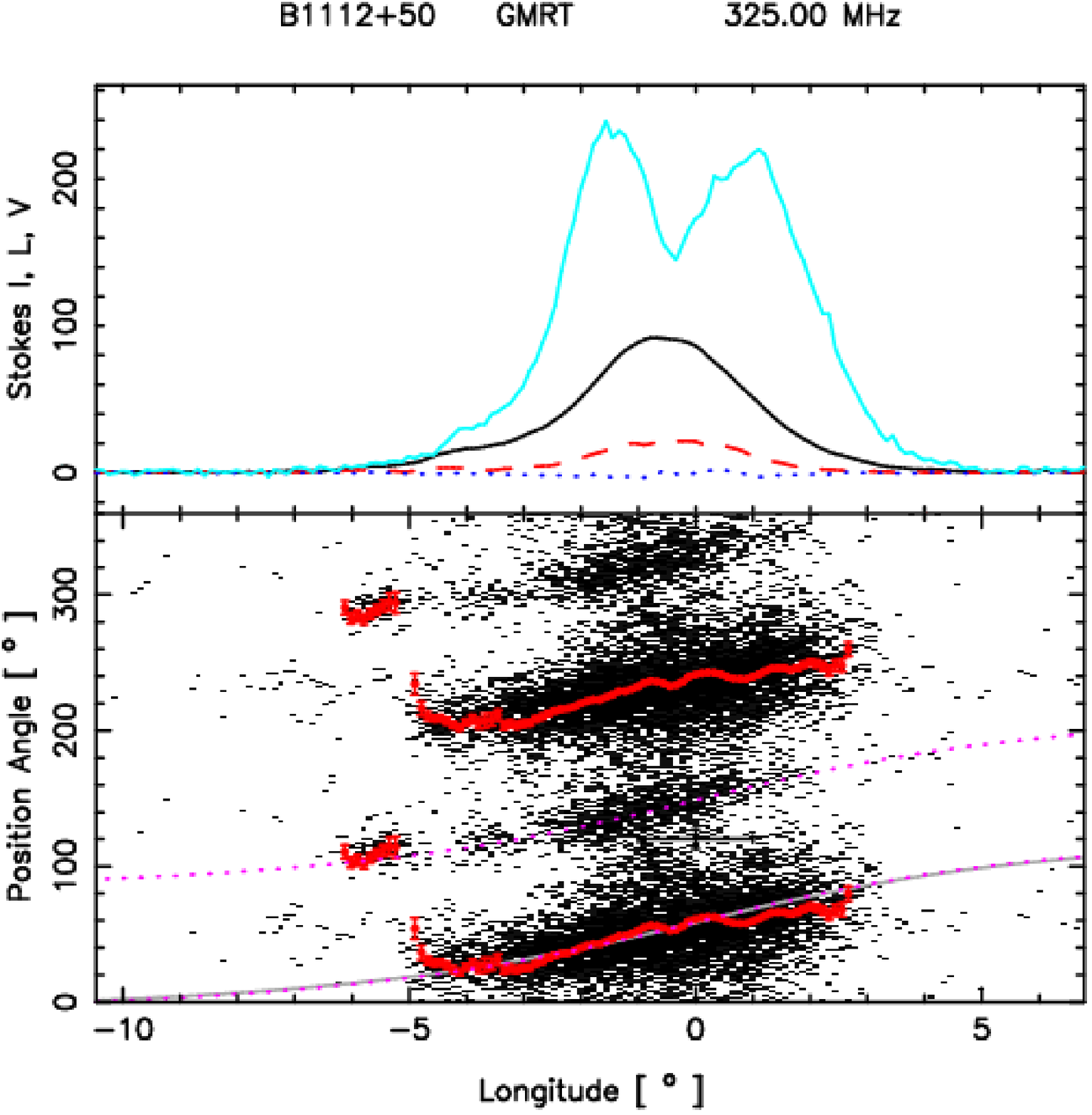}}}\\
{\mbox{\includegraphics[width=78mm,angle=-90.]{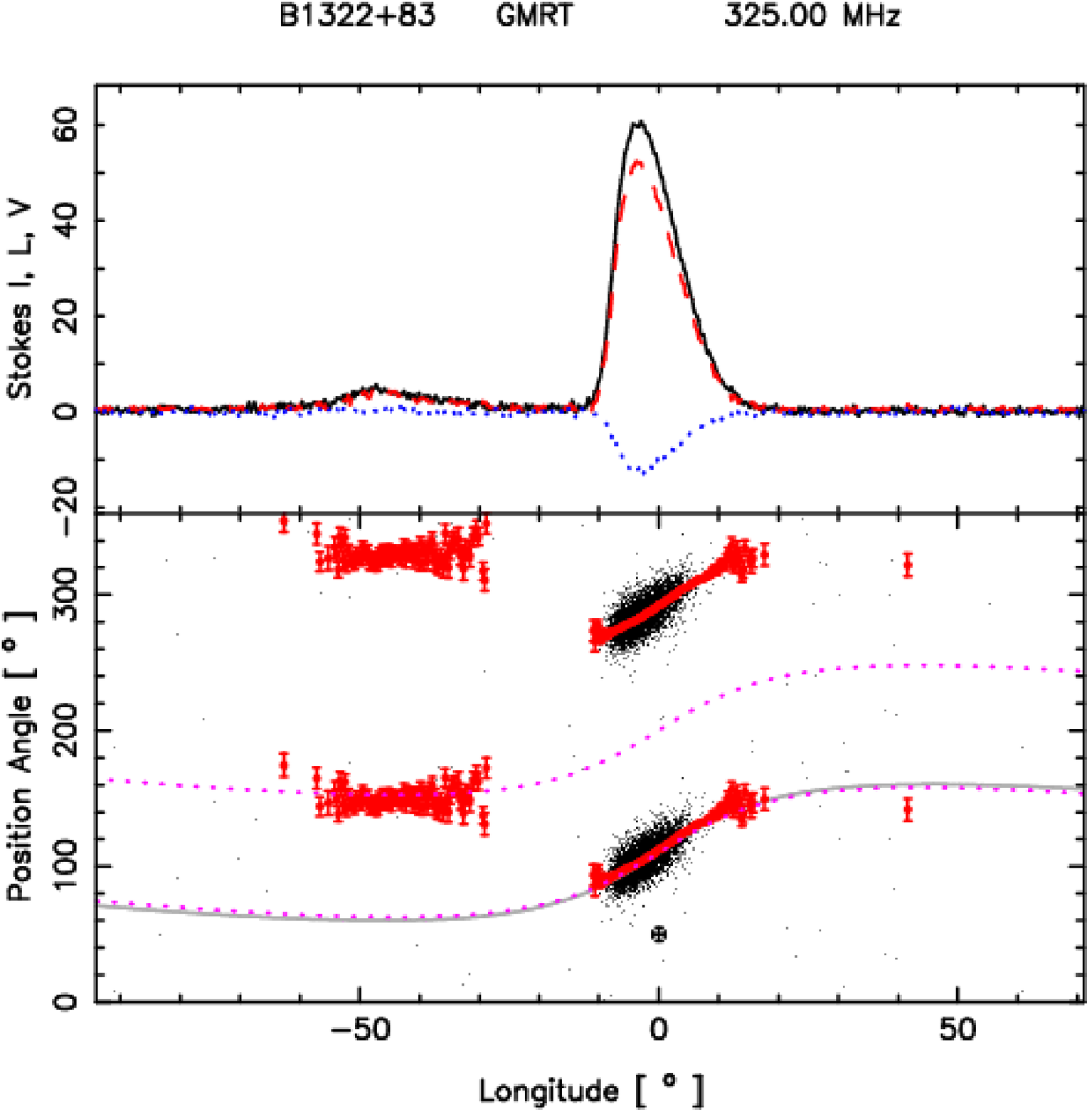}}}& \ \ \ \ \ \ 
{\mbox{\includegraphics[width=78mm,angle=-90.]{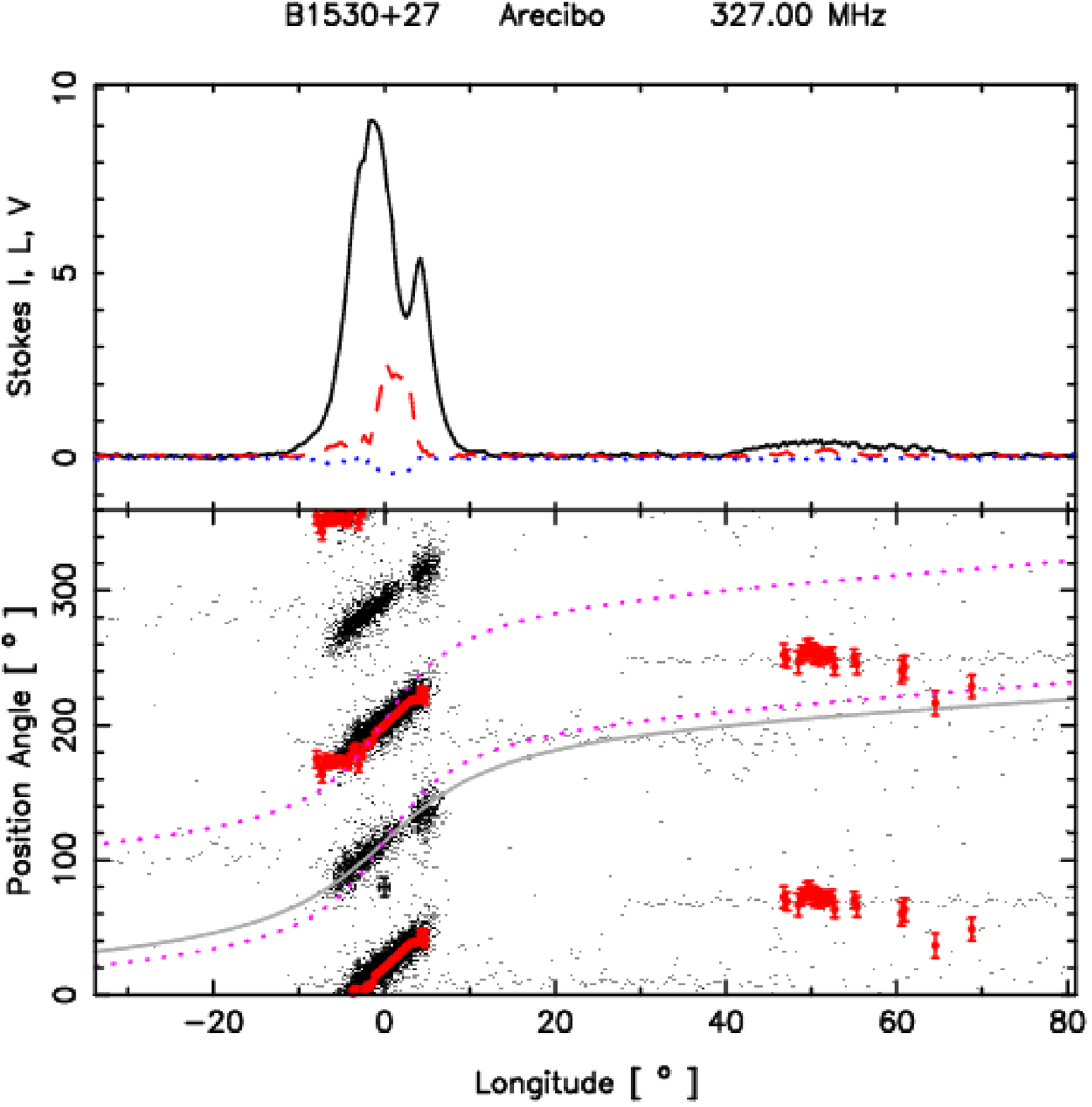}}}\\
\end{tabular}
\caption{PPA histograms and ``flared''-emission profiles as in Fig.~\ref{figA1} 
for pulsars B1055--52, B1112+50, B1322+83 and B1530+27.}
\label{figA3}
\end{center}
\end{figure*}

\begin{figure*}
\begin{center}
\begin{tabular}{@{}lr@{}}
{\mbox{\includegraphics[width=78mm,angle=-90.]{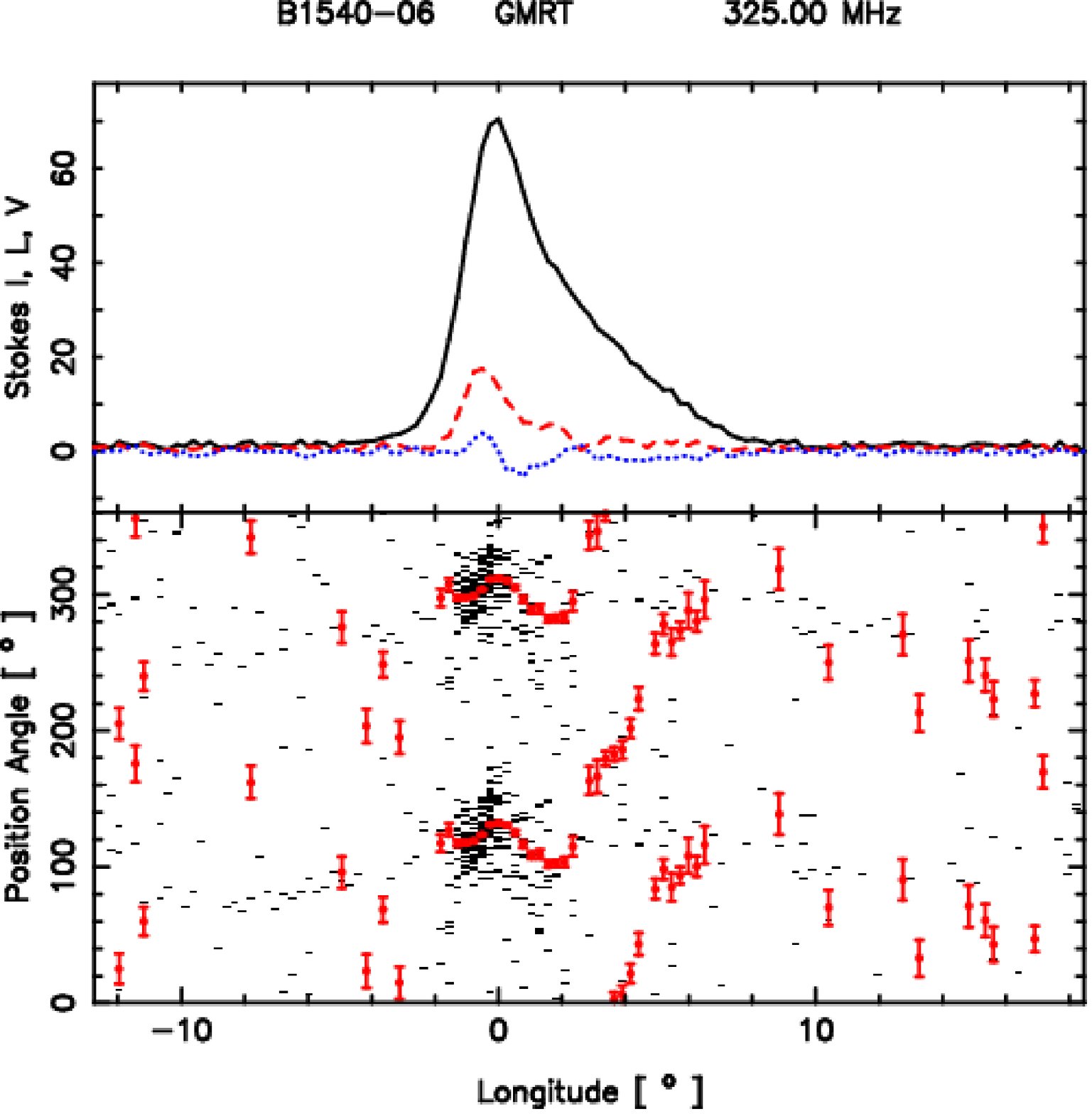}}}& \ \ \ \ \ \ 
{\mbox{\includegraphics[width=78mm,angle=-90.]{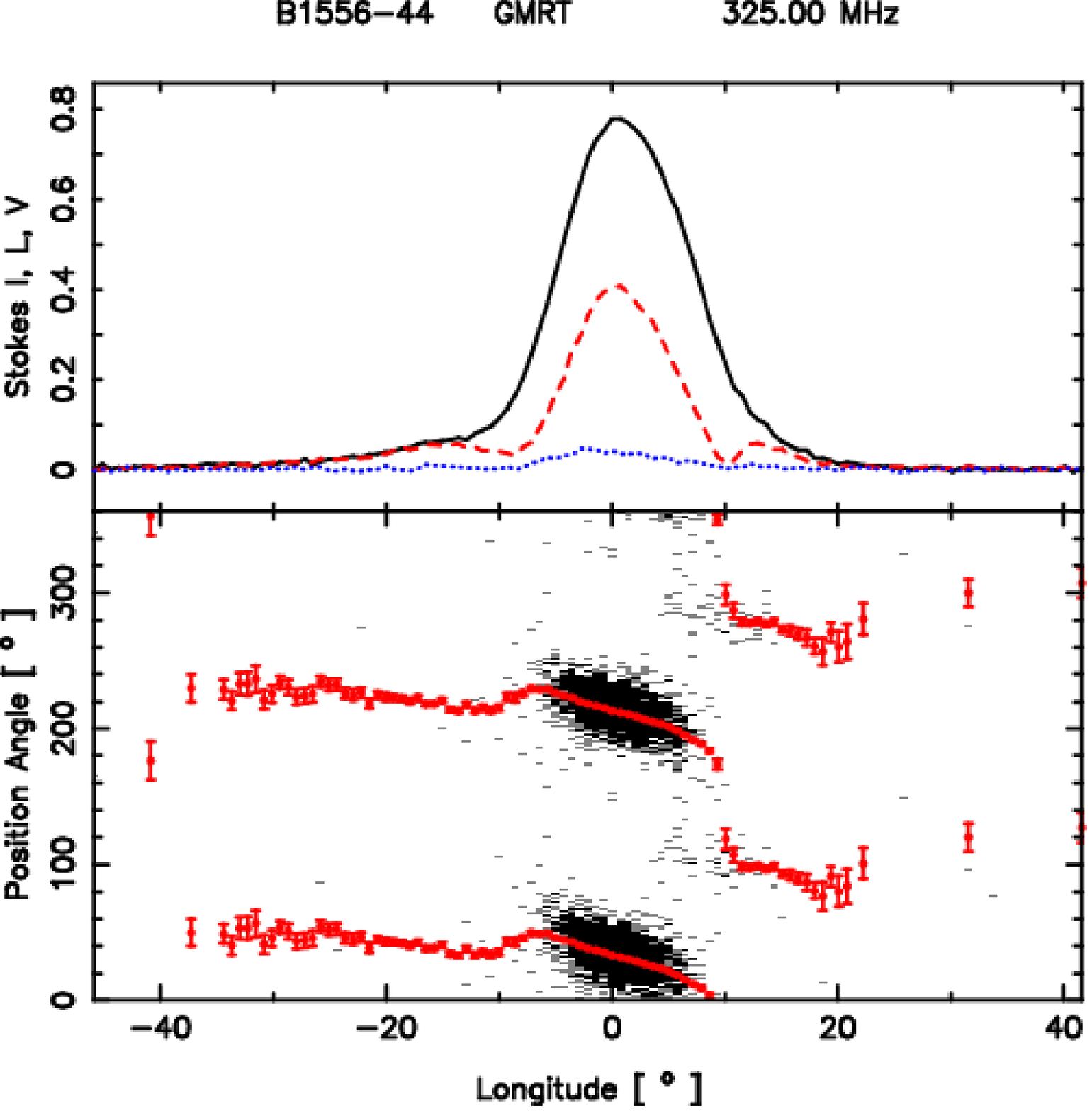}}}\\
{\mbox{\includegraphics[width=78mm,angle=-90.]{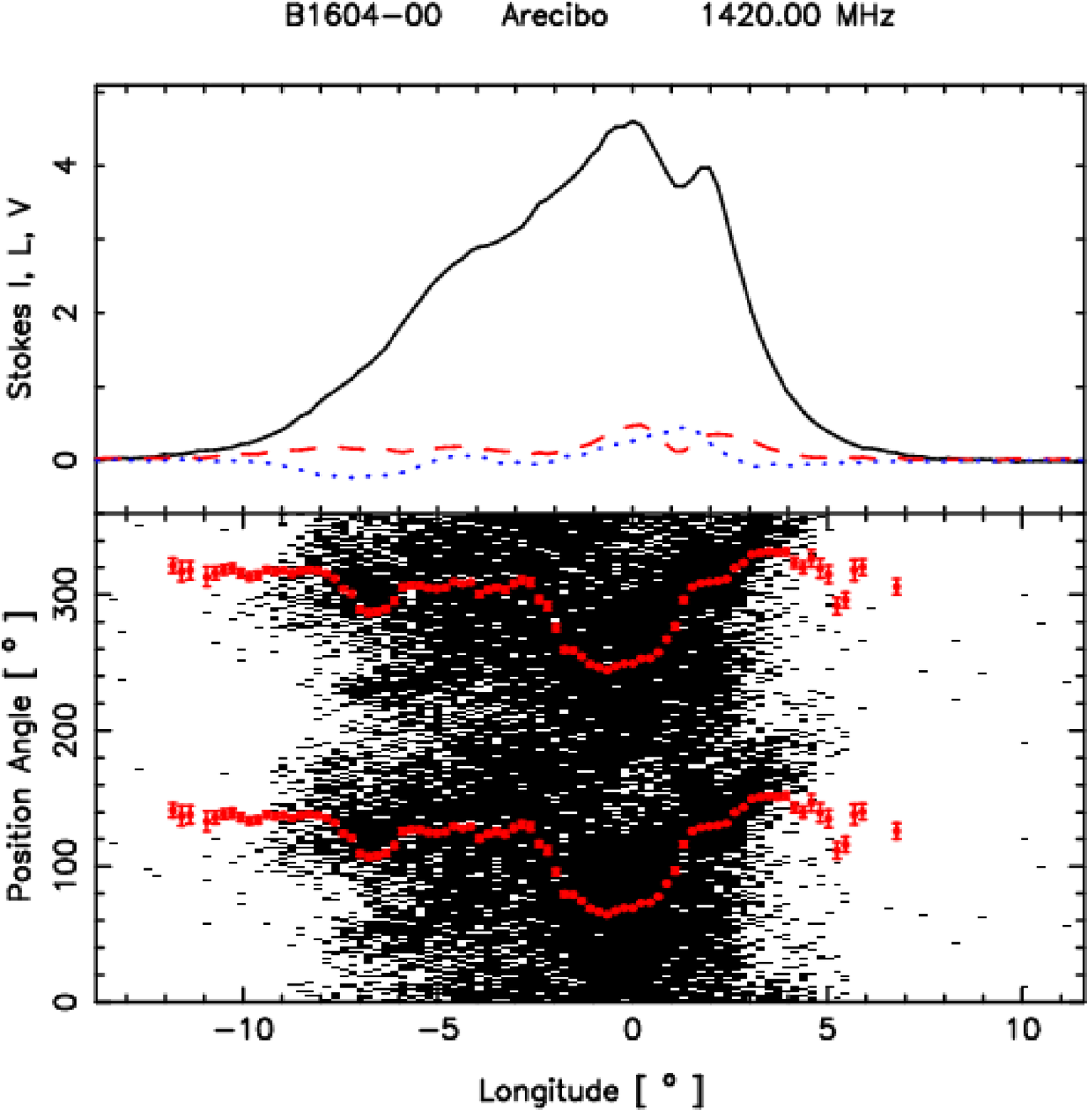}}}& \ \ \ \ \ \ 
{\mbox{\includegraphics[width=78mm,angle=-90.]{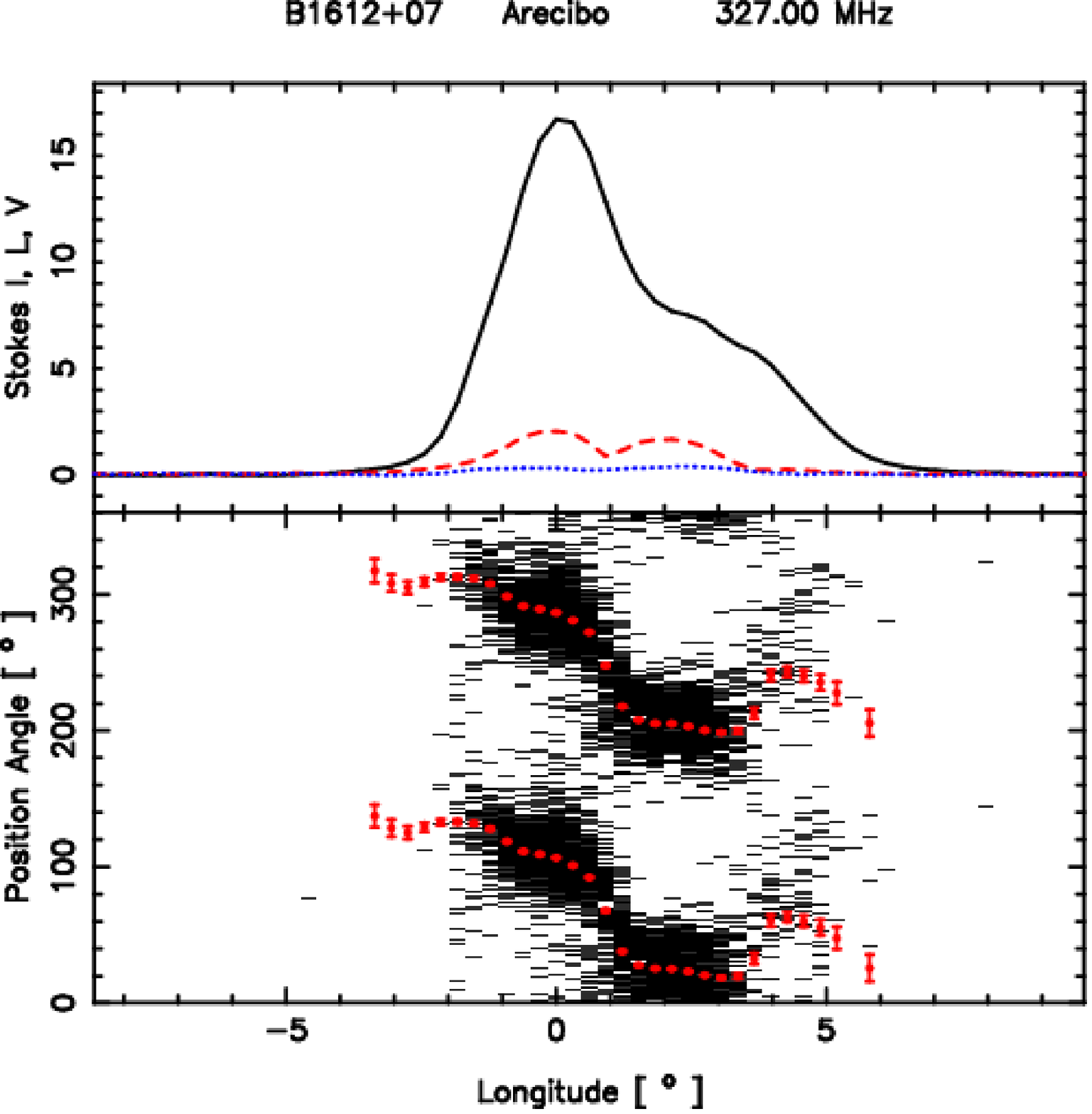}}}\\
\end{tabular}
\caption{PPA histograms and ``flared''-emission profiles as in Fig.~\ref{figA1} 
for pulsars B1540--06, B1556--44, B1604--00 and B1612+07.}
\label{figA4}
\end{center}
\end{figure*}

\begin{figure*}
\begin{center}
\begin{tabular}{@{}lr@{}}
{\mbox{\includegraphics[width=78mm,angle=-90.]{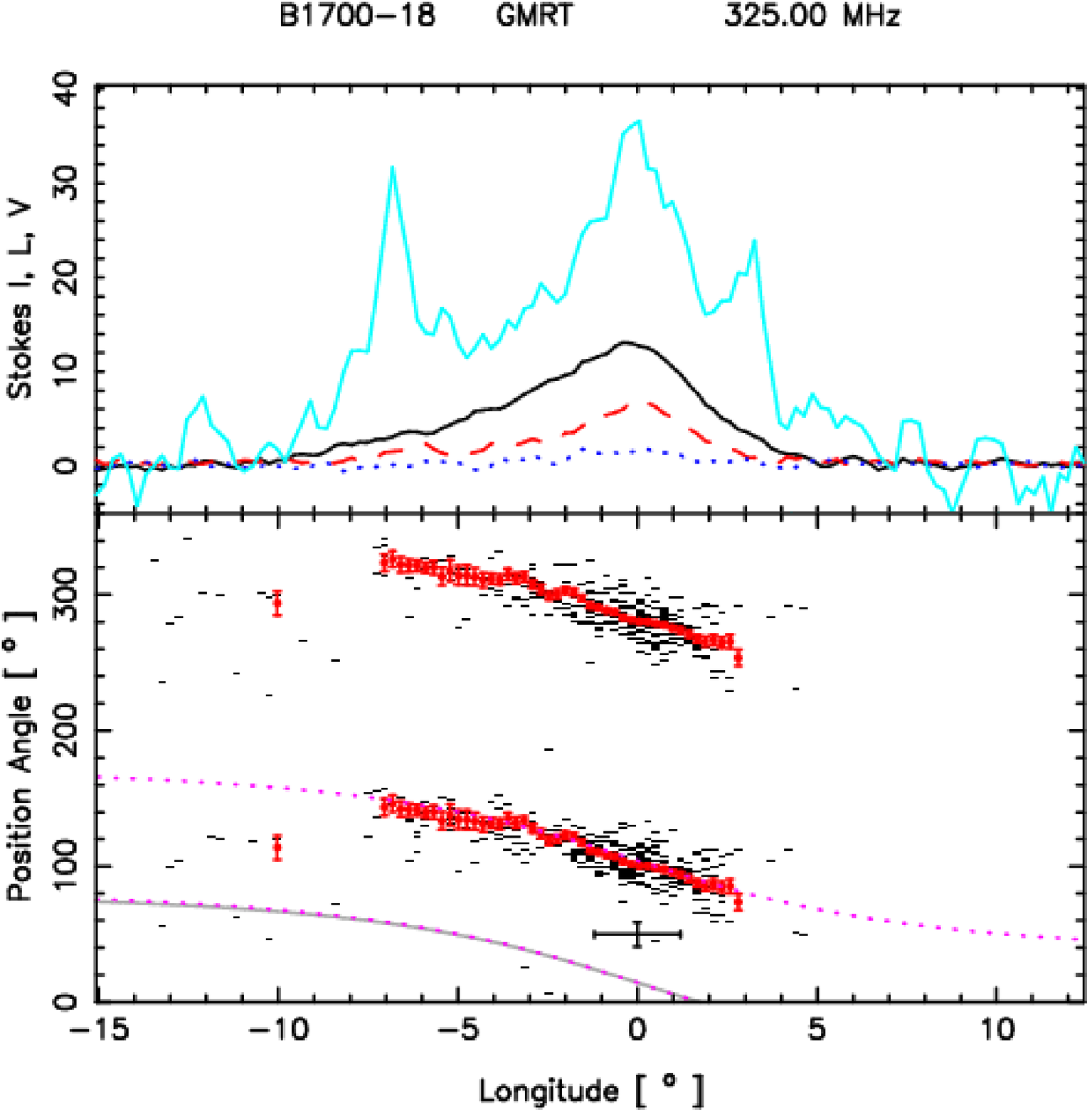}}}& \ \ \ \ \ \ 
{\mbox{\includegraphics[width=78mm,angle=-90.]{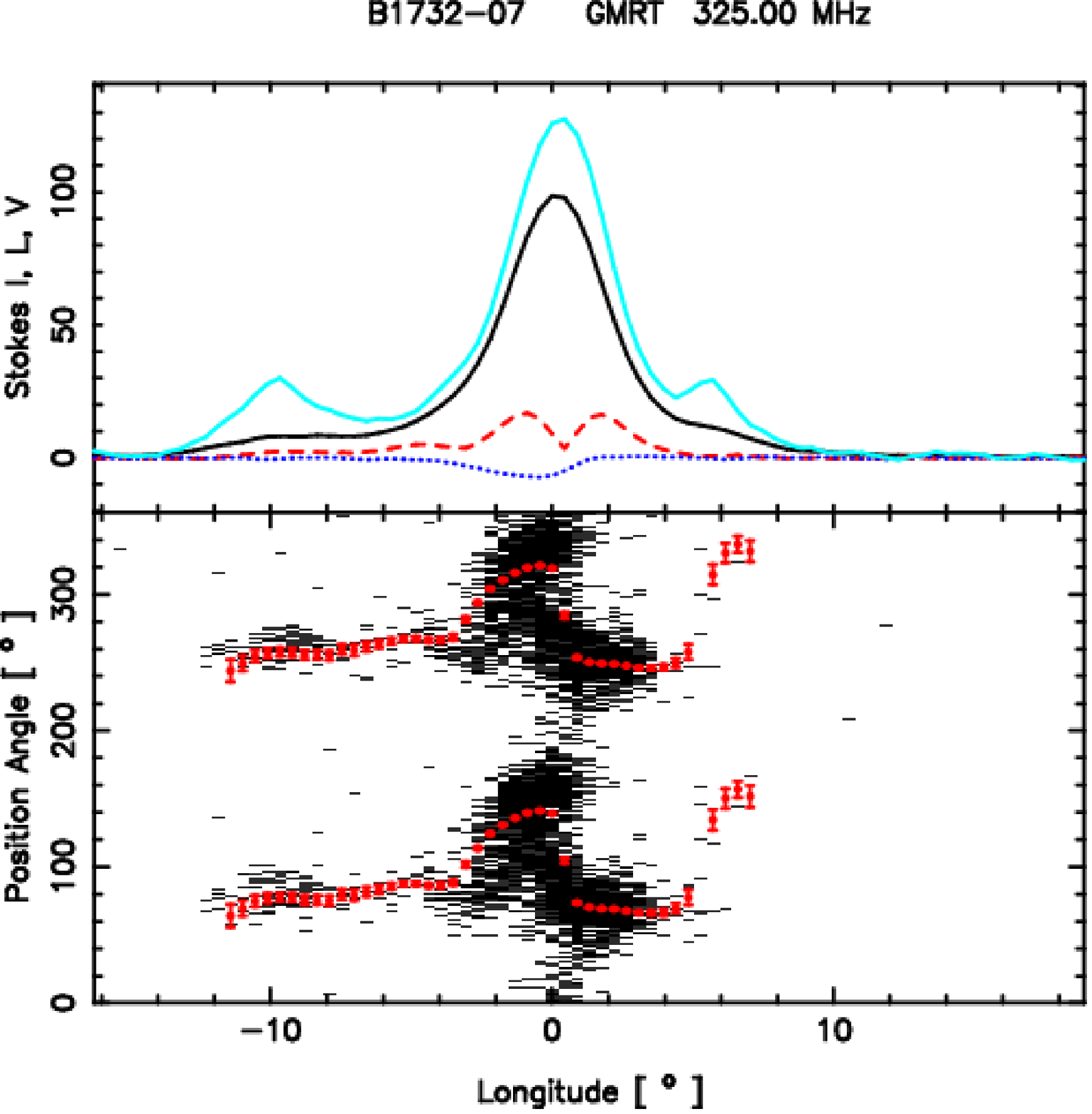}}}\\
{\mbox{\includegraphics[width=78mm,angle=-90.]{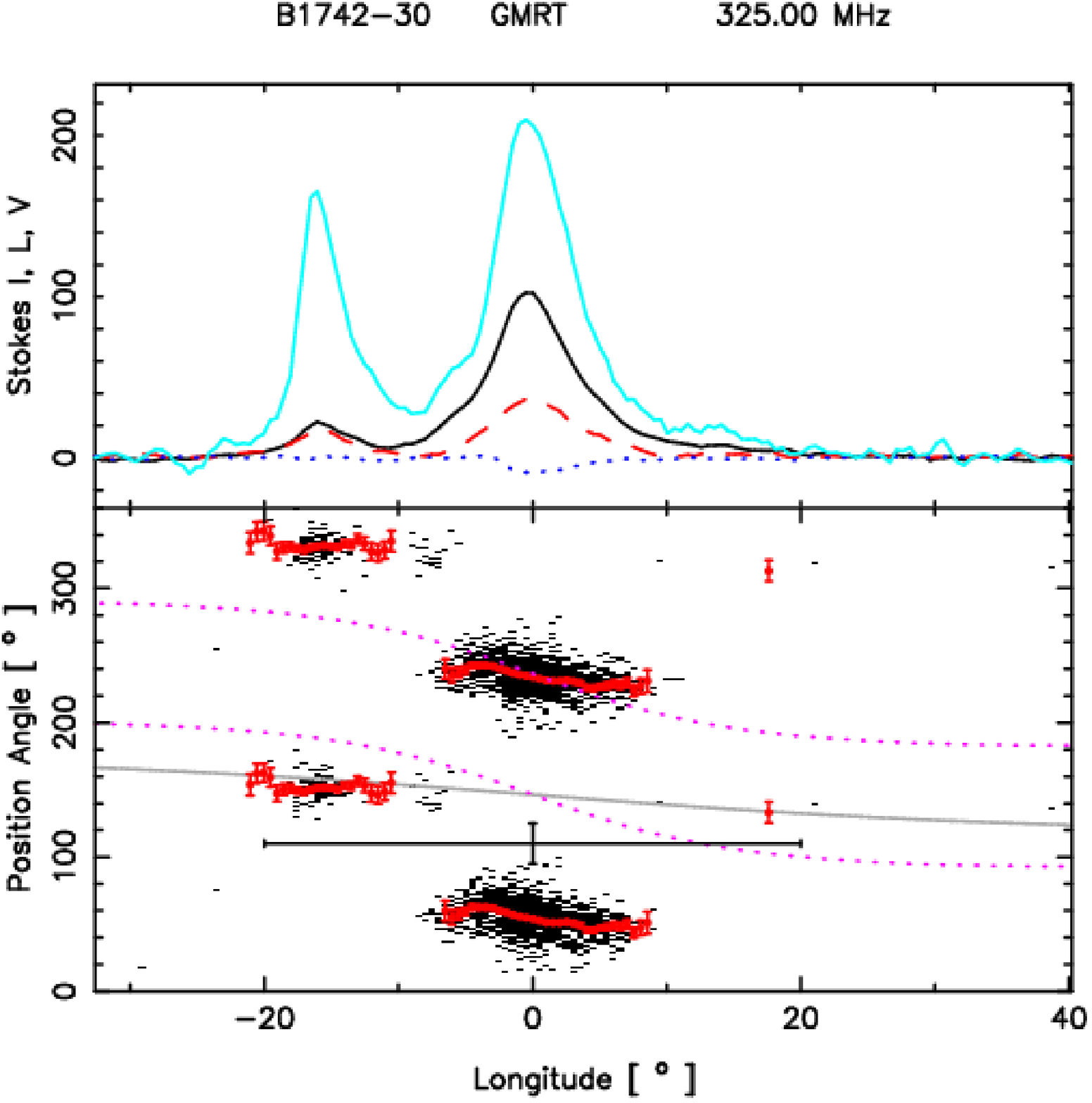}}}& \ \ \ \ \ \ 
{\mbox{\includegraphics[width=78mm,angle=-90.]{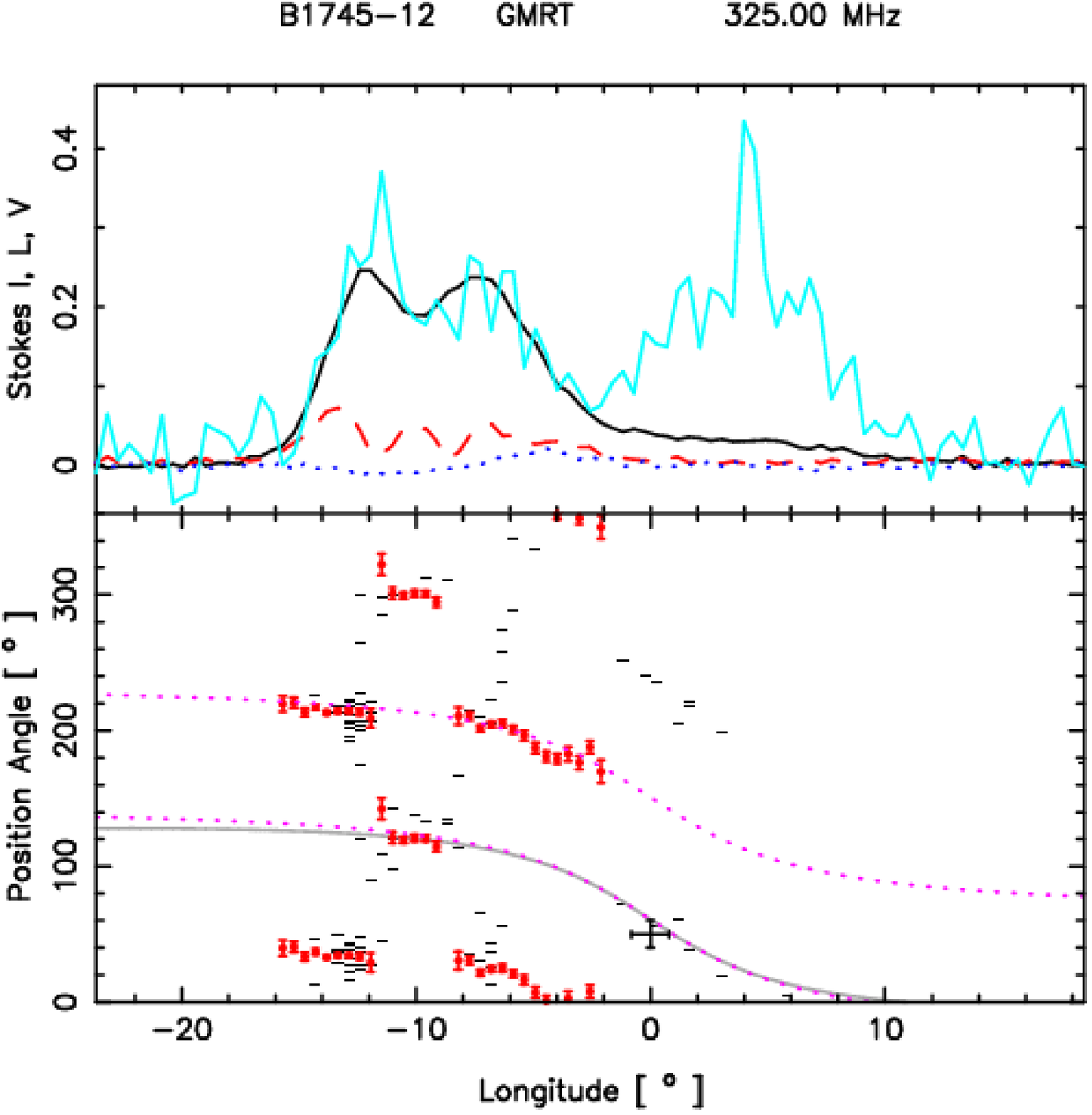}}}\\
\end{tabular}
\caption{PPA histograms and ``flared''-emission profiles as in Fig.~\ref{figA1} 
for pulsars B1700--18, B1732--07, B1742--30 and B1745--12.}
\label{figA5}
\end{center}
\end{figure*}

\begin{figure*}
\begin{center}
\begin{tabular}{@{}lr@{}}
{\mbox{\includegraphics[width=78mm,angle=-90.]{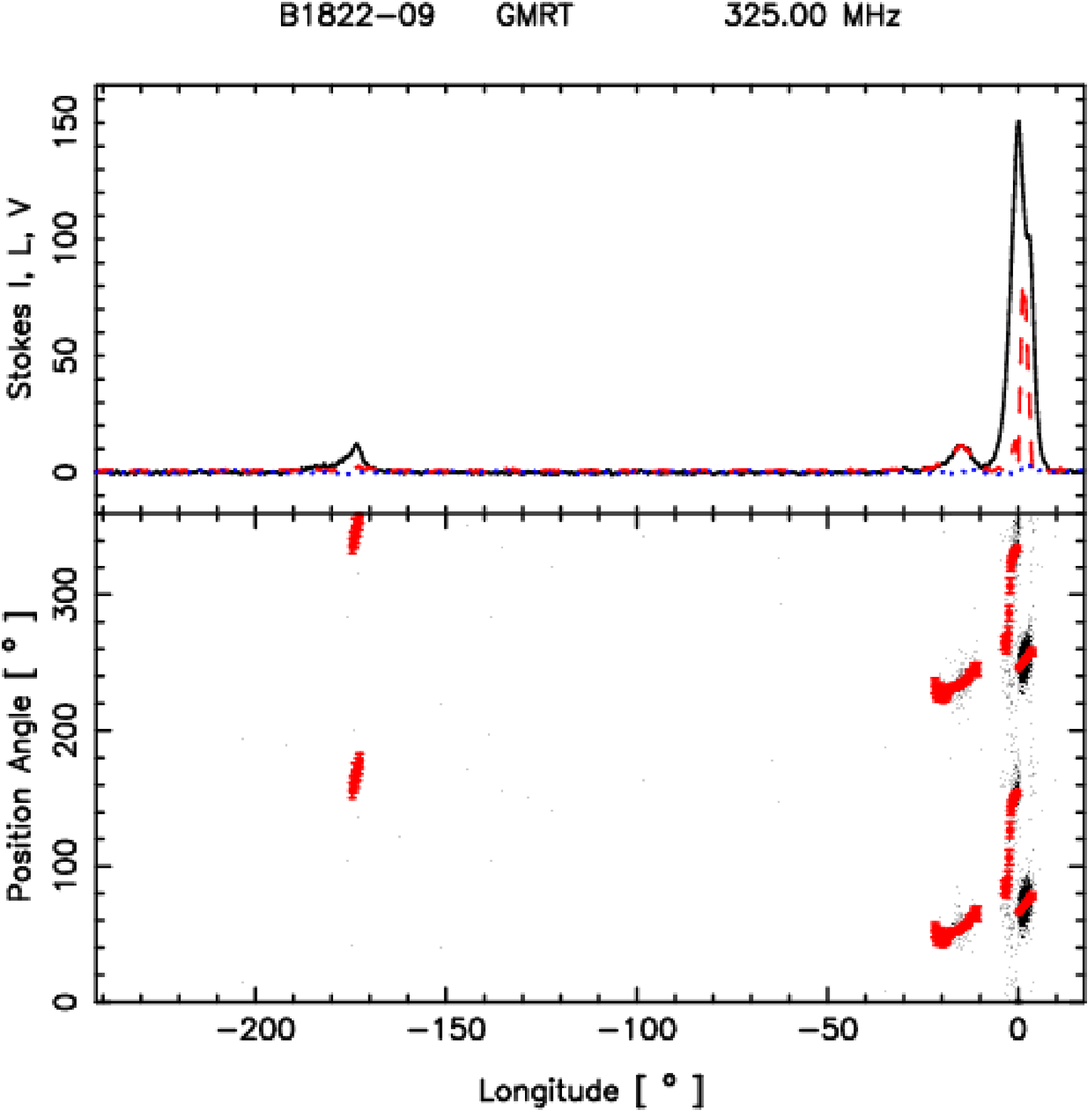}}}& \ \ \ \ \ \ 
{\mbox{\includegraphics[width=78mm,angle=-90.]{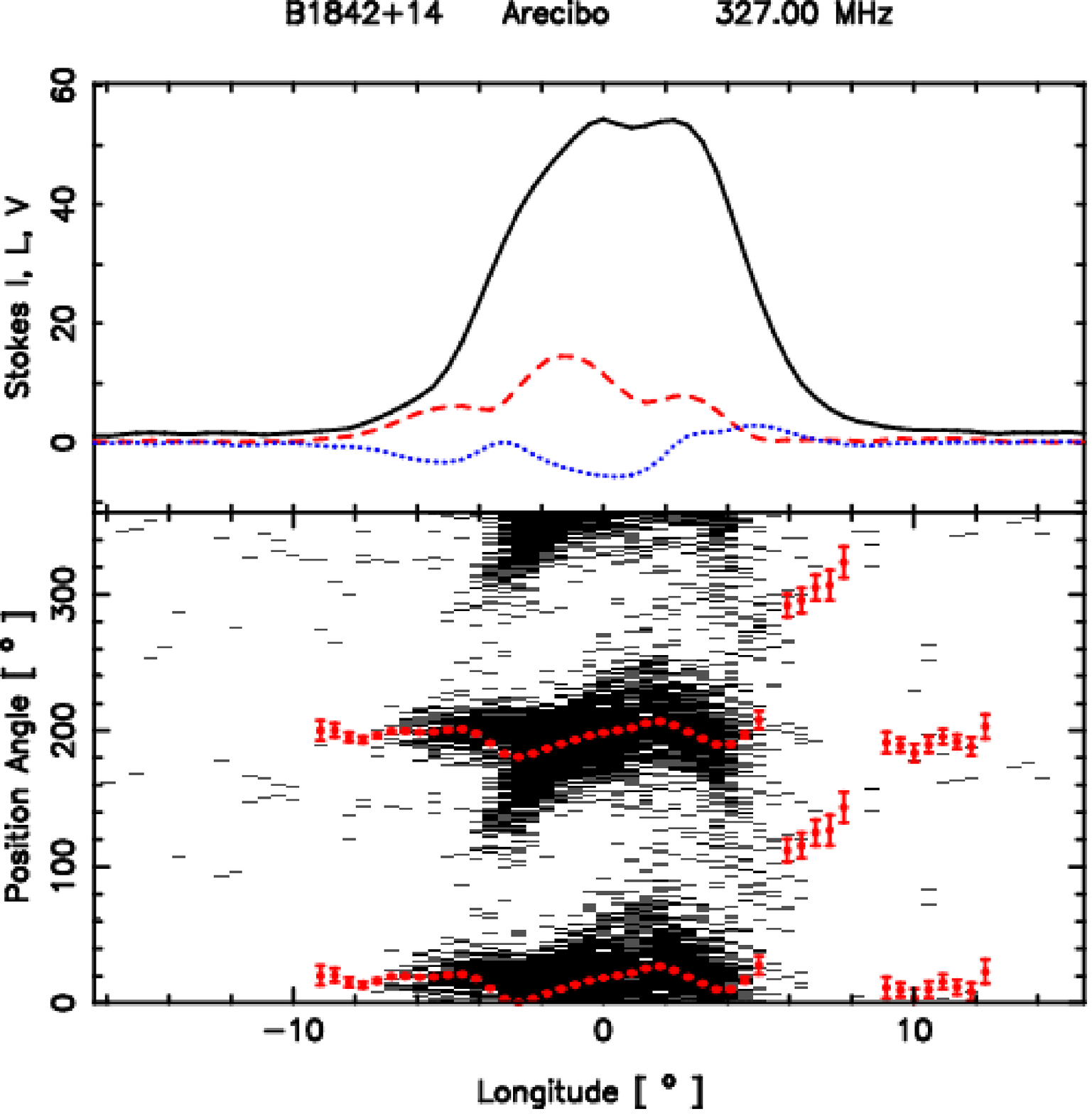}}}\\
{\mbox{\includegraphics[width=78mm,angle=-90.]{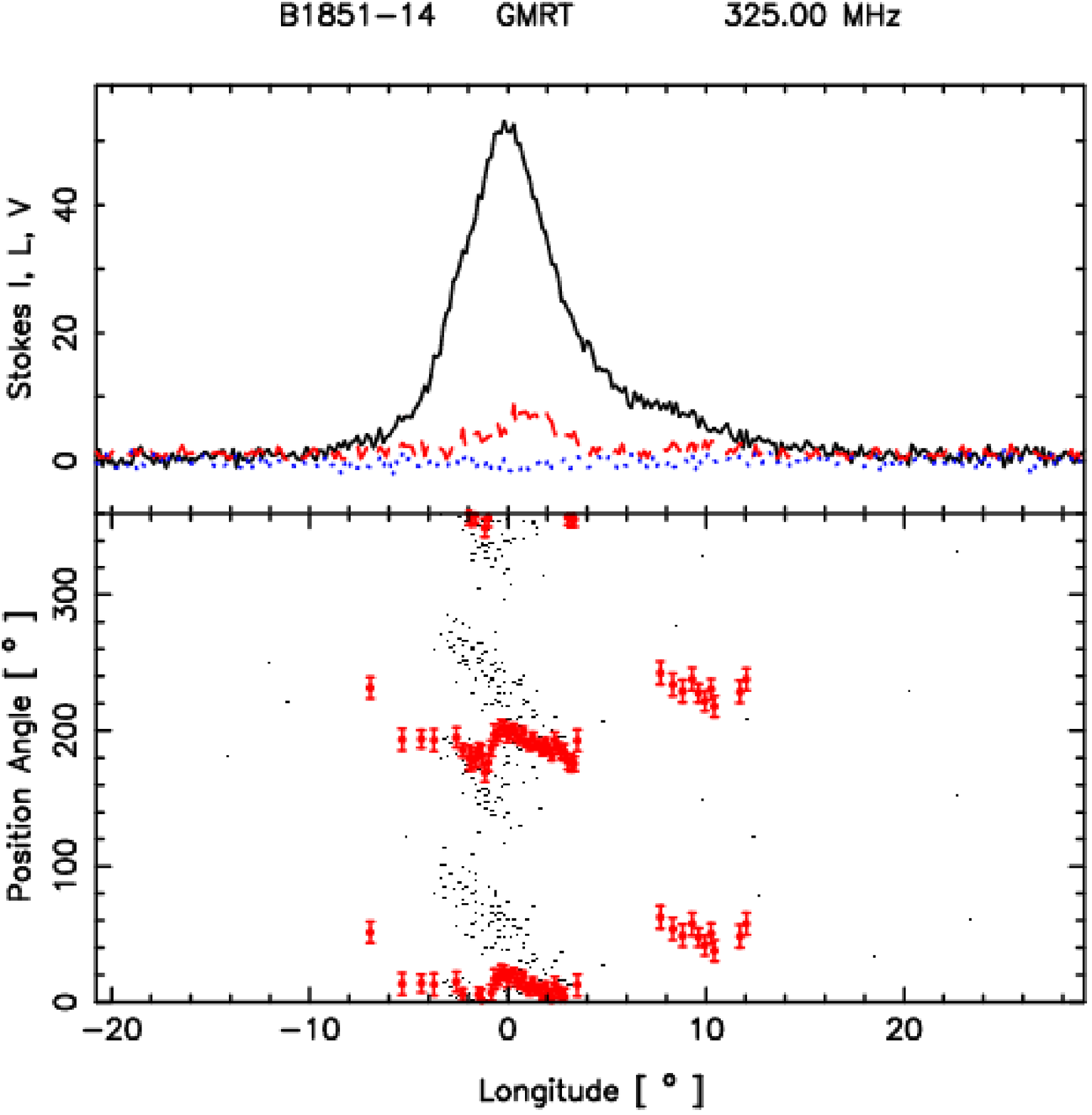}}}& \ \ \ \ \ \ 
{\mbox{\includegraphics[width=78mm,angle=-90.]{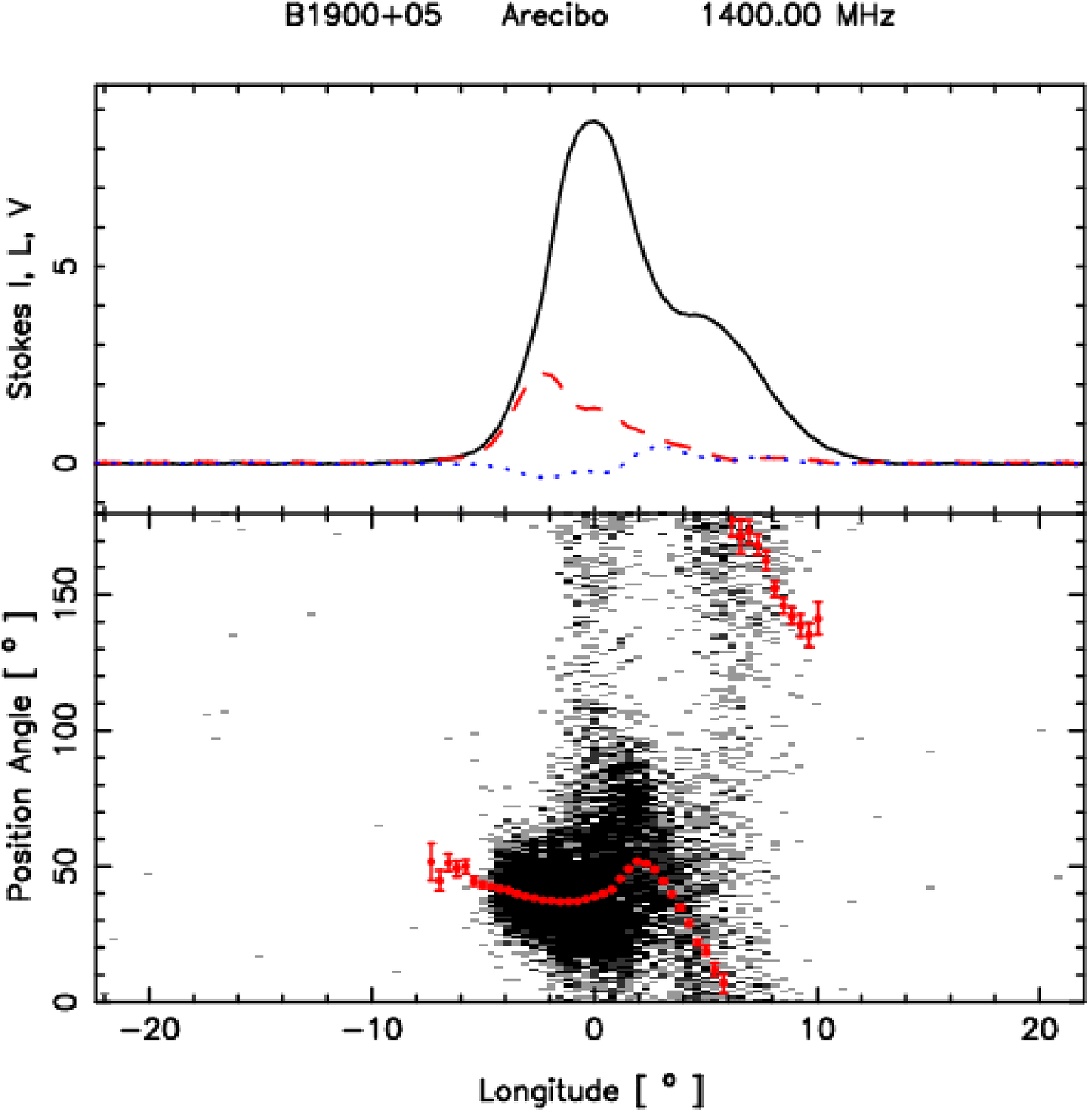}}}\\
\end{tabular}
\caption{PPA histograms and ``flared''-emission profiles as in Fig.~\ref{figA1} 
for pulsars B1822--09, B1842+14, B1851--14 and B1900+05.}
\label{figA6}
\end{center}
\end{figure*}

\begin{figure*}
\begin{center}
\begin{tabular}{@{}lr@{}}
{\mbox{\includegraphics[width=78mm,angle=-90.]{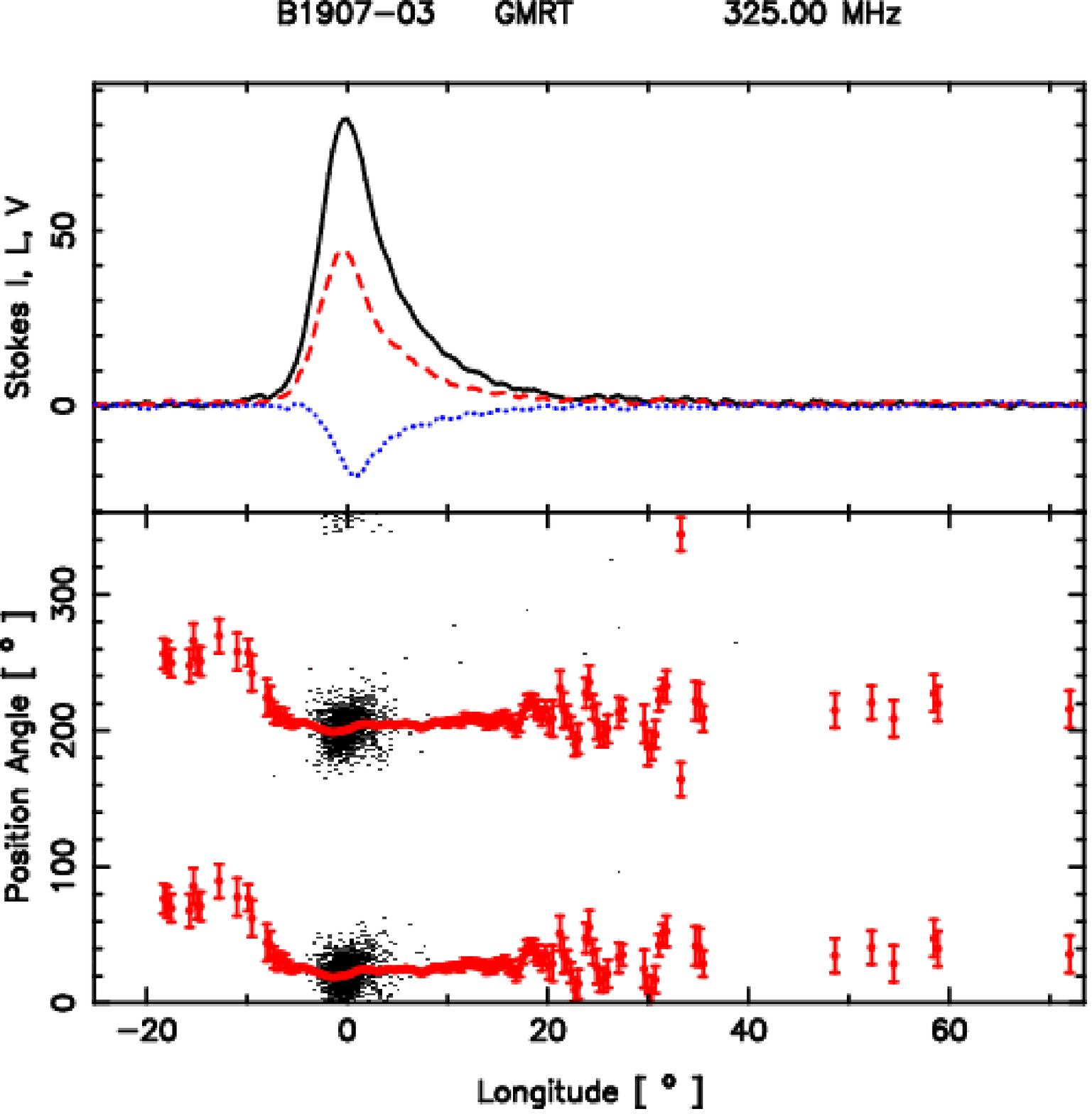}}}& \ \ \ \ \ \ 
{\mbox{\includegraphics[width=78mm,angle=-90.]{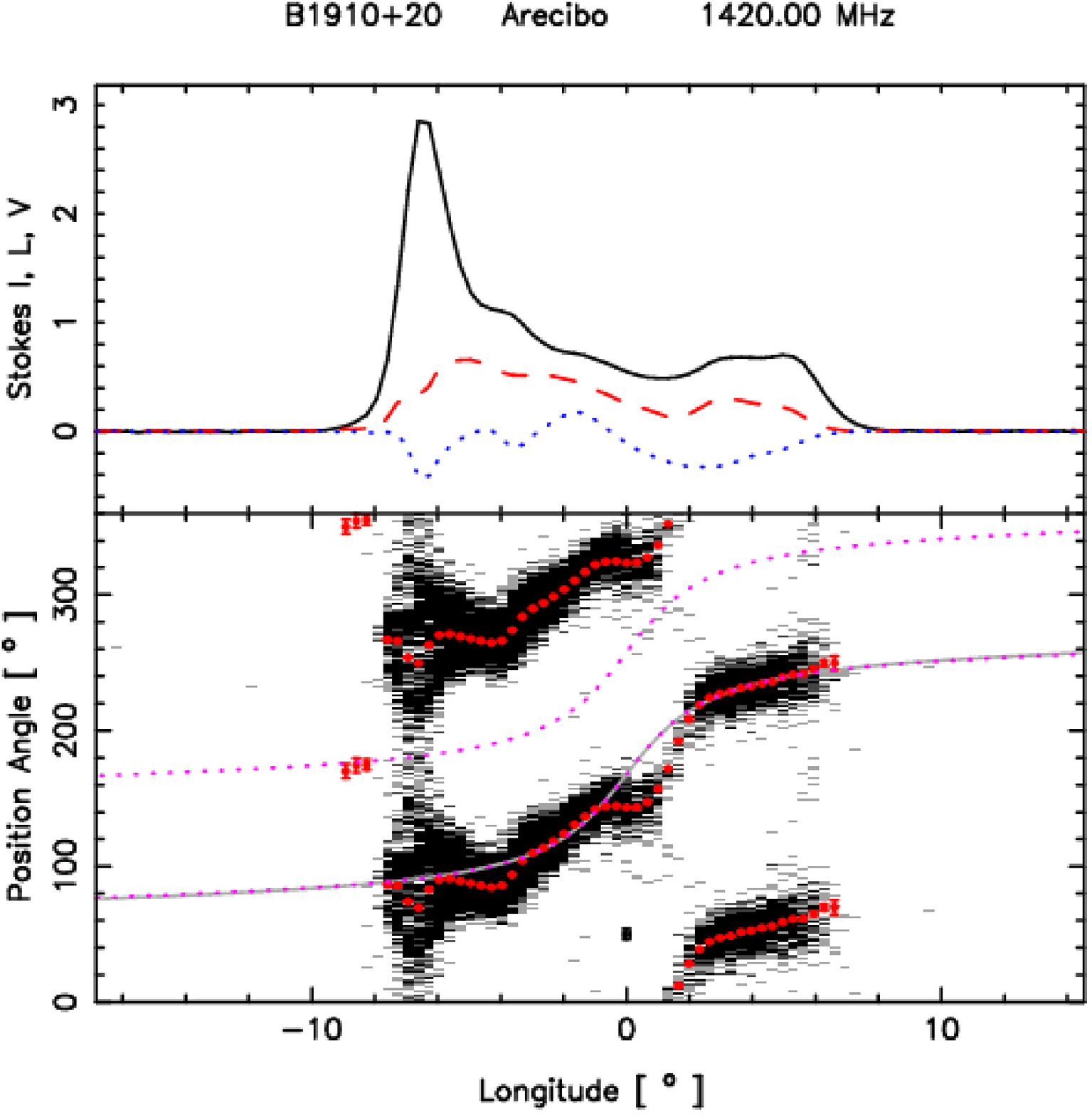}}}\\
{\mbox{\includegraphics[width=78mm,angle=-90.]{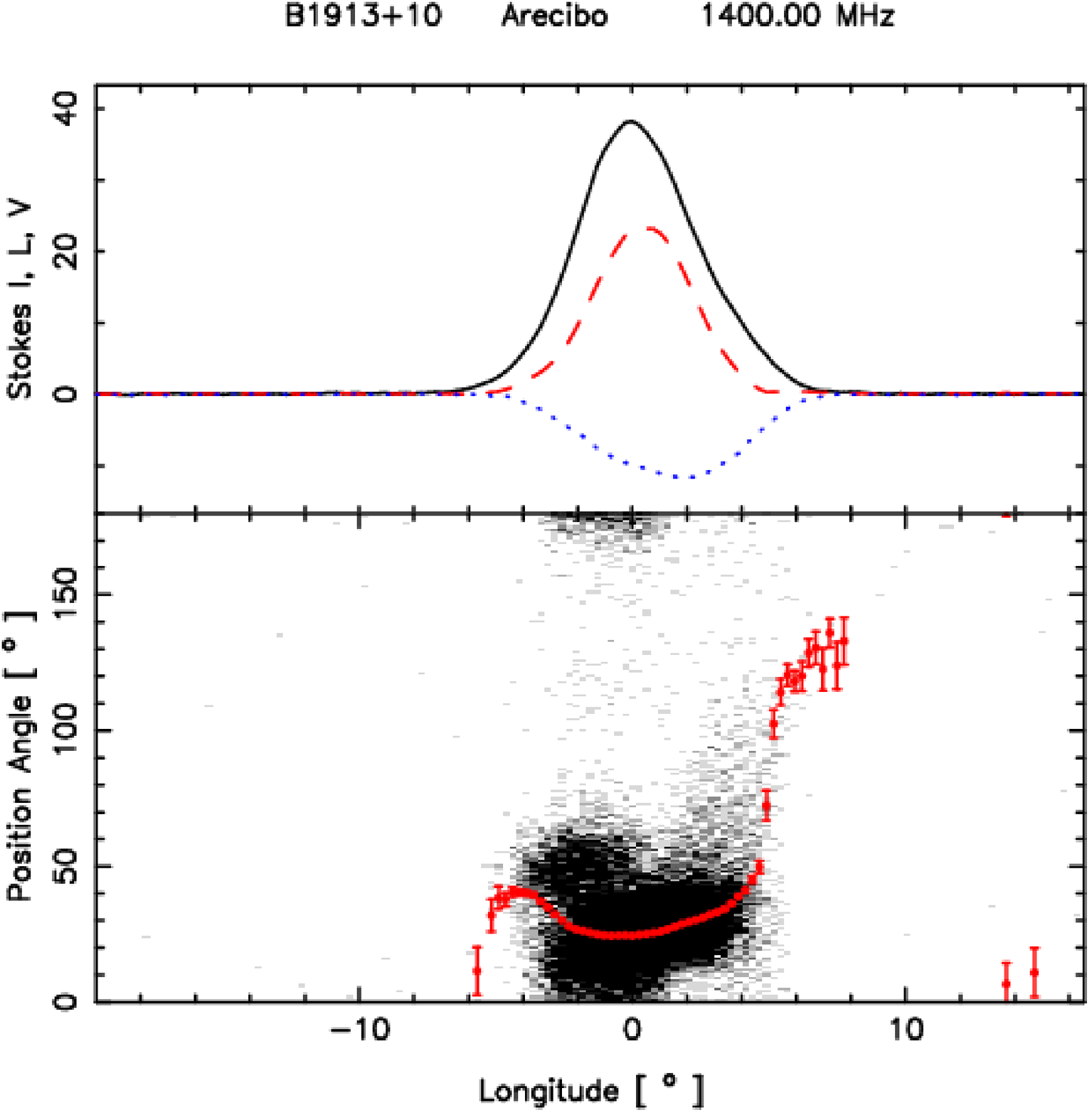}}}& \ \ \ \ \ \ 
{\mbox{\includegraphics[width=78mm,angle=-90.]{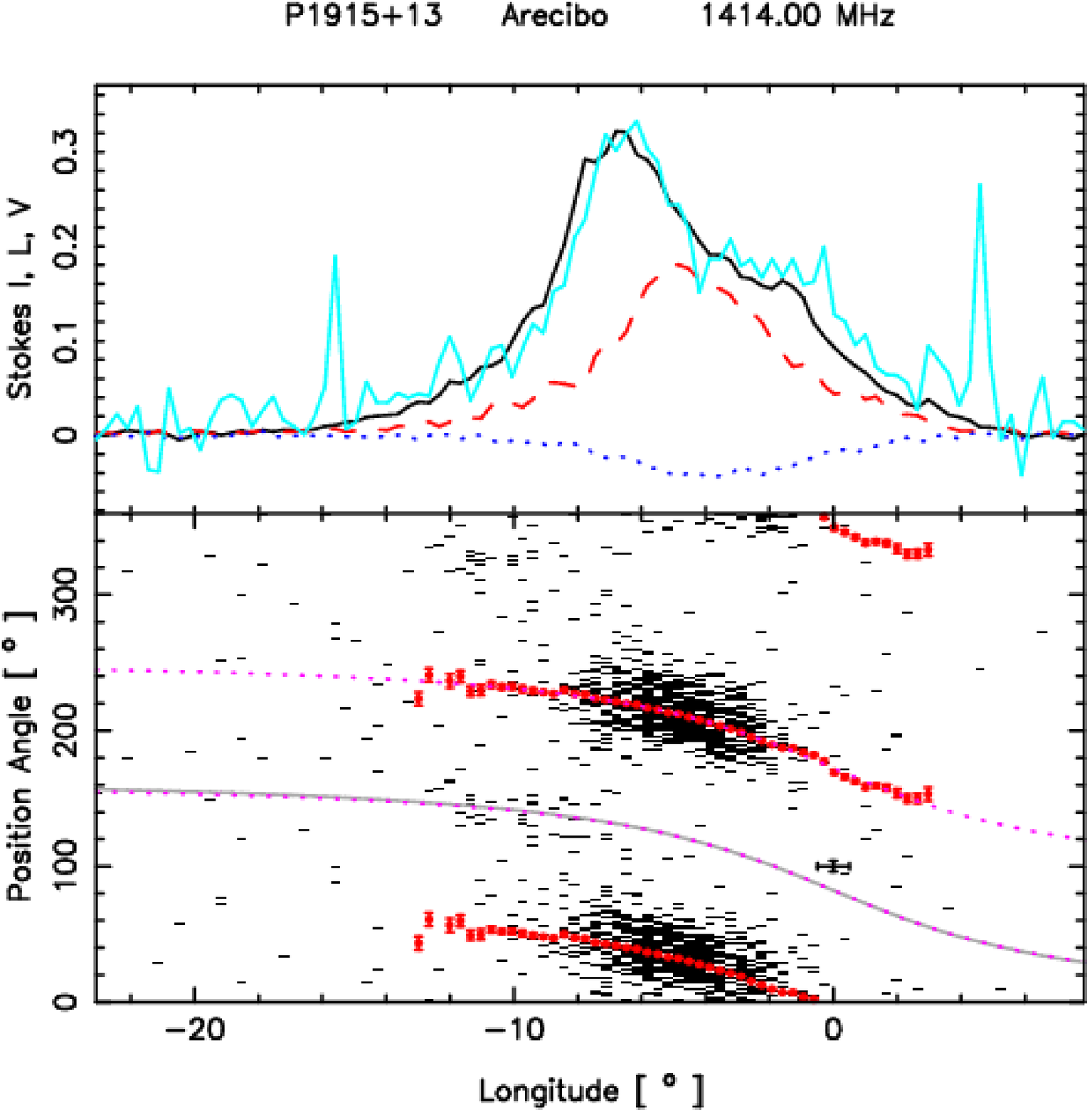}}}\\
\end{tabular}
\caption{PPA histograms and ``flared''-emission profiles as in Fig.~\ref{figA1} 
for pulsars B1907--03, B1910+20, B1913+10 and B1915+13.}
\label{figA7}
\end{center}
\end{figure*}

\begin{figure*}
\begin{center}
\begin{tabular}{@{}lr@{}}
{\mbox{\includegraphics[width=78mm,angle=-90.]{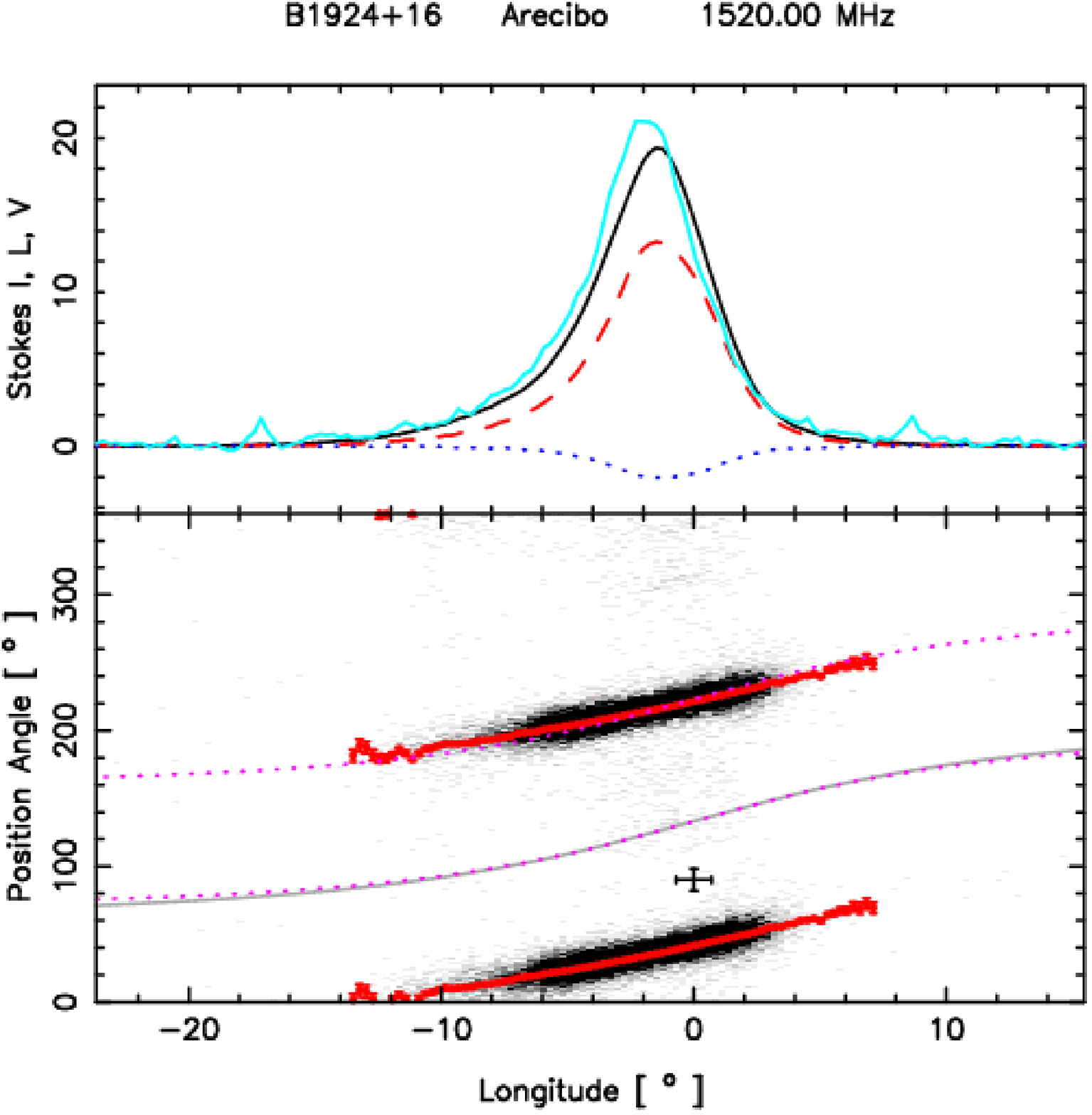}}}& \ \ \ \ \ \ 
{\mbox{\includegraphics[width=78mm,angle=-90.]{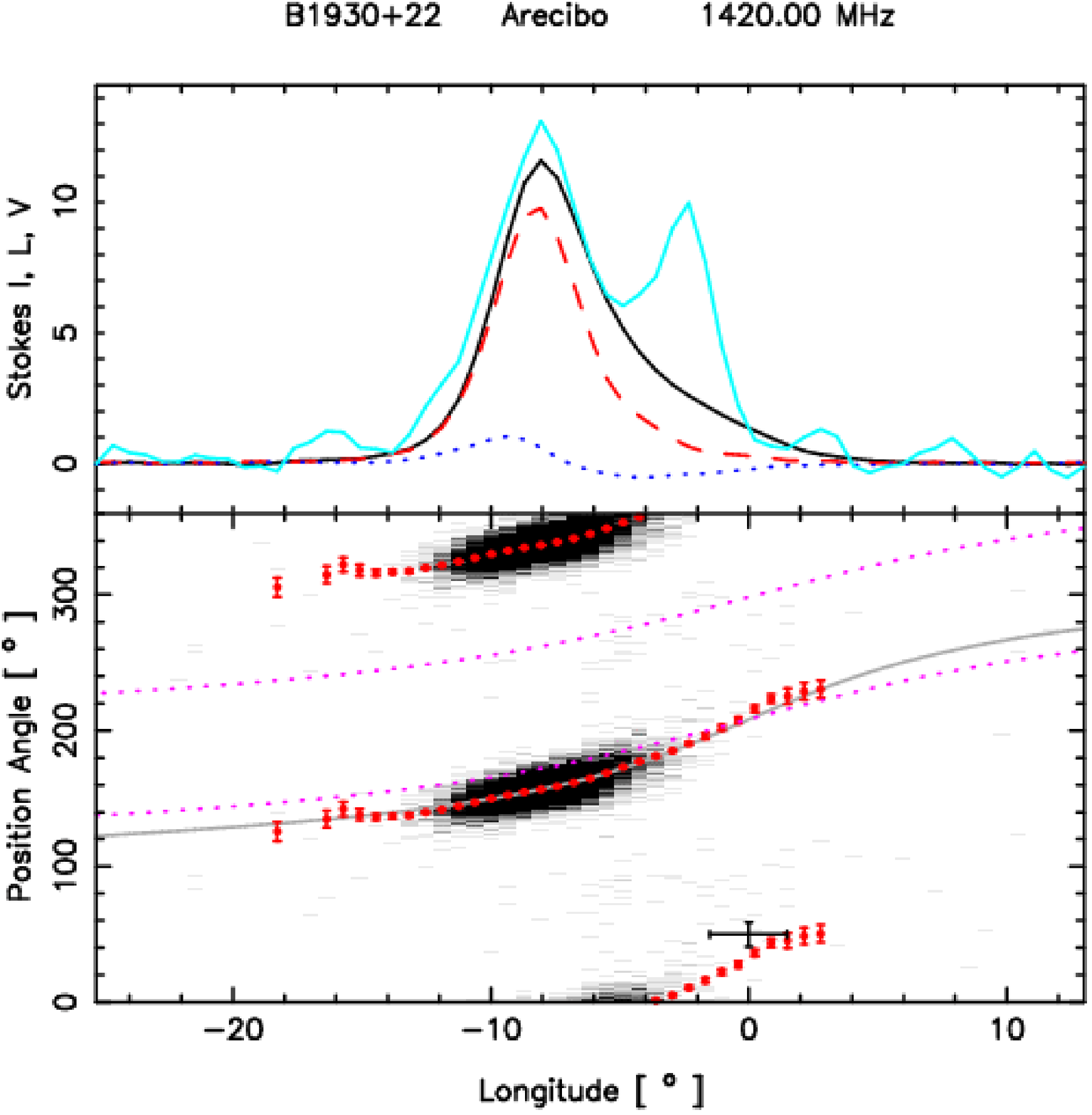}}}\\
{\mbox{\includegraphics[width=78mm,angle=-90.]{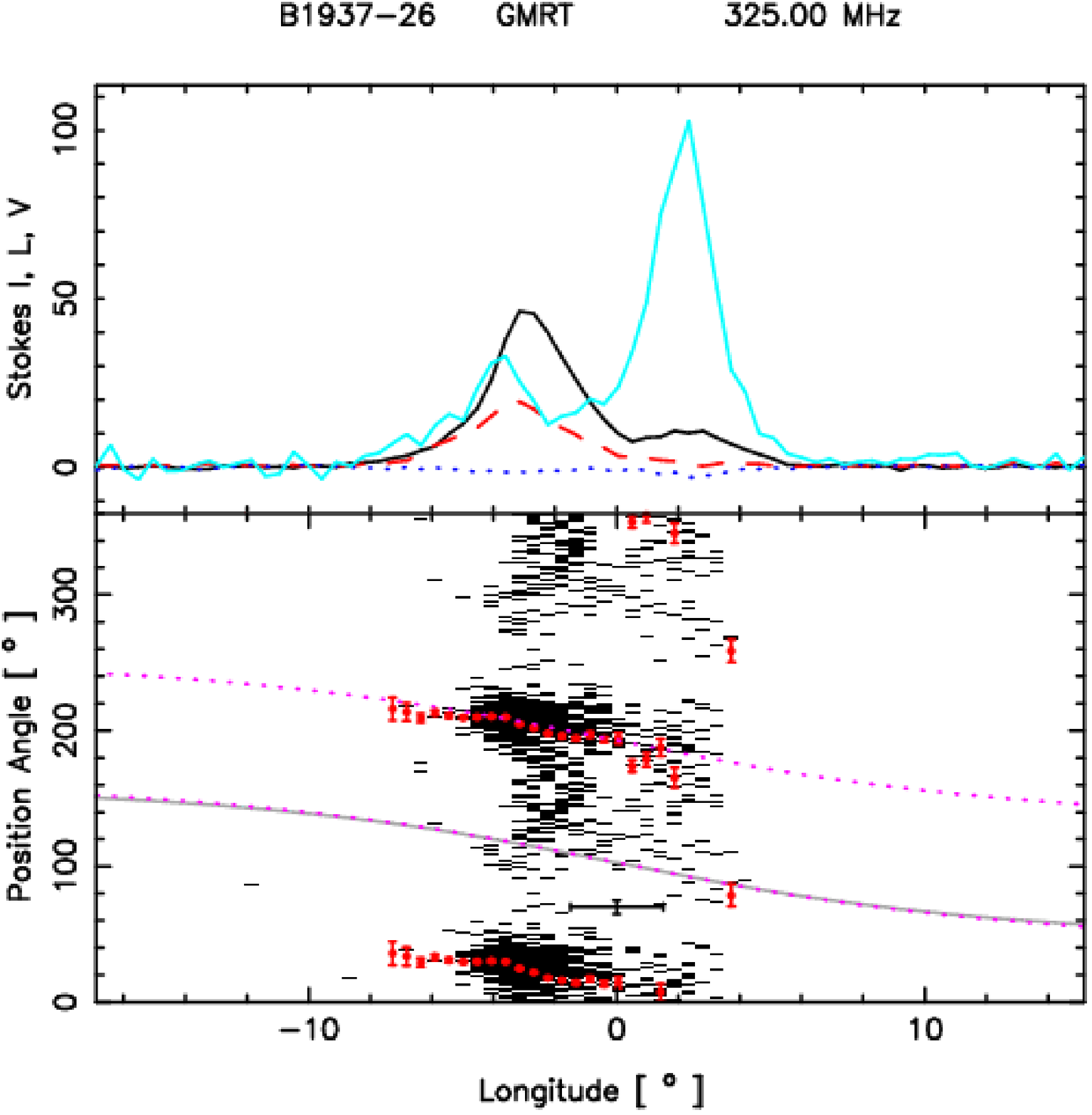}}}& \ \ \ \ \ \ 
{\mbox{\includegraphics[width=78mm,angle=-90.]{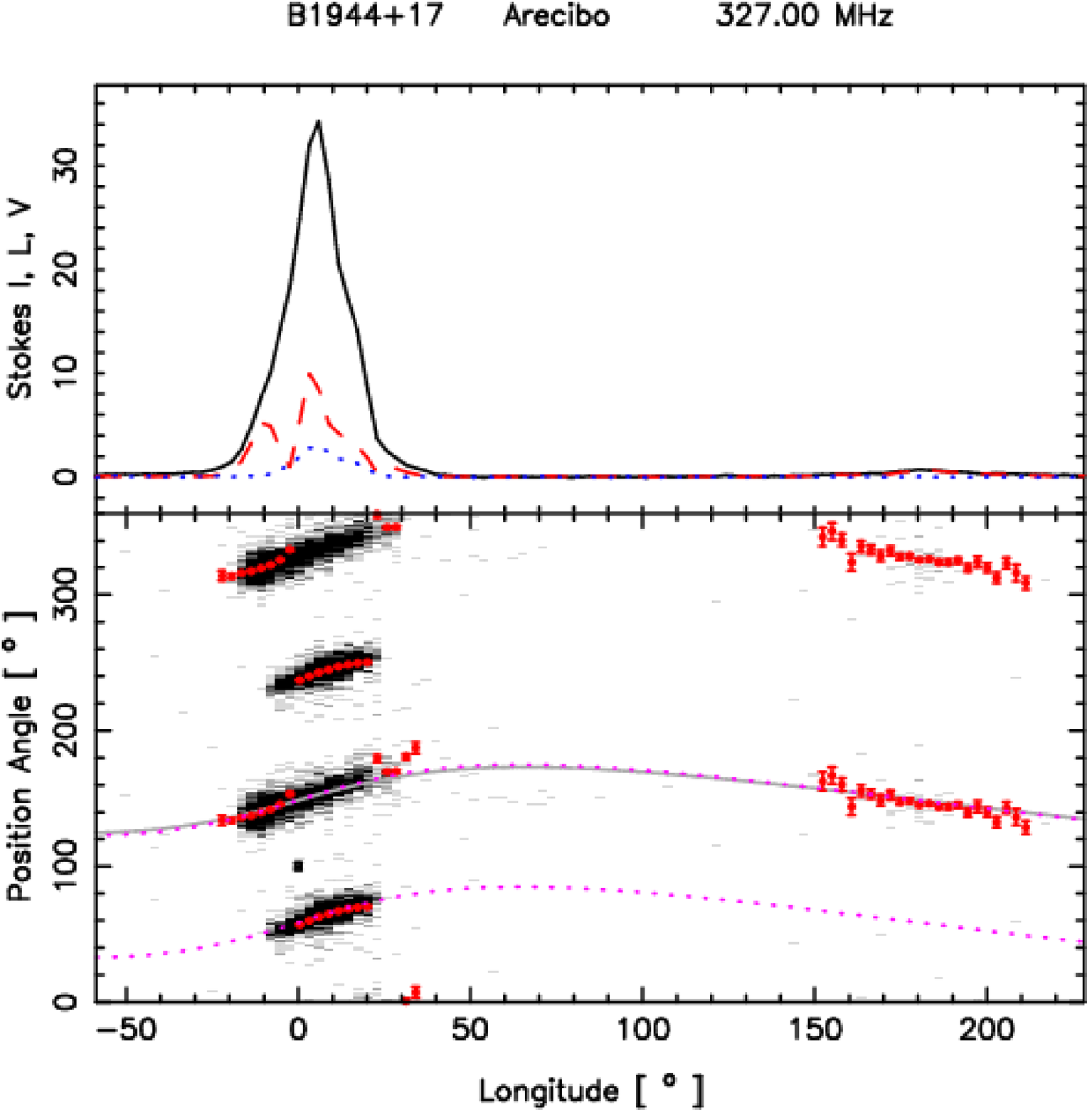}}}\\
\end{tabular}
\caption{PPA histograms and ``flared''-emission profiles as in Fig.~\ref{figA1} 
for pulsars B1924+16, B1930+22, B1937--26 and B1944+17.}
\label{figA8}
\end{center}
\end{figure*}

\begin{figure*}
\begin{center}
\begin{tabular}{@{}lr@{}}
{\mbox{\includegraphics[width=78mm,angle=-90.]{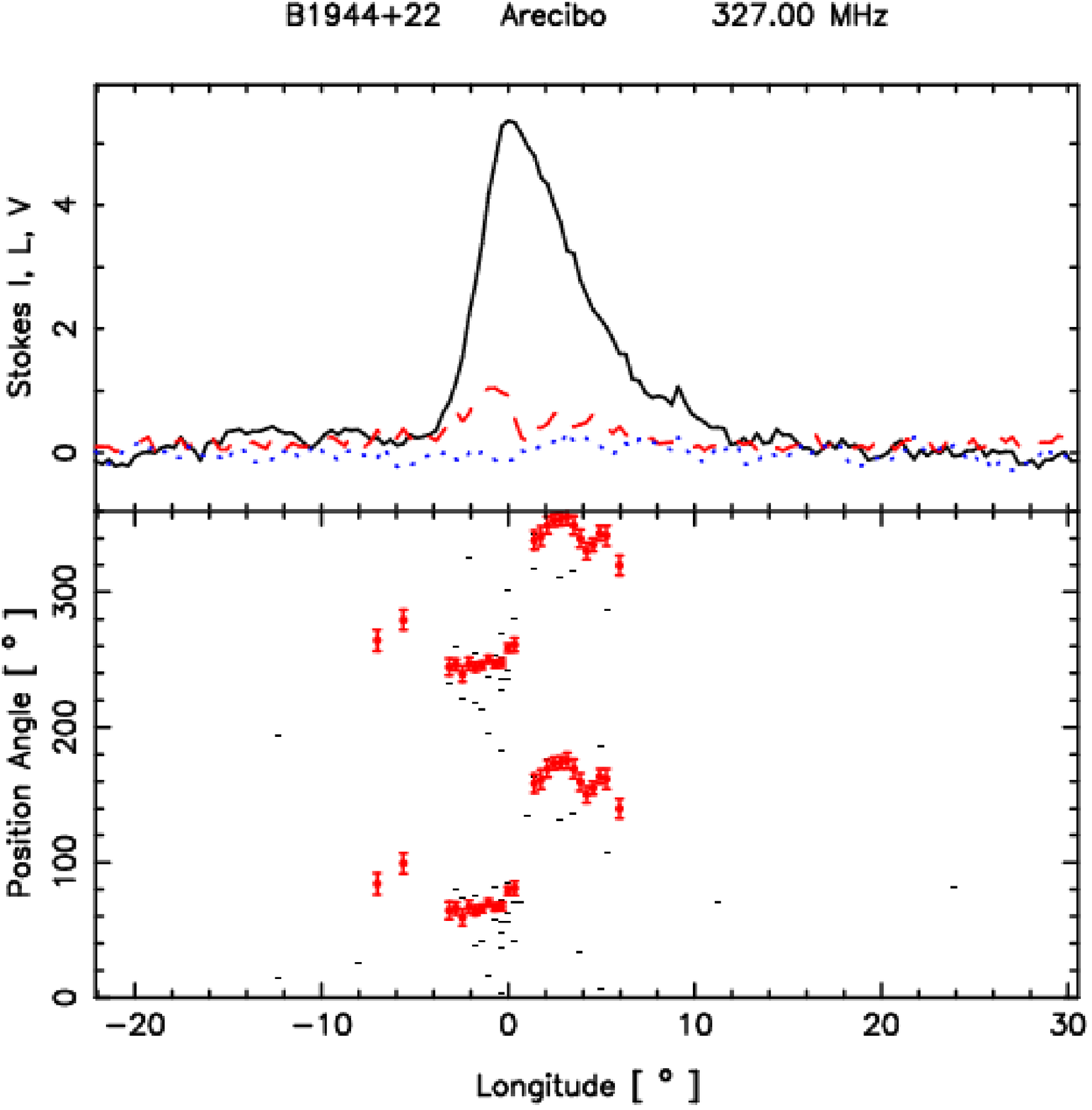}}}& \ \ \ \ \ \ 
{\mbox{\includegraphics[width=78mm,angle=-90.]{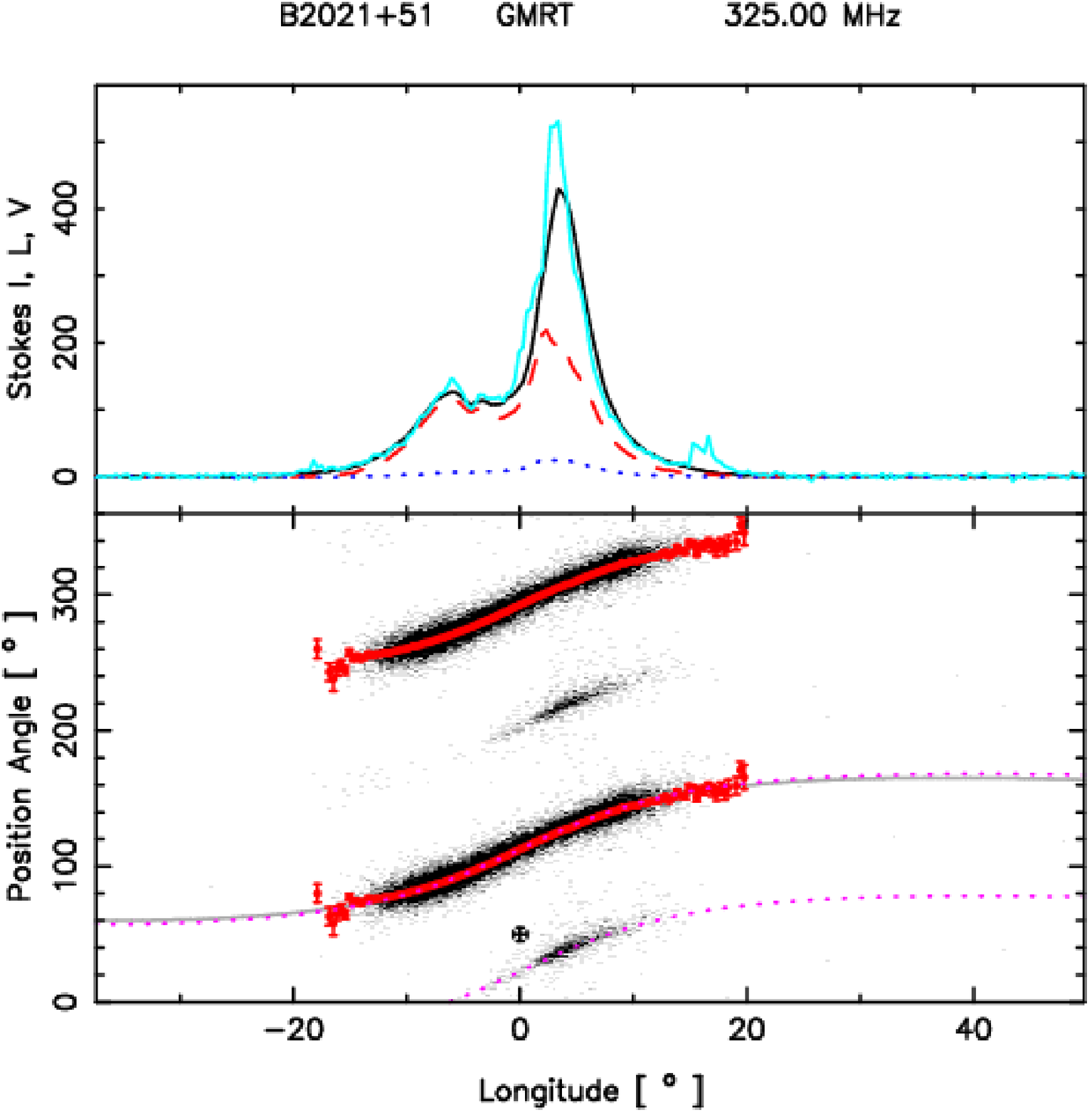}}}\\
{\mbox{\includegraphics[width=78mm,angle=-90.]{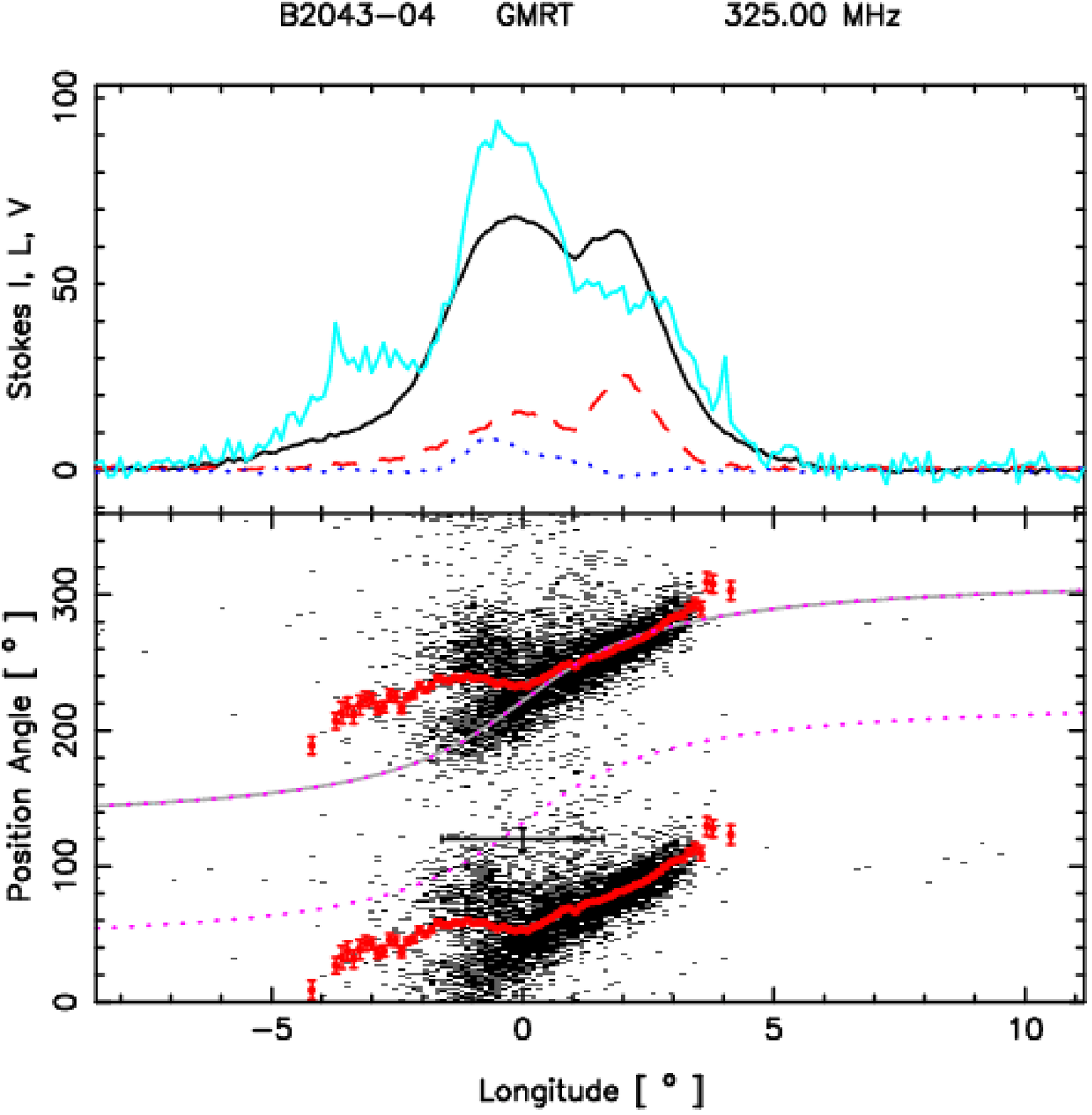}}}& \ \ \ \ \ \ 
{\mbox{\includegraphics[width=78mm,angle=-90.]{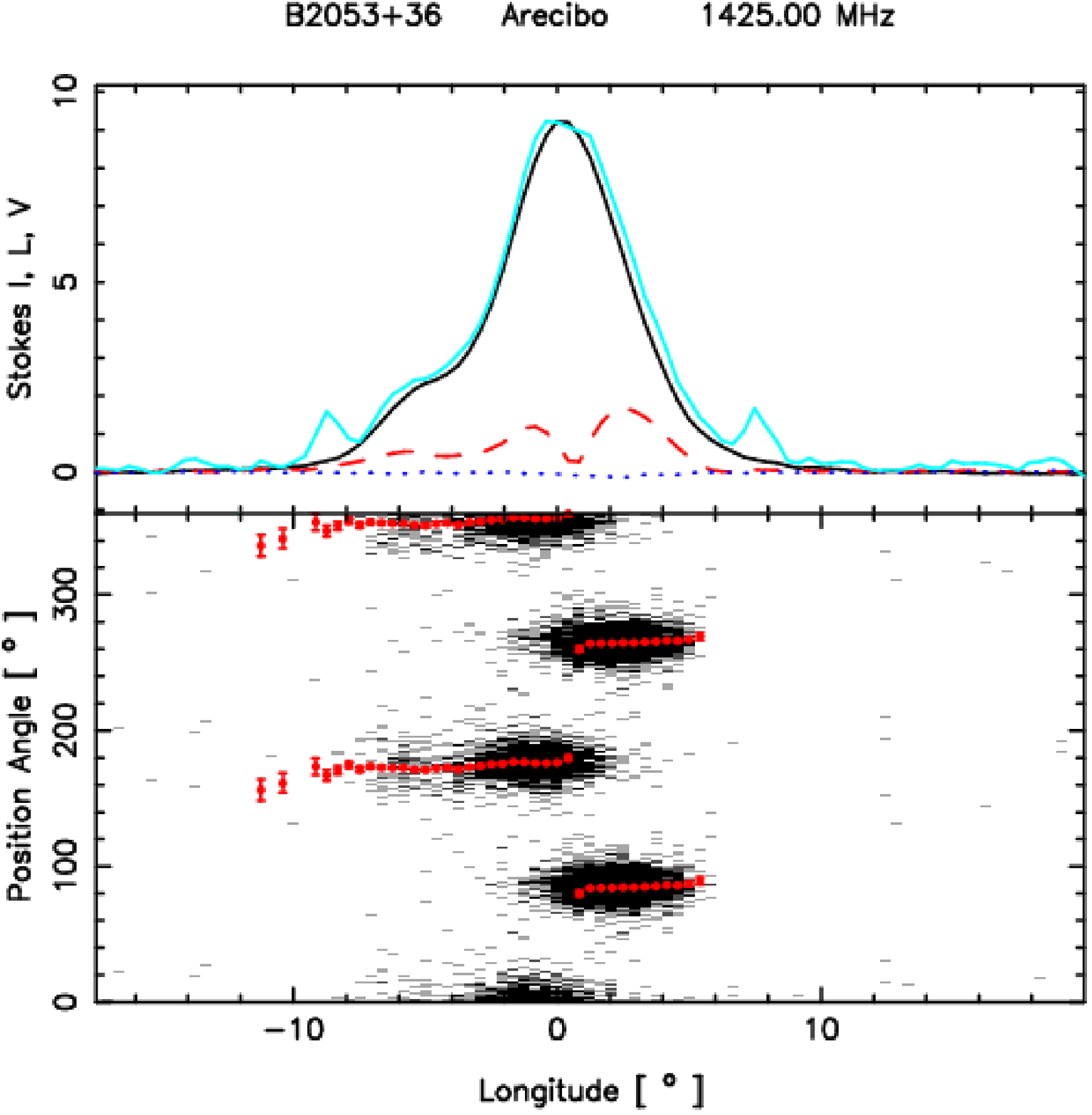}}}\\
\end{tabular}
\caption{PPA histograms and ``flared''-emission profiles as in Fig.~\ref{figA1} 
for pulsars B1944+22, B2021+51, B2043--04 and B2053+36.}
\label{figA9}
\end{center}
\end{figure*}

\begin{figure*}
\begin{center}
\begin{tabular}{@{}lr@{}}
{\mbox{\includegraphics[width=78mm,angle=-90.]{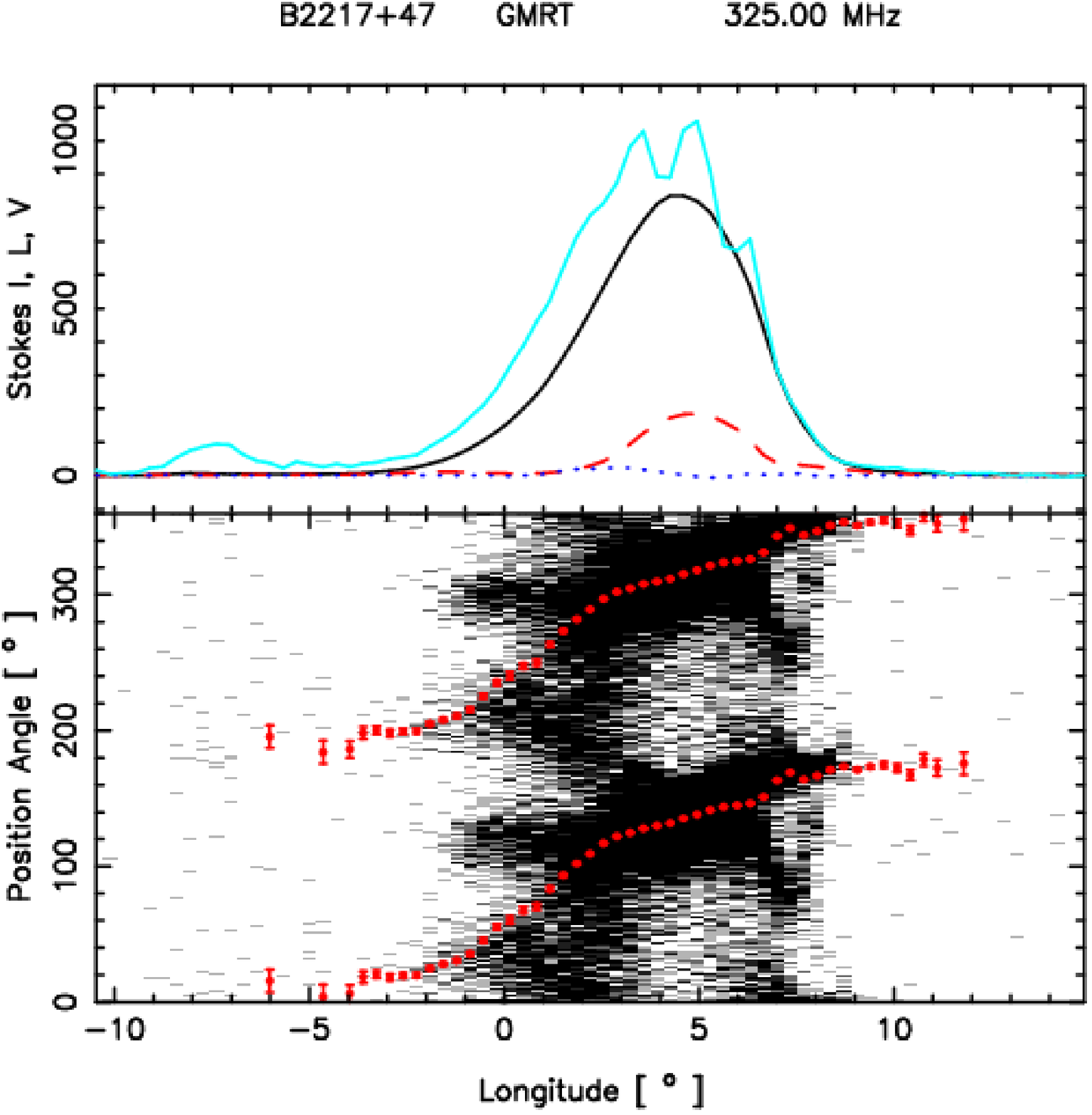}}}& \ \ \ \ \ \ 
{\mbox{\includegraphics[width=78mm,angle=-90.]{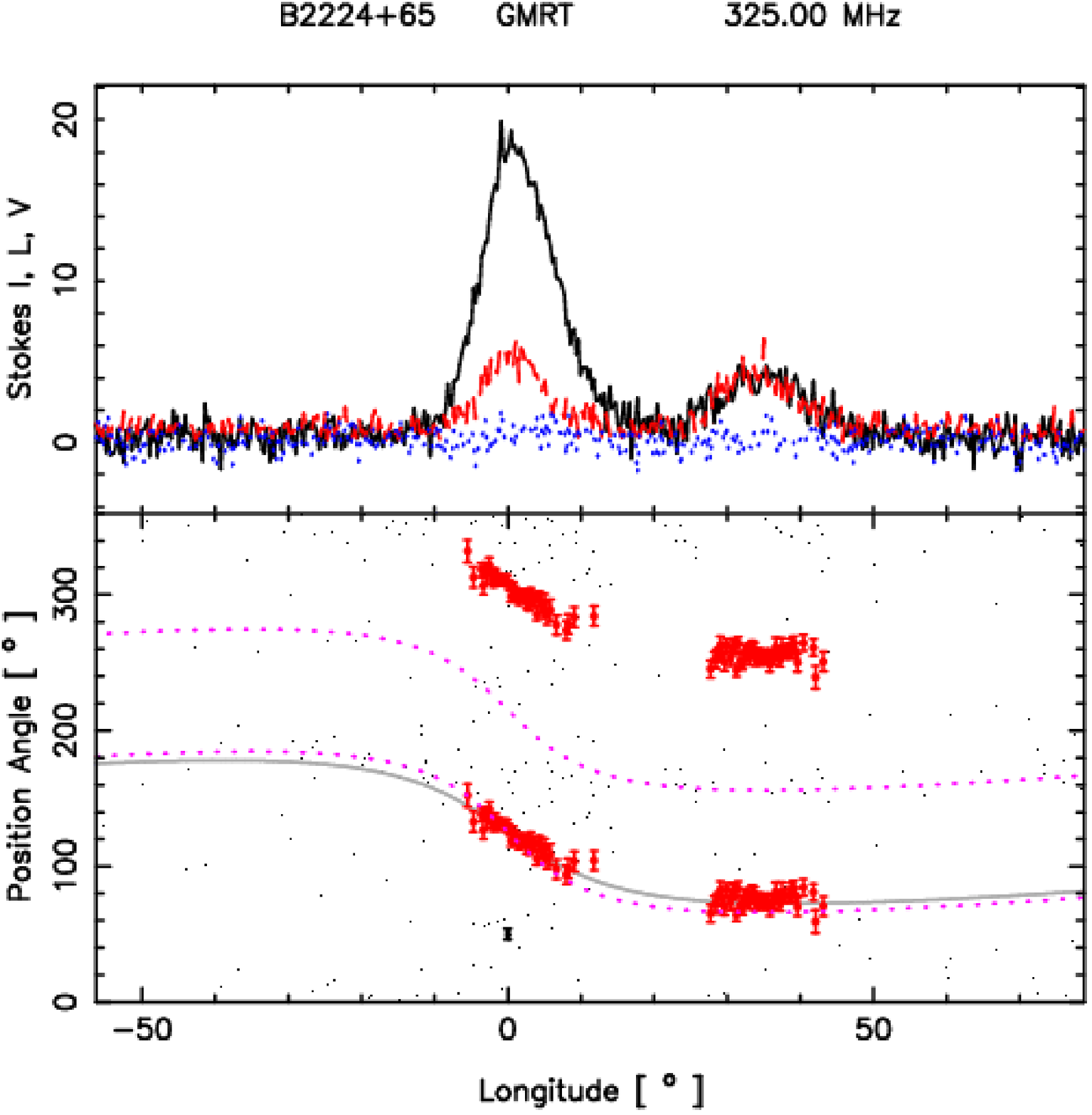}}}\\
{\mbox{\includegraphics[width=78mm,angle=-90.]{B2327-20_PAHIST.ps}}}& \ \ \ \ \ \  \\
\end{tabular}
\caption{PPA histograms and ``flared''-emission profiles as in Fig.~\ref{figA1} 
for pulsars B2217+47, B2224+65 and B2327--20.}
\label{figA10}
\end{center}
\end{figure*}

\begin{table*}
\caption{Table of RVM fitting results}
\label{tab5}
\begin{center}
\begin{tabular} {llll}\hline \hline
{\bf PSR} & $\sigma_{\chi_{\circ}}$ & $\sigma_{\phi_{\circ}}$   & {\bf R}\\
          & ($^\circ$)              &  ($^\circ$)            & (\degr/\degr)\\
\hline
B0138+59 &  3 &  0.1  & --11.2 $\pm$ 0.1\\
B0355+54  & 2 &  0.5  & --9.2 $\pm$ 0.1\\
B0450+55 & 12 &  0.1  & --8.5 $\pm$ 0.1\\
B0540+23 & 15 &  1.5  & --3.4 $\pm$ 0.2\\
B0740--28 &  5 &  1.3  & --5.5 $\pm$ 0.4\\
B0809+74 &  7 &  2    &  --3.4 $\pm$ 0.5\\
B0906--17 &  9 &  2    &  --2.3 $\pm$ 0.4\\
B0919+06 &  6 &  1    &  11.8  $\pm$ 1\\
B1112+50 &  7 &  0.7  & 10.1 $\pm$ 1\\
B1322+83 &  5 &  1    &  2.8 $\pm$ 0.1\\
B1530+27 &  7 &  0.6  & 5.8 $\pm$ 1\\
B1700--18 &  9 &  1.2  & --8.2$\pm$ 1\\
B1742--30 & 15 & ???   &   ???\\
B1745--12 & 10 &  0.8  & --11.7 $\pm$ 0.5\\
B1910+20 &  4 &  0.1  &  30 $\pm$ 1\\
B1915+13 &  4 &  0.5  & --9.8 $\pm$ 0.4\\
B1924+16 &  8 &  0.7  &  5.2 $\pm$ 0.4\\
B1930+22 &  9 &  1.5  &  8.6 $\pm$ 1\\
B1937--26 &  5 &  1.5  & --4.5 $\pm$ 0.5\\
B1944+17 &  3 &  0.9  & 0.8 $\pm$ 0.1\\
B2021+51 &  5 &  0.6  & 3.9 $\pm$ 0.4\\
B2043--04 &  8 &  1.6  & 27.1$\pm$ 3\\
B2224+65 &  4 &  0.2  & --3.6 $\pm$ 0.3\\
B2327--20 &  2 &  0.1  & 43  $\pm$ 1.5\\
\hline
\end{tabular}
\end{center}
\end{table*}

\noindent{\bf B0138+59} presents an excellent example of a partial-cone profile
in LM's intended sense.  As in the 325 MHz PS in Figure~\ref{figA1} and LM:fig.
6, we mainly see the central and trailing parts of what could be a conal double
({\bf M}) or quadruple ({\bf Q}) structure.  Only at 100 MHz does the leading feature
fully reveal itself in the Pushchino profiles (SVS, MIS, K-98, KL), making the full
half-power width nearly 40\degr.  This suggests an outer cone with a 1-GHz width
of about 27\degr, and a core width can be estimated as some 6-7\degr\ constraining
$\alpha$ to some 20\degr.  Moreover, SVS's elegant 102.5-MHz Faraday polarimetry 
clearly counters any easy argument that this star's SG point leads its profile center; 
the star's leading profile region is not visible at higher frequencies, but here the SG 
point lies near the profile midpoint.

The PPA traverse using the GMRT PS at 325 MHz is fitted with the RVM to obtain 
the SG point (see Table~\ref{tab5}) which is well constrained by the PPA fit
is taken as the zero longitude in Figure~\ref{figA1}.
We see no flared emission towards the profile edges; however the fluctuation spectrum
shows a low frequency excess as has been reported by the WES/WSE analyses. Given
that the existing 100-MHz profiles seem to reveal the bright leading feature, sensitive
new observations at low frequencies are needed to investigate this missing area of
emission.

\noindent{\bf B0254--53} seems to have a narrow, conal double ({\bf D}) profile (MHM,
MHMA, MHMb, MHQ, vO97) with a slightly stronger leading component above 1 GHz
and the reverse below.  Its profiles are nearly depolarized and the PPA information
difficult to interpret.  The sweep value given by L\&M seems too steep; rather we use
a value of --8\degr/\degr\ from the 278-MHz MHMb profile.  In short, it is not clear why
L\&M regarded this pulsar as a ``partial cone''.

\noindent{\bf PSR B0355+54:} The pulse profile at various frequencies (\eg, LM,
GL, Xilouris \etal\ 1998) clearly show three components, and the pulsar is classified
as a core single by R93 due to the domination of the bright central component
over the weak conal outriders.  L\&M identifies this pulsar as a ``partial cone''
owing to its asymmetric profile at high frequencies with the SG point of PPA traverse
lying towards the trailing edge of the profile.  This is also apparent from the GMRT
PS at 325 MHz in Figure~\ref{figA1}, where the PPA track is clearly delayed with
respect to the putative core-component peak.  Now, we interpret this displacement
of the SG as indicating that A/R plays a strong role in this pulsar's profile form.  To
fit the RVM to the displaced-PPA track, we use a two-way mode-separation technique
(\eg, Gil \etal\ 1993). The RVM fit yields the SG point of the PPA track with good 
accuracy (see Table~\ref{tab5}) and corresponds to the longitude origin in Fig.~\ref{figA1}.  
Moreover, the width of the central core component is measured to be just over 9\degr, 
indicates that $\alpha$ is just over 40\degr.

Changes in PSR B0355+54's profile shape were noted earlier by Morris \etal\
(1980) at 11 cm, where short averages were seen to change slowly from one
profile mode to the other over an interval of about 1000 pulses. We see no such
mode change in our observation.  The pulsar, however, shows sudden ``flarings''
towards the pulse edges for about 2\% of the total time.  Flarings on the leading
and the trailing edges of the profile are generally uncorrelated and without any
obvious periodicity.  The flared profile in Fig.~\ref{figA1} clearly shows what seem
to be conal outriders, and its overall form can be well described in terms of the
triple (T) or perhaps M class---therefore we use the hybrid designation arT/M.

Further the midway point of the peak of the outer components clearly leads the
steepest gradient point by about 15\degr, and hence the core peak lags the
midway point by about 3\degr.  The overall geometrical evidence here can be
understood by invoking effects of A/R (BCW as corrected by Dyks \etal\ 2004).
If we assume that the flared pulse profile illuminates the full polar cap, then the
BCW model gives an emission height for the outer cone of around 494 km.
Assuming that the central feature is of the core type, then its peak leads the SG
point of the PPA traverse by 12\degr\ yielding a core emission height of about
390 km. These 325-MHz height estimates of a few hundred km's are quite
reasonable when compared to the radio emission heights estimated in other
pulsars.

\noindent{\bf PSR B0450+55:} We have generally viewed this pulsar as having
a triple ({\bf T}) profile, and its bright component as a core feature marked by
sense-changing circular polarization especially at high frequency (\eg, vH,
GL, MIS, LM, KL, K-98).  However, L\&M were correct to note its forward-shifted
PPA traverse at meter wavelengths, such that A/R seems to displace its core
well toward the leading edge of its profile.  The RVM-fitted PPA traverse 
(see Table~\ref{tab5})  gives the SG point at the longitude origin in Figure~\ref{figA1}
with an error of about 1\degr.  The about 8\degr\ width of the core at near 1 GHz
also constrains $\alpha$ to some 30\degr.   Therefore, here we  designate 
the pulsar as having an arT profile.

Fig.~\ref{figA1} shows the pulsar's profile at 325 MHz, where the green curve
shows the ``flaring'' character of the emission on the far trailing edge of the
profile, something also seen via the large modulation index in the WES/WSE
analysis.  The regions immediately adjacent to the bright feature show strong
stationery 9.5-$P_1$ modulation, suggesting that they are conal.  The flared
profile gives clear indication of the leading and trailing conal emission, and
we can use the edges of the outermost cone to estimate the midway point of
the profile which leads the SG point by about 3.9\degr, in turn giving an emission
height of about 275 km.  This reasonable height estimate for the conal emission
supports the conclusion that A/R plays a strong role in the profile  evolution.

\noindent{\bf PSR B0540+23}, with its steeply rising profile, long trailing ``tail''
and flat to steep PPA traverse, is one of LM's classic ``partial cone'' objects.
Moreover, this behavior is progressive over a very broad band from some 0.3
to 10 GHz (RSW, GL, W-99, W-04, vHX, TR, X93, J-07, BCW), such that at
very high frequency the profile has a nearly Gaussian form and an ever more
extended ``tail'' at longer wavelengths.  Further, the star's PPA traverse is
consistently flatish on the leading side of the profile and rotates ever more
steeply downward in the trailing region---perhaps an indicator of A/R, but 
also perhaps simply the usual steep PPA rotation under the leading portion 
of a conal double profile.  

Careful inspection reveals that the PPA SG point lags the profile peak ever 
farther at lower frequencies; for example, at 10.5 GHz the SG point falls under 
the symmetrical profile, whereas at 327 MHz the peak leads the SG point by
more than 20\degr!  In some low frequency profiles, the trailing ``tail'' does 
suggest an unresolved second component (\eg, see GL's profiles at 408 and 
234 MHz); however, our 327-MHz observation in Figure~\ref{figA1} shows 
little hint of this feature, so this behavior may not be consistent.  Overall, we 
find few signs of conal emission in the star's profiles:  no ``outriders'' are seen
at high frequency, no periodic features are seen in its fluctuation spectra
(WES/WSE), and the ``flaring'' on the extreme profile edges is weak.

Fortunately, B0540+23's PSs have been studied carefully at 430 MHz by 
Nowakowski (1991).  Using several different analyses including intensity-fractionated 
profiles, he finds evidence for two or three other regions of 
emission following the bright component.  In Paper VI this pulsar was 
classified as having a core-single ({\bf S}$_t$) profile; now we tentatively 
classify it as triple ({\bf T}).  The PPA traverse in Fig.~\ref{figA1} is well fitted 
by giving an $R$ value of --3.4\degr/\degr (see Table~\ref{tab5})

Estimating the outside half-power width of the star's full 430-MHz profile as 
38\degr\ (see Nowakowski's fig. 4) and scaling to 1 GHz using BCW's profile 
dimensions measured from the fitted SG points, one obtains the 29\degr\ 
width value used in Table~\ref{tab3}.  While a satisfactory solution of the 
emission geometry can be obtained from this width estimate and fitted $R$ 
value in the above table, it is interesting to note that the SG point appears 
to fall well after the profile center.  Specifically referring to the 327-MHz 
profile in Fig.~\ref{figA1}, we might expect the scaled half-power width of 
the putative triple profile to be some 40\degr, such that the SG point would 
lag the center by 9\degr---very likely suggesting that A/R is operative in this 
pulsar's emission. 

\noindent{\bf\em B0643+80:}  Profiles for this pulsar have been published
spanning 100 MHz to 5 GHz (GL, vH, MIS, MM, S95).  Most of these show a
bright leading and weak trailing component; however, at 102 MHz this
configuration is reversed, perhaps due to a second mode (MIS).  The  profile 
exhibits a nearly constant half power width of about 9\degr\ over this large 
frequency range, and the PPA sweep rate can be estimated from GL's 1.4-GHz 
profile.  Overall, the pulsar appears to show a conal single evolution that is
quite reminiscent of B0943+10.  WSE report that the star's fluctuation spectra
are featureless at 92 cm.  It is thus unsurprising that L\&M classed this pulsar 
as having a ``partial cone'' profile.

\noindent{\bf PSR B0740--28} is yet another example of a ``partial cone'' in
the L\&M sense, apparently owning to the displacement of its PPA SG point on
the trailing side of its profile---probably indicative, we now know, of A/R effects.
Its average profiles show significant pulse-shape evolution with frequency
(\eg, vH, MHQ, GL, J-05, KJ, JKW),\footnote{Some of the older published profiles
(\ie, MHMA, MHM, MHMb, MGSBT, vO97, even GL) show little detail and thus
do not seem to have been resolved adequately.} and as many as seven
components are needed to fit its profile at 1.4 GHz (Kramer 1994).

In Figure~\ref{figA2} we show the average Stokes profile and grey-scale PPA
histogram of the 325-MHz pulse from the GMRT PS.  Using a similar analysis
as for B0355+54, we failed to find any evidence for ``flaring'' on the profile
edges.  Unfortunately, this high S/N profile does resolve the PPA traverse any
farther into the ``wings'' than the previous observations.

More clarifying are the LRF spectra for this pulsar (Fig.~\ref{fig2}), computed 
from the above PS, that exhibit a narrow 3.6 c/$P_1$ fluctuation feature on the 
edges of its profile---suggesting that the pulsar illuminates the entire annular 
region around its magnetic axis and that its profile edges correspond to the 
outer edges of this region.  No such feature was reported by WES/WSE; but it 
is possible that such ``drifting'' intervals are episodic (as for pulsar B1944+17 
below), and this may account for this star's profile instability as well.

The RVM fitted PPA gives a maximum sweep rate of --5.5\degr/\degr\ with the 
resultant SG point at the indicated longitude origin (see Table~\ref{tab5}).  
The star's several components (\ie, see KJ's profiles at 1.4 and 3.1 GHz) can 
only be understood quantitatively as a core/double structure if its magnetic 
geometry is nearly orthogonal, such that the inner and outer conal components 
have outside dimensions of something like 9 and 18\degr, respectively.  We 
then designate this pulsar as having an arM profile.

\noindent{\bf B0809+74} presented the defining example of how profile 
``absorption'' (Bartel \etal\ 1981, Bartel 1981) could result in ``partial cone'' 
emission.  A review of the many consequences as well as modern efforts to 
understand the effects appears in RRS/RRvLS, which show that the longitude 
of the magnetic axis at meter wavelengths falls on the leading edge of the profile 
at about the half-power point; whereas at both higher and lower frequencies 
the star's profiles appear to be complete.  Our GMRT profile in Figure~\ref{figA2}
shows the ``absorbed'' 325-MHz form, such that the full profile would have a
half-power width of some 17\degr.  Here, the nearly linear and very highly 
correlated PPA fit (see Table~\ref{tab5}) does not properly locate the fiducial 
longitude---which here would fall about --7\degr---and the PPA ``jump'' is modal 
in origin.  Despite its ``absorption'' the star's profile and frequency evolution is 
characteristic of the conal single ({\bf S}$_d$) class, and A/R effects seem to play 
no significant role.  We retain the model of RRS/RRvLS in Table~\ref{tab3}.  

\noindent{\bf B0906--17:}  As seen in Figure~\ref{figA2} the pulsar exhibits a
sharply rising profile with a long weak trailing tail as well as a PPA traverse
that steepens to the SG point only on the trailing edge of the profile, prompting
L\&M to regard it as having ``partial cone'' emission.  We also see evidence of
sporadic emission across the entire profile.  More recent work (XRSS, vH, GL)
often resolves a trailing component, and the 21-cm profiles (see J-05) show a 
leading-edge inflection that suggests a third.  The asymmetrically curved PPA
traverse (with a prominent 90\degr\ ``jump'') steepens steadily with longitude
and strongly suggests A/R.   Unfortunately, the WES/WSE analyses are not
very revealing in this case.   The overall evidence then suggests that the star's
profile might be regarded as an A/R triple (ar{\bf T}), such that this structure is 
obscured at low frequency, perhaps because A/R moves some central 
(putative core-component) emission to ever earlier longitudes.  The PPA fit
(see Table~\ref{tab5}) in Fig.~\ref{figA2} is neither able 
to measure the maximum sweep rate nor to locate the PPA inflection point.  That 
the sweep rate here is far too shallow is clear by reference to the 1.4-GHz profile 
of J-05.

\noindent{\bf B0919+06:} Figure~\ref{figA2} L\&M classify this pulsar as a
``partial cone'' with its SG point lying towards the trailing part of its profile. The
star's average emission shows a long dim ramp proceeding its bright trailing
component, and it exhibits very similar profiles in form and dimensions over
the 0.1 to 10.6-GHz range of the existing observations.  A recent single pulse
polarimetric study (Rankin \etal\ 2006a) found that the dimmer leading parts of
the profile can suddenly brighten up for several tens of pulses and then revert
back to their normal faintness (see their Figs. 1 and 3).  The effect is similar to
the ``flaring'' event seen for PSR B0355+54 discussed earlier, and here the
above study demonstrated that the overall profile is triple (T) with both core
and conal dimensions scaling in terms of the polar-cap size.  The fluctuation
spectra provide (see WES/WSE) little insight for this star.

The 325-MHz PS in Rankin \etal\ (2006): fig. 3 clearly shows that the PPAs
exhibit strong OPMs mostly towards the leading parts of the profile.  The 
average PPAs thus show a complicated behavior which probably led 
LM88 to conclude that the SG point is towards the leading edge of the 
profile.  Here, we have used the same 1400-MHz PS as in the above study 
to fit the RVM to the PPA traverse. The PPA at this frequency is mostly 
dominated by a single OPM.  Our RVM--fitted PPA yields a maximum sweep 
rate of +11.8\degr/\degr (see Table~\ref{tab5}) as shown in Fig.~\ref{figA2} 
with the longitude origin falling at the SG point towards the trailing edge of 
the profile.  The above sweep rate is steeper and more linear than that seen 
at meter wavelengths;  therefore, we have retained the model values from the 
above study in Table~\ref{tab3} (see also BCW's profiles).  Clearly the star's 
PPA behavior is consistent with an A/R signature as predicted by the BCW 
model---so we designate it as having an arT profile---although its overall effect 
is not at all clear.

\noindent{\bf B1055--52I:}   This prominent southern interpulsar has been
studied by many investigators (HMAK, MHMA, vO97, Biggs 1990, MHMb,
LM, CMH) and the configurations of its main pulse and interpulse widely
debated.  An interesting comprehensive treatment has been given in the
recent paper by Weltevrede \& Wright (2009).  These authors find a nearly
orthogonal geometry ($\alpha$=75\degr) as have several other groups
including ourselves (Paper VI).  They also support the idea that a trailing 
portion of the star's interpulse is missing, as did L\&M in arguing that it was a 
``partial cone''. We do not see any flaring towards the leading or trailing 
edge of the pulsar's interpulse, and also did not manage to get reliable RVM 
fits to the PPA.  However, given the profound differences between the star's 
main pulse and interpulse properties---together with their large widths---we 
suspect that a solution with a small value of $\alpha$ will ultimately be fully 
demonstrated.  An illustrative such outer cone model for the interpulse is 
given in in Table~\ref{tab3}.

\noindent{\bf\em B1112+50:}   At meter wavelengths this pulsar has an asymmetric
single profile, and it is apparently on this basis that L\&M regarded it as a ``partial
cone''.  Above 1 GHz the star's profile consists of two components which are at times
well resolved and sometimes not, indicating several modes.  Profiles and polarimetry
are available by a number of authors (MGSBT, GL, L90, XRSS, KL, MIS), and both
modal and fluctuation studies are available by Wright \etal\ (1986) and WES/WSE.
Our 325-MHz GMRT observation is shown in Figure~\ref{figA3}, which shows both
its asymmetric single profile and ``flared'' double form.  The LRFs show only weak
periodic modulation, but the star's PS are highly modulated at both frequencies in
the WES/WSE analyses.  Overall, the profile evolution appears conal, though the
forms may entail some core emission in the profile center at lower frequencies.  
In any case, the PPA fitting in Fig.~\ref{figA3} yields 
a sweep rate of 10\degr/\degr\ and a poorly determined SG point (see Table~\ref{tab5}).  This 
together with the profile width and an estimate of the putative core dimension 
suggest the inner cone geometry in Table~\ref{tab3}.

\noindent{\bf B1221--63:}   Here, we do not understand why L\&M regarded this
pulsar as a ``partial cone''.  Profile polarimetry of uneven quality is available over
a band from 0.27 to 1.6 GHz (MHM, MHMA, MHMb, vO97, WMLQ).  Overall, the
pulsar seems to exhibit a triple form (MHM), and estimates of the profile and core
widths together with the sweep rate suggest an inner cone geometry as seen in
Table~\ref{tab3}.

\noindent{\bf B1240--64:}   This pulsar has a symmetrical single profile below 1
GHz, though some of the observations are poorly resolved (MHMA, CMH, vO97).
It was probably the leading ``ramp'' on MHM's 1.6 GHz observation that pushed
this star into L\&M's ``partial cone'' category.  Surely, KJ's recent 1.4 and 3.1-GHz
profiles are the best quality available, and these show perhaps a central notched
core flanked, by a leading conal outrider, and just a hint of the trailing one.  Further,
the PPA traverse above 1 GHz exhibits a perplexing rotation through more than
180\degr.  Despite these difficulties, the star exhibits what is essentially a core-single
profile evolution, and the rough 8.4-GHz detection of JKW may show the surviving
pair of conal components.  If the vO97 profile provides a reliable sweep rate, then
the profile dimensions can be roughly squared with an inner cone geometry as
shown in the table.

\noindent{\bf\em B1322+83m:}  Little can be gleaned about this star's emission 
from the published profiles (GL, KL); however, the high quality GMRT 325-MHz 
profile in Figure~\ref{figA3} is more scrutable.  The star has two regions of emission, 
one in the form of a highly polarized ``precursor'' with a completely flat PPA traverse,
and then a second region of emission which is also highly linearly polarized but
with a positive sweep rate.  We take the position that the precursor is unrelated to 
the polar-cap core/cone emission structures.  Then, the ``main pulse'' is very likely 
a conal single profile.  This configuration would then be very similar to what is 
observed in the B0943+10 `Q' mode (Backus \etal\ 2010). The RVM fit (see 
Table~\ref{tab5}) yields a somewhat poorly determined SG point, and its location 
within the main pulse is consistent with its being close to the profile center.

\noindent{\bf B1356--60:}   Some published profiles are useless for our purposes
because of scattering or poor resolution  (vO97, WMLQ, MHQ).   However, the two
recent polarized profiles (KJ, JKW) suggest a core-single evolution without conal
outriders.  Interestingly, KJ find a significant, apparently A/R shift, between 1.4 
and 3.1 GHz when the profiles are aligned using their SG points.  

\noindent{\bf B1426--66:}   Many published observations are available for this
southern pulsar (HMAK, MHMA, MHM, MHMb, vO97, J-07, JKW), and most are
of good quality.  Apart from its odd profile shape, we cannot see why L\&M saw
this star as a ``partial cone''.  Again, it is the Johnston \etal\ (J-05) work
that is most insightful.  The bright narrow feature marked by antisymmetric $V$
is clearly a core component, and it is flanked by a broad leading component
and a weak trailing one.  Using the core width to determine $\alpha$, the conal
dimensions and the PPA sweep rate, it is clear that an inner cone geometry obtains.

\noindent{\bf B1449--64:}   An identical set of observations is available for this
prominent southern pulsar (HMAK, MHMA, MHM, MHMb, vO97, J-05, JKW),
and while it seems likely that this 180-msec pulsar would generate a core
feature, no clear circularly polarized signature is apparent.  We do see evidence 
of  conal outriders in both the 1.6 (MHM) and 1.4-GHz (J-05) profiles, and the 
width of the central (putative) core constrains $\alpha$ to some 43\degr.  A rough 
estimate of the conal outrider dimension then strongly suggests an inner conal 
geometry.

\noindent{\bf B1530+27:}  It is easy to see why LM placed this pulsar in their
``partial cone'' category with its bright leading component and weaker, barely
resolved trailing one (RSW, BCW, GL, W-99, W-04, MM)---not to mention its weak 
postcursor.  As we see Figure~\ref{figA3}, neither its profile nor shallow PPA 
traverse readily indicate that this could be a conal single or double ({\bf D}) 
profile.  However, the PRAO profiles (MIS, K-98, KL) show that the trailing 
component becomes as strong as the first at 100 MHz, and HR's time-aligned 
profiles show how this comes about (properly aligned with a little smaller DM).  
Moreover, both D86 and WES exhibit the star's prominent correlated subpulse 
motion, showing that the profile is basically conal. 

None of this, though, accounts for the star's weak, highly linearly polarized
``postcursor'' component, which trails its main emission components by some
50\degr.  Please also note that W-04 shows that the PPA traverse under this
feature is nearly constant.  We could fit the RVM (see Table~\ref{tab5}) to the main pulse
however the postcursor emission could not be fitted with the same RVM.  The 
``flaring'' analysis did not yield any significant sporadic emission at the profile 
edges. The SG point appears to be coincident with the profile center as measured 
with respect to the 10\% outer widths. 

\noindent{\bf B1530--53}  has received no recent study, but indeed it appears
to present another good example of the ``partial cone'' emission envisioned
by LM.  Just as in the case of B0906--17 above, it shows a bright leading and
faint trailing component over a broad frequency band (HMAK, MHMA, MHM,
MHMb, vO97), and we see also some evidence for both a ``90\degr\ jump'' on
the leading edge and a steep rotation of the PPA across the middle of the profile.
The lowest frequency profiles (MHMb) suggest that the trailing component may
increase in relative strength at low frequency.  Overall, this pulsar's profile
seems to represent a very asymmetric conal double ({\bf D}) profile.

\noindent{\bf\em B1540--06}  exhibits an asymmetric single profile with a steep
rise and slow fall-off over a very broad frequency range (MHMb, MGSBT, MIS,
GL, KL) as also seen in Figure~\ref{figA4}.  Its linear polarization is small,
especially in the trailing part of the profile, and its PPA behavior disorderly and
inconsistent---both frequent properties of conal single ({\bf S}$_d$) profiles.  It
is the WES/WSE work, however, which provides evidence in the form of narrow
0.32-c/$P_1$ modulation features at both 92 and 21 cms.  GL's high frequency
profiles suggest a sweep rate of about --14\degr/\degr, and the constancy of its
profile width over at least four octaves suggests an inner cone geometry.

\noindent{\bf B1556--44:} This is a well studied southern pulsar, and the profile
has an asymmetric triple form around 21 cms. with a broad central component
and weak conal outriders (WMLQ and MHQ)---probably prompting L\&M to see 
it as a ``partial cone''.  At meter wavelengths its profile has a symmetric single
form (LM, MHMA, vO97), and at higher frequencies the central (putative core)
component is seen to be composed of two overlapping components (MHQ,
J-07).  The high frequency PPA traverse shows an orthogonal jump below the
conal components.  Nonetheless, using L\&M's $R$ value and the core width to 
constrain $\alpha$, we find that an inner cone {\bf S}$_t/${\bf T} geometry fits 
very well.  We did not find any ``flaring'' in the PS.

L\&M considered the star as an example where the SG point is located towards
leading side of the profile, and all the above profiles below 1 GHz show this PPA
curvature strongly.  Interestingly, the PPA histogram observed with the GMRT at
325 MHz, shown in Figure~\ref{figA4}, does not.  The PPA traverse for the central
and trailing components is very similar to that at higher frequencies, but shows
a non-orthogonal jump (by about 50\degr) below the leading component.  This
could result from OPM averaging, and the single pulses are not strong enough
to distinguish the modes for the conal components.  As a consequence no
reasonable RVM fit to the PPA swing was possible.  However, based on the
average PPA traverse, one can readily see the downward trend from the leading
to the trailing edge of the profile.

\noindent{\bf B1604--00}  has been studied extensively and can be observed
down to 50 MHz and up to at least 5 GHz (MGSBT, MHMA, MHMb, RSW, GL, 
HW, vO97, vH, MIS, W-99, KL, K-98, MM).  L\&M apparently regarded this pulsar
has having a ``partial cone'' profile because of the asymmetric slow rise and
steep falloff of its higher frequency profiles---\eg, see Figure~\ref{figA4}.  We
have earlier regarded this pulsar has being a triple ({\bf T}), but its profile does
not evolve in the usual manner (\cf, HR), and there is no strong indication to the
effect that the middle feature is a core component (R88).  Its profile evolution is
more suggestive of the conal triple class, as the central component's strength 
diminishes at low frequency and never dominates the profile.  Moreover, while 
the star's PSs exhibit no clear drift, its subpulses seem to show a kind of ``moding'' 
and a long period fluctuation feature (WES/WSE).  Overall, we can now best 
regard B1604--00 as having an inner-conal triple (c{\bf T}) profile, such that our 
sightline at meter wavelengths cuts close to the boundary between its two 
polarization-modal subcones.  It is then likely that the weak leading-edge
emission is associated with the outer cone as seen in other pulsars with similar
geometries [\eg, B0834+06 and B1919+21 (ET VII)].  As in Paper VI, the values
in Table~\ref{tab3} are taken from the mode-separated profiles in R88.

\noindent{\bf B1612+07}  has been observed over a broad band from 0.1 to 5 GHz,
and overall it exhibits a barely resolved two-component profile with the leading
component consistently brighter; see Figure~\ref{figA4} as well as GL, vH, W-99,
W-04, MIS, KL and MM.  It was this consistent asymmetry that probably caused L\&M
to regard it as a ``partial cone''.  Evidence for subpulse drift comes from D86 and
WES/WSE.  Moreover, the tendency for the star's low frequency profile to have
better resolved components further suggests a conal single ({\bf S}$_d$) evolution.

\noindent{\bf B1641--45:}  This bright, distant, southern pulsar has been observed
repeatedly, but at frequencies below 1 GHz its profiles are corrupted by scattering
(MHMA, MHM, MHMb, vO97).  Only in the 1.4/3.1-GHz profiles of KJ do we begin
to see some profile structure, but the conal outriders are far from clear, and the
PPA traverses are impossible to decipher.  However, the 8.4-GHz polarized profile
recently measured by JKW clarifies matters completely.  Here we see that the PPA
traverse is essentially central, and the outside dimensions of the outrider pair can
be reliably determined.  This is the basis of the inner-cone {\bf S}$_t$ geometry 
determination in Table~\ref{tab3}.

\noindent{\bf B1648--42:}  Only two observations (vO97, WMLQ) are available
for this wide profiled pulsar, and both show two components with a prominently
steepening PPA traverse---indeed, probably it was on this basis that L\&M came
to regard the pulsar as a ``partial cone''.  Here, we have no basis to decide on
whether a trailing portion of the profile is ``missing'', or whether the profile is
complete as it is.  In either case a simple geometric model can be assembled
to suit the situation:  in the first case, probably a c{\bf T} would be invoked, and in
the latter situation a conal double ({\bf D}) configuration.  Table~\ref{tab3} gives
values for the latter case.

\noindent{\bf\em B1700--18:}  In addition to our Figure~\ref{figA5}, profiles have
been published for this somewhat weak pulsar only by GL, S95 and MM.  The star's
asymmetrical single profile undoubtedly accounts for its ``partial cone'' status
in L\&M's effort.  The strongest evidence, however, comes from WSE, who find
drift-associated modulation with a $P_3$ of about 3.5 $P_1$ as well as a strong
low frequency modulation feature.  The star's profile must then be of the conal
single type, and indeed, many such profiles are quite asymmetric.  We find some
``flaring'' in the star's PSs as seen in Fig.~\ref{figA5}, and either this pattern or
the average profiles can be used to obtain a half-power width of about 12\degr.
Similarly, an RVM fit to the PPA traverse yields 
a maximum sweep rate of --8.4\degr/\degr\ (see Table~\ref{tab5}).  The SG point 
lags the center of the outer conal ``flared'' profile by about 1.7\degr---apparently 
due to A/R---giving a very reasonable radio-emission height of 279 km.

\noindent{\bf\em B1732--07:}  Figure~\ref{figA5} gives our 325-MHz GMRT profile,
and other published observations are available from GL, vH, J-07, and S95.  At
meter wavelengths the star's profile is somewhat asymmetric, and perhaps this
is why L\&M regarded it as a possible ``partial cone''.  In fact, there can now be
little doubt but that this star has a triple ({\bf T}) profile with a central core component.
WES/WSE find no evidence of conal modulation features, the star's PPA traverse
is highly central, and $\alpha$ can be estimated from the core width.  Significant
``flaring'' can be seen in Fig.~\ref{figA5} that appears to coincide with the three
profile components.  All these circumstances square in the outer conal geometry
of Table~\ref{tab3}.  The midway point of the ``flared'' profile leads the peak of the 
central core component by 2.3\degr. This gives an A/R conal emission height of 
205 km with respect to the core (see G\&G).

\noindent{\bf\em B1742--30:}  This pulsar's geometry has long presented something
of a mystery---and indeed it appears to have been to L\&M who listed this star as
a possible ``partial cone''.  Several of the older published profiles (MHMA, vO97,
XRSS) do not show its full extent, but the long, weak trailing portion is visible in
all of LM's observations, that of WMLQ, and the GMRT 325-MHz polarimetry of
Figure~\ref{figA5}.  Nor is it easy to interpret the PPA rotation across the various
profiles, but apart from several ``90\degr\ jumps'' and the ``hat'' above the bright,
central component, one can interpret the traverse as basically flat and central.
The trailing part of the merged main feature thus appears to be a core, with two
components preceding it and the two trailing components merged in the long
``tail''---reminiscent of B1237+25 in its ``abnormal'' mode.  Fig.~\ref{figA5} further
shows most of the five components in the ``flaring'' analysis, and here the core
appears independently enough to measure its half-power width.  With all this
information and interpretation, the double cone/core geometry of the pulsar is
assembled quantitatively in Table~\ref{tab3}.  Our RVM fits to the PPA does not
constrain the SG point at all, as the PPA traverse 
was essentially a linear slope. We assume that the SG point is close to the 
peak of the central core component, and hence can quote and error for the PPA 
offset (see Table~\ref{tab5}). An A/R height can be estimated 
for B1742--30's outer cone with respect to its central core feature (see G\&G).  
The outer cone's midway point the core peak by some 1.14\degr, yielding an 
emission height of 87 km.

\noindent{\bf\em B1745--12:} The pulsar has been observed at several frequencies
by GL, XRSS, S95 and MM.  The pulsar's highly asymmetric profile at meter wavelengths
surely led L\&M to see it as a possible ``partial cone''; however, three (or possibly
four) components can be discerned at higher frequencies.  The GMRT observation
at 325 MHz in Figure~\ref{figA5} clearly shows two components as well as a long 
weak trailing ``tail''.  The PS polarization is weak, but the average shows that the 
PPA traverse is flat under the leading components (with an OPM ``jump'' ) but 
steepens to an SG point near the middle of the overall profile center.  We have
fitted the OPM-corrected average PPA traverse to the RVM (see Table~\ref{tab5}), and the SG point is the
longitude origin in the figure.

A search for ``flaring'' in the trailing portion of the profile identified 14 occasions
as indicated by the green curve in Fig.~\ref{figA5}.  No evidence for periodic
modulation was found in the PSs, and indeed the WES/WSE fluctuation spectra
are completely featureless.  We cannot then be sure about this star's classification,
but the weak spreading of its outer conal components appear to reflect an outer
cone, and this dimension together with the resultant --11\degr/\degr\ sweep rate
provide a basic quantitative geometry in Table~\ref{tab3} which seems compatible
with the available observational evidence.  Also, the SG point lags the midway 
point between the outer conal component pair of the ``flared'' profile by 2.6\degr\ 
giving an A/R emission height for the outer cone of 215 km.   

\noindent{\bf\em B1756--22:}  Apart from GL's five polarimetric profiles, only the
1.4-GHz total power observation of S95 has been published; and both spectra
of WES/WSE are entirely featureless.  The PPA traverse does seem to be nearly
flat, and the width of the bright putative core component roughly compatible with
an orthogonal magnetic colatitude.  Putting this interpretation into Table~\ref{tab3},
we find that it is compatible quantitatively with an inner cone/core ({\bf S}$_t$) 
emission geometry.

\noindent{\bf B1822--09:}  With both its interpulse and precursor component, (see 
Figure~\ref{figA6}), this pulsar's geometry has been debated actively since near 
the time of its discovery.  A great many published studies are available (MHMA, 
MHM, SVS, vO97, MGSBT, MHMb, GL, vH, MIS, KL, K-98, X-95, KJ, J-07, WES/WSE), 
and surely L\&M had adequate reason by virtue of its apparent main-pulse asymmetry 
to view it as a ``partial cone''.  However, we take the position that the precursor and 
main pulse are separate entities (Backus \etal\ 2010), so our geometric analysis here 
follows that in this study and applies only to the main pulse.  Our PS analyses of it 
indicate that this structure represents an inner cone/core triple ({\bf T}) profile with 
a nearly central sightline trajectory. An RVM fit to the PPA appeared to be rather
complicated for this pulsar, however based on Backus \etal's (2010) fig. 8, we argue 
that the SG point under the main pulse is consistent with being close to the profile 
center.

\noindent{\bf\em B1842+14:}  Figure~\ref{figA6} gives our 327-MHz Arecibo profile,
and many other published observations are available (RSW, vH, GL, HR, MIS,
W-99, J-07, MM, WES/WSE).  Again, it is not clear what caused L\&M to regard this
star as a possible ``partial cone'', but its flat PPA traverse and steep upturn (see
W-99 and J-07) might well have suggested that a further trailing component was
missing.  Perhaps.  The core-single evolution and quantitative geometry fits the
pulsar rather well (ET VI)---that is, apart from the unusually flat initial PPA, too
sharp upturn, and disparity between the two putative conal components above
1 GHz.  The delayed upturn might be caused by an A/R shift, but this idea does
not seem to fit.  One other possibility is that the flat PPA represents emission
from a highly polarized precursor, and that the remaining parts of the profile
represent a core-single structure.  However, without quality higher frequency
profiles to draw on, this possibility cannot be evaluated.  Therefore, we retain
the first model in the table.

\noindent{\bf\em B1851--14:}  For this pulsar we have little to go on apart from the
few published profiles (L90, XRSS, GL, WSE) and our GMRT 325-MHz profile
in Figure~\ref{figA6}.  Our ``educated'' guess is that the profile is of the conal
single ({\bf D}$_d$) type, and the quantitative geometry in the table is compatible 
but not well constrained.  Some confirmation might come from the fluctuation 
spectra,but both our own and that of WSE are featureless.

\noindent{\bf (B1859+07):}  This pulsar was not among the L\&M ``partial
cone'' grouping, but probably would have been included had they known of it.
Its asymmetric profile is subject to occasional ``events'' during which emission
in single pulses moves to earlier longitude (see RRW).  Otherwise, we found
few published references to this pulsar (GL, W-04, WSE), which unfortunately
provide little further insight.  The quantitative geometry in Table~\ref{tab3} is
taken verbatim from RRW.

\noindent{\bf\em B1900+05:}  Again, this pulsar's asymmetric profiles (MGSBT,
RSW, GL, W-99, WES/WSE), also Figure~\ref{figA6}, probably encouraged L\&M 
to regard it as a possible ``partial cone''.  Beyond this, there is little clear evidence 
to go on.  Both Paper VI and W-99 classified it as a core-single star, but no core 
signature can be discerned in any of the existing profiles.  This classification still 
seems likely, but the profile width---and thus the geometrical model---in Paper 
VI are incorrect.  We have repaired this error above in Table~\ref{tab3} on the 
basis of revised estimates. 

\noindent{\bf\em B1907--03:}  Here the published studies clearly show a core/cone
triple ({\bf S}$_t$) profile at 21 cms. (GL, S95), whose core is even marked by 
sign-changing circular polarization, and a single profile at 408 MHz (L90).  Our 
GMRT 325-MHz profile in Figure~\ref{figA7}, unfortunately, is useless owing to its 
distortion by scattering, and no useful information comes from the fluctuation 
spectra in WSE.  Most observations suggest a flat, central PPA traverse, and this 
together with the profile dimensions fixes an inner cone/core geometry quantitatively.

\noindent{\bf B1910+20:} The pulsar was classified as a ``partial cone'' by L\&M
with the SG point lying towards the trailing edge of the profile.  Average profiles
at 610 and 1410 MHz (GL) show a strong leading and weak trailing component.
These PPA traverses appear complex such that no clear interpretation can be
made.  However, the 1.4-GHz profiles of W-99, RSW and this paper, Figure~\ref{figA7},
show the full PPA behavior in some detail.  The latter in particular is complex and
cannot be described by the smooth RVM curve, but the expected underlying `S'
shape is evident---and our efforts to fit the RVM to the PPM traverse 
yield the SG point with resonable precision (see Table~\ref{tab5}).

The available pulse-modulation studies (D86, WES/WSE) leave no doubt that
the profile represents a double conal structure:  the outer conal components show
a fairly regular stationary modulation with a $P_3$ of about 2.7 $P_1$, and we see
evidence (\eg, DHCR) that this modulation is shared by the inner conal components
as well.  We have found no evidence of flared emission at the profile edges.  Our
LRF spectra in Fig.~\ref{fig2} yields signatures of the 2.7-$P_1$ $P_3$ modulation
in the outer conal component pair.

As no clear signature of a core component can be seen in any profile, its width
cannot be determined.  We do see hints of core activity including antisymmetric
$V$ in some profiles, but overall we cannot resolve whether this profile is of the
{\bf M} or c{\bf Q} type.  However, using the PPA fit above and the profile dimensions 
from Paper VI, a slightly revised quantitative model of the emission geometry can
be found in Table~\ref{tab3}.  Further our measurements show that the midway point 
leads the SG point by only 0.5\degr, giving an emission height of about 228 km.

\noindent{\bf\em B1913+10:}  Little more can be said about this pulsar's geometry
than was possible in Paper VI.  The 4.85-GHz profile is so poorly resolved that
no structure can be seen, and the 400-MHz profiles have scattering ``tails''.  The
recent, well measured profiles of J-07, W-99 and Figure~\ref{figA7} resolve a 
feature on the profile's trailing edge at 1.4 GHz that becomes very pronounced 
at 3.1 GHz.  The one available LRF (WSE) is featureless.  It still may be that this 
is a core-single star, but no geometrical solution bears this out.  The two resolved 
components at 3.1 GHz cannot be interpreted as a conal outrider pair:  their 
outside dimension is much too small for them to be an inner cone.

\noindent{\bf B1915+13:}  In slighly poorer observations this pulsar exhibits only
a single narrow Gaussian-shaped component, but when resolved optimally it
has an unresolved feature on its trailing edge.  This structure is clearly seen in
the 1.4-GHz profiles of BCW, EW, W-99, HR and Figure~\ref{figA7}---and these
and many other observation also show an accelerating PPA rotation such that
the SG point falls far on the trailing edge of the profile (GL, RSW, RB, vH)---and
very like that of B0540+23 above where the shift increases with wavelength.
PPA fits by BCW and EW as well as ourselves at 1.4 GHz consistently show that
the SG point falls far on the trailing edge of its profile, and GL's lower frequency
profiles suggest even greater displacements, such that A/R effects provide a
natural explanation.  The star's fluctuation spectra are featureless (WES/WSE),
and the weak ``flaring'' in the above figure does not seem indicative of conal
emission.  Interestingly, the pulsar has been detected down to 100 MHz (KL,
MIS).  For all these reasons we classify this star as having an arSt profile.

\noindent{\bf B1924+16:}  This pulsar exhibits a single component with a long
slow rise on its leading edge.  The published profiles give a mixed impression
regarding the curvature of PPA traverse (RB, BCW, RSW, GL, vH, W-99), but the
fits by BCW and ourselves in Figure~\ref{figA8} concur in showing a slight upward
acceleration and thus placing the SG point toward the trailing edge of the profile
(the longitude origin in the above figure).  Weak indication of a long (about 60
$P_1$ modulation) is seen in the fluctuation spectra (WES/WSE), but overall
there is little indication of conal activity.  On this basis we designate the pulsar
as having an arSt profile. The RVM fit to the PPA yields the SG point with resonable
accuracy (see Table~\ref{tab5}).  The midway point of the profile calculated 
using the outer peaks of the ``flared'' profile leads the SG point by roughly 
4.1\degr, giving an A/R height of about 506 km.   

\noindent{\bf\em B1930+22:}  This fast, highly dispersed pulsar is difficult to 
observe at lower frequencies (although MM report a 100-MHz detection), 
and only the 1.4-GHz profiles of GL, BCW, W-99  and the AO observation in 
Figure~\ref{figA8} fully show its fast rise and slower falloff.  Several of the 
observations suggest an upwardly curved PPA traverse, and our RVM fit
places the SG point towards the trailing side of the profile.  The significance of this 
placement is not yet clear:  One might attribute this configuration to A/R, 
however, the ``flaring'' on the two edges of the profile suggests some conal activity there.
Possibly A/R does shift some high-altitude core emission earlier so as to 
overlie the leading conal feature, but new high quality observations are 
needed at lower frequencies to assess this possibility.  We then retain the 
{\bf S}$_t$ designation of Paper VI but amend it to show the probable role of 
abberation/retardation.  Using both the ``flared'' and the average profiles, the 
center of the conal peaks lead the SG point by 8.3\degr. This shift provides an 
A/R-height estimate of 250 km for the conal emission.

\noindent{\bf B1937--26}  shows a consistently asymmetric profile that prompted
L\&M to regarding it as a ``partial cone'' (GL, WMLQ, vO97, vH, MHQ).  A bright
leading and weak trailing feature are seen in all the star's profiles, but in several
of the higher frequency profiles (including the J-05 that is best resolved), we see 
a suggestion of a third feature on the leading edge.  Further, the fluctuation
spectra (WES/WSE) suggest conal emission as does the ``flaring'' in our GMRT
325-MHz observation in Figure~\ref{figA8}.  We cannot then resolve just how the
profile should be classified, but using the fitted PPA sweep rate and conjecturing
that the high frequency profile width reflects the core width, the geometry in the
table is compatible quantitatively with an inner cone.  RVM PPA fitting to this 
shallow PPA traverse does not constrain the SG point well enough (see Table~\ref{tab5}).  
Nonetheless, using the ``flared'' and average profiles 
the midway point of the profile leads the SG point by 0.8\degr\ giving an A/R height 
estimate of about 70 km.  

\noindent{\bf\em B1944+17:}  The pulsar's main-pulse profile (see Figure~\ref{figA8})
superficially resembles some of the conal ``partial cone'' objects we have identified 
above (\eg, B1540--06) with weak emission on their trailing sides (\cf, HR), but the 
detailed published studies leave little doubt that this star is correctly classed as 
having a conal triple/quadruple (c{\bf T}/c{\bf Q}/) profile.  The pulsar has several 
modes, some with orderly drift (Deich \etal\ 1986; WES/WSE), and these together 
with its $\gtrsim$60\% null pulses make its profiles somewhat unstable (MHMA, 
MHM, MGSBT, RSW, vO97, vH, W-99, MIS).  Its shallow, linear PPA rotation and 
orderly profile evolution (\eg, HR) further support this understanding of its emission 
geometry.  Indeed, in a recent study by Kloumann \& Rankin (2010) the pulsar's 
geometry has been analyzed in detail;  the values in Table~\ref{tab3} are taken 
from this work. Our RVM fittig results are given in Table~\ref{tab5}.

\noindent{\bf\em B1944+22:}  The two existing AO profiles (RB, W-99) of this weak
pulsar reveal only that it has two unresolved components---much as seen in 
Figure~\ref{figA9}---the second of which is much weaker.  The profile is almost 
certainly conal, and thus its behavior is very likely akin to that of the many conal 
single (or inner-cone double) stars with weak or missing trailing emission.  If the 
W-99 PPA rate is reliable, then we can easily compute a model for the star's 
geometry as in Table~\ref{tab3}.

\noindent{\bf B2021+51:}  This bright pulsar has been studied for many years,
and most evidence points to its having a conal single ({\bf S}$_d$) profile that 
shows the characteristic low frequency bifurcation with a much weaker leading 
component (M71, MGSBT, vH, GL, X-95, KL, MM, K-98); see especially K-98.  This 
behavior is thus very similar to that of B0809+74 (\eg, RRS).  Both its SVM PPA 
traverse and subpulse-drift modulation (ETIII; WES/WSE) are also largely 
compatible with this understanding.  Interestingly (and unusually) the pulsar's 
leading edge emission is fully linearly polarized at meter wavelengths (as is 
B0809+74's at frequencies above 1 GHz) such that the one active polarization 
mode here must be completely linearly polarized.  Note in Figure~\ref{figA9} 
that both OPMs are active only under the trailing component.  In Table~\ref{tab3} 
we revise slightly our earlier emission model in Paper VI:  surely the pronounced 
conal spreading seen in K-98 argues for an outer-cone geometry. In Table~\ref{tab5}
we give the RVM fitted parameters, and the SG point seems to be well constrained. 
The ``flaring'' analysis for this star 
shows only weak, occasional emission on the profile edges.  These do provide 
a means of estimating the profile's midpoint---which leads the fitted SG point by 
about 1.1\degr---giving an A/R emission-height estimate of 113 km.   

\noindent{\bf\em B2043--04:}  The published profiles of this pulsar all show a
symmetrical single form with a ``soft'' leading edge (GL, vO97, MIS).  It is thus
not very clear why L\&M regarded it as a possible ``partial cone''.  The 325-MHz
GMRT profile in Figure~\ref{figA9} is the only one that it well enough resolved 
to indicate two features as well perhaps as ``flaring'' on its leading edge.  There
can be little doubt that the profile is conal, probably an inner cone {\bf S}$_d$
with occasional outer conal subpulses on the far leading edge (as seen in some
other such pulsars, \eg, see B1604--00).  This understanding is corroborated
strongly by WES/WSE's analyses showing a strong narrow fluctuation feature
at 0.37 c/$P_1$ that is clearly indicative of subpulse drift.  The SG point can be 
fixed by an RVM PPA fit (see Table~\ref{tab5}) as shown in Fig.~\ref{figA9}. The 
midpoint of the outer conal peaks of the ``flared'' profile coincides with the SG 
point within the measurement errors, suggesting that A/R is not significant in 
this slow pulsar.  

\noindent{\bf B2053+36:}  This pulsar's asymmetric single profile seems to have
been the reason for L\&M's ``partial cone'' categorization (RSW, GL, W-99).  And,
indeed, it is also problematic from our perspective.  Its flat segmented PPA traverse
is unusual and apparently indicative of a central sightline traverse, even if the
``jump'' is due to an OPM dominance transition as indicated in Figure~\ref{figA9}.
Entirely conal profiles are rare in pulsars with such a short period, but we see no
hint of core action.  Moreover, WES/WSE make a claim for subpulse drift without
direction!  However, they find no consistent behavior at their two frequencies.  No
consistent geometrical model can be computed from the available information.

\noindent{\bf B2217+47}  exhibits a somewhat asymmetrical profile over a broad
band, and this apparently led L\&M to see it as  a ``partial cone'' (GL, SVS, MIS, MM).
Our GMRT 325-MHz profile in Figure~\ref{figA10} shows a similar form.  We see
some hint of conal outriders at 21 cm (K-98 and MGSBT)---and these appear to
dominate the profile at 4.9 GHz (SRW)---arguing strongly that the star is a member
of the core-single ({\bf S}$_t$) class.  Interestingly, WES/WSE find some evidence 
for systematic subpulse motion at 1.4 GHz, but without a fuller study their result is 
hard to interpret.  These results then provide the needed information to construct 
the quantitative emission model given in Table~\ref{tab3}.

B2217+47, however, shows further unorthodox behaviors that need further study.
Downs (1979) found that there was a strange truncated-exponential baseline 
emission that decayed after the pulse, and MGSBT's profiles were not sensitive 
enough fully confirm or refute it.   Moreover, SS find that the star has a postcursor 
feature was variable in its intensity and position over a few years.  Our search however did 
not show such a feature in our data.

\noindent{\bf B2224+65m:}  This pulsar has two well separated Gaussian-shaped
components (MGSBT, GL, LM, vH, K-98, KL, MM) as seen in Figure~\ref{figA10}.  
The trailing one, however, has a flat PPA, apparently causing L\&M to see the profile
as a ``partial cone'' with a missing leading component.  Indeed, we classified it as 
having a T$_{1/2}$ profile in Paper VI.  Clearly, we must now view the fully linearly 
polarized trailing component as being a ``postcursor'' feature, and the much less 
polarized leading component as a ``main pulse'' in its own right.  The PPA fit 
(see Table~\ref{tab5}) in the figure does 
seem to fit both features well, but their separation is large---some 35\degr.  This said, 
we can only estimate $\alpha$ from the width of the putative core main pulse.  We 
see no hint of conal outriders, and WES/WSE report featureless fluctuation spectra.  
We note that vH finds this pulsar similar to B0355+54; however, we see no evidence 
of the A/R which is prominent in that star's PPA traverse. The SG point in 
this pulsar is consistent with being coincident with the profile center of the
main pulse.

\noindent{\bf B2327--20:}  As shown in Figure~\ref{figA10} the pulsar has a triple
profile with a weak trailing component.  This is much clearer in the well resolved
GMRT profile than in many of the published ones (MHMA, MHMb, MGSBT, vH,
vO97, GL, CMH), especially at high frequency where the weak component can
be gleaned only as an inflection on the trailing edge.  Only J-07's 691-MHz
profile provides comparable clarity.  Clearly, this star became one of L\&M's best
examples of ``partial cone'' profiles.  The remaining question is whether the star
has an entirely conal triple profile or a core-cone triple one, and this question is
difficult to fully resolve.  However, the intensity dependence of the central feature
and its aberrant PPA behavior tilts in favor of it being a {\bf T} pulsar.  WES/WSE
find a 50-$P_1$ feature shared by both the leading and middle components that
could be null-related, whereas the weaker 0.39-c/$P_1$ modulation seems to be
present only in the leading component.  Apart from the now much better measured
PPA rate, taken from the fit in the figure, the quantitative geometry in Table~\ref{tab3}
follows the earlier analysis in Paper VI.  Our RVM fit to the PPA traverse yields 
well determined SG point (see Table~\ref{tab5}) and the center of the outer conal peaks 
coincide with the SG point within the measurement errors. This behavior is consistent 
with other slow pulsars showing little or no A/R effect.  

\end{document}